\documentclass[11pt,twoside]{article}
\usepackage{rotating}
\usepackage{epsfig}
 
\usepackage[latin1]{inputenc}
\usepackage[T1]{fontenc}
\usepackage{graphics,vmargin,french}
 
\newcommand{\U}[1]{\mbox{#1}}                                                   
\newcommand{\UU}[2]{\mbox{$\mbox{#1}^{\mbox{\scriptsize #2}}$}}                 

\newcommand{\Uk}[1]{\mbox{k#1}}                                                 
               
\newcommand{\UM}[1]{\mbox{M#1}}                                                 
               
\newcommand{\UG}[1]{\mbox{G#1}}                                                 
               
\newcommand{\UT}[1]{\mbox{T#1}}                                                 
               
\newcommand{\Uc}[1]{\mbox{c#1}}                                                 
\newcommand{\UUc}[2]{\mbox{$\mbox{c#1}^{\mbox{\scriptsize #2}}$}}               
\newcommand{\Um}[1]{\mbox{m#1}}                                                 
               
\newcommand{\Uu}[1]{\mbox{$\mu$#1}}                                             
           
\newcommand{\Un}[1]{\mbox{n#1}}                                                 
               
\newcommand{\Up}[1]{\mbox{p#1}}                                                 
               
\newcommand{\Uf}[1]{\mbox{f#1}}                                                 
               
\newcommand{\di}{\displaystyle}

\newcommand{\text}[1]{\mbox{\rm #1}}
\newcommand{\dd}{\mbox{d}}
\newcommand{\boul}{$\bullet$}
\newcommand{\bc}{\begin{center}}
\newcommand{\ec}{\end{center}}
\newcommand{\be}{\begin{equation}}
\newcommand{\ee}{\end{equation}}
                                               
\newcommand{\e}{\text{e}}

\begin{document}

~

\bc \bf \large Acc\'el\'eration  de Particules dans un Plasma Excité par un Laser \ec

\bc D.~Bernard \ec

\bc Laboratoire Leprince-Ringuet
 Ecole Polytechnique, IN2P3 \& CNRS, 91128 Palaiseau, France \ec


\vspace{12cm}

Mémoire de ma thèse d'habilitation, soutenue le 11 mai 2005 à Orsay.

\vspace{2cm}

\begin{tabular}{lllllll}
Jury : \\
 & François Amiranoff \\
 & Brigitte Cros (Invitée) \\
 & Terry Garvey \\
 & Tom Katsouleas \\
 & Wim Leemans \\
 & Gérard Mourou \\
 & François Richard (Président)
\end{tabular}

\vspace{2cm}

\mbox{\tiny denis.bernard@in2p3.fr}

 \clearpage

La physique des particules utilise des accélérateurs d'énergie
croissante pour pouvoir engendrer des particules de masse de plus en
plus grande.
L'énergie perdue par rayonnement synchrotron dans un anneau 
  variant comme $\gamma^4 /\rho$, où $\gamma$ est le facteur de
 Lorentz de la particule
et $\rho$ le rayon de courbure de l'anneau, le budget énergétique devient rapidement prohibitif :
 pour les électrons, LEP atteint certainement la limite.
Des particules plus massives sans constituants internes comme les
muons peuvent être employées.  Leur utilisation pose des problèmes
(production, durée de vie) en cours d'étude.
Les protons sont composés de partons ayant une densité de probabilité
de porter une fraction $x$ de l'impulsion du proton, s'étalant
largement sur le segment $[0,1]$, le spectre de luminosité
différentielle en fonction de $\sqrt{s}$ est extrêmement large, effet
délétère pour le potentiel de physique de la machine, ainsi que les
produits des interactions des partons spectateurs.

Ainsi, pour les futurs accélérateurs, 
on revient donc aux électrons, et on s'oriente vers un collisionneur
linéaire multi-TeV.  Le ``gradient'', le 
champ électrique longitudinal maximal disponible
dans une cavité accélératrice est de l'ordre de $0.1$ GV/m :
Un collisionneur multi-TeV devient donc long, à un prix du 
``mètre linéaire'' élevé : à nouveau prohibitif.

On recherche donc de nouvelles méthodes de production de champ
électrique longitudinal, visant un gain de plusieurs ordres de
grandeur sur le gradient.

Un laser peut  générer des impulsions lumineuses
cohérentes ultra-courtes et ultra-intenses, dont le champ électrique est
extrèmement élevé.
Mais le champ est transverse à la direction de propagation.

Nous allons examiner dans la suite de ce mémoire l'emploi d'un {\bf
 plasma} comme {\bf transformateur} pouvant porter une onde plasma
 électronique de champ électrique longitudinal, sous l'excitation
 d'une impulsion lumineuse de champ électrique transverse.
Un plasma, milieu ionisé, n'est pas soumis aux limitations d'intensité
de champ des milieux moléculaires, comme la surface d'une cavité.

\clearpage

Après l'article fondateur de Tajima et Dawson \cite{tajima} en 1979,
plusieurs expériences ont été conduites  
 dans le but de démontrer le principe de
l'accélération de particules dans un plasma excité par un laser. 
%
%
Ces expériences de première génération, initiées à la fin des années
80, ont atteint leur objectif au cours des années 90, avec des gains en énergie de quelques MeV.

La communauté s'est ensuite orientée vers la production de bouffées  d'électrons 
courtes par le déferlement d'une  onde plasma  intense, dans un plasma dense
($\approx 10^{19} e^-/\text{cm}^3$), créée par une impulsion laser courte et très intense.
Ces résultats ont été obtenus dans de  nombreux laboratoires, avec
un spectre  thermique, et une queue à haute énergie qui peut
s'étendre à quelques centaines de MeV.
Au cours de l'année 2004, trois groupes ont découvert la possibilité
d'obtenir un pic raisonnablement mono-énergétique, à une énergie
typique de l'ordre de 100 MeV, avec une largeur relative de quelques
pour-cent, par l'ajustement fin des paramètres expérimentaux.

D'autre part,
une collaboration utilisant le faisceau de 30 GeV de SLAC étudie
l'accélération de particules dans une onde plasma créée par un paquet
de particules (PWFA).
Cette même année, un gain en énergie de 4 GeV a été observé,
franchissant la barrière du GeV rêvée depuis longtemps (en fait la
tête du paquet crée l'onde plasma, et ralentit, et la queue du paquet est
accélérée par cette onde).

Une réflexion est en cours pour développer des expériences de deuxième
génération d'accélération de particules dans un plasma excité par un
laser visant à démontrer la faisabilité d'une ``cavité'' à plasma,
avec en particulier des performances quantitatives précises, non
seulement quant à la valeur du champ longitudinal, mais aussi la
conservation des propriétés d'un faisceau, par exemple son émittance,
sa dispersion en impulsion, la longueur du paquet, la charge limite.

~

Je me propose, dans le cadre de ce mémoire d'habilitation, de faire le point
de l'acquis de la décennie passée, et de présenter  quelques pistes qui
peuvent \^etre explorées dans le futur.
\begin{itemize}
\item
Après une présentation rapide du principe de 
l'accélération de particules dans un plasma excité par un laser
(section \ref{sec:intro}), je décris les expériences réalisées à
ce jour, m'attardant plus particulièrement sur leurs limitations et sur les
enseignements qu'elles nous ont apportés (section \ref{sec:prems}) .
\item
J'explore ensuite ce que pourrait \^etre une expérience de seconde
génération, sans prendre en compte de fa\c{c}on explicite les contraintes
d'un accélérateur (section \ref{sec:deux}) .
\item
Je présente enfin les quelques efforts récents de conception d'un
accélérateur multi-TeV utilisant une ``cavité à plasma'' (section \ref{sec:multitev}).
\end{itemize}

Un appendice (Sections \ref{sec:las}--\ref{sec:transverse}) 
donne une présentation plus détaillée de
l'accélération de particules dans un plasma excité par un laser.

Au fil du texte, il sera fait mention de mes travaux dans l'expérience
d'accélération laser de particules de l'Ecole polytechnique, et en
particulier : 
\begin{itemize}
\item  de l'injection et du contrôle du faisceau à accélérer,  analyse en impulsion 
du faisceau accéléré  \cite{nim95}, dont j'avais la responsabilité,
\item de l'observation de l'accélération par battement d'ondes  \cite{prl95},
\item de l'analyse et simulation de l'expérience d'accélération par sillage \cite{prl98}.
\end{itemize}

La présentation des différentes contraintes qui limitent l'accélération en limite du mode linéaire
(section \ref{subsec:dim}) est basée sur une de mes publications \cite{kyoto_one}.

\section{Introduction}\label{sec:intro}

Un laser est un outil de choix pour générer un champ
électrique intense. 
L'amplitude du champ\footnote{La plupart des variables sont
 utilisées avec leur sens habituel~: celui-ci n'est pas toujours
 rappelé. Un tableau récapitatif peut \^etre trouvé dans la section \ref{sec:tableau}.}  est 
 ${\cal E}=\sqrt{2I/\epsilon_{0}c}$ o\`u
 $I$ est l'intensité lumineuse.
Pour un dispositif existant actuellement, 1 J, 25 fs, 40 TW, focalisé
sur $(20\mu\text{m})^2$, l'intensité vaut $10^{19} \text{W/cm}^2$ et le champ
10~TV/m.
Mais en champ lointain\footnote{A une distance de tout dioptre plus grande que quelques longueurs d'onde.}, le champ électrique d'une onde
électromagnétique est {\bf transverse}, $\vec{{\cal E}}.\vec{k}=0$, ce
qui fait qu'une particule voyageant parallèlement au champ sur le même
axe que l'impulsion laser ne peut rester en phase avec le champ
pendant plus d'une demi-longueur d'onde, soit une fraction de micron.

%


Un plasma peut \^etre utilisé comme un {\bf
transformateur}, le laser générant une onde plasma {\bf longitudinale}
--- modulation de la densité
électronique du plasma --- dont le champ électrostatique associé sera
utilisé pour l'accélération de particules chargées.

De plus, les  oscillateurs harmoniques que constituent
différentes régions d'un plasma froid n'étant pas couplés,
l'onde plasma a simplement pour vitesse de phase la vitesse de groupe
du laser dans le plasma, proche de la vitesse de la lumière.
\footnote{Lorsque la densité électronique est beaucoup plus petite que la densité critique, ce qui est le cas ici.}
Le champ électrique produit vaut~(eq. \ref{eq:Ez:delta}, en annexe) :
\begin{equation}
E_{z,max} = \frac{2\pi mc^{2}}{e\lambda_{p}}\delta
\end{equation}
soit $E_{z,max} =3.2 ~\delta ~\text{MV}/\lambda_{p}$, 
o\`u $\lambda_{p}$ est la longueur d'onde de l'onde plasma et o\`u
$\delta \equiv \delta n/n$ est l'amplitude relative de la modulation de la
densité électronique $n$ du plasma.
On préfèrera donc une longueur d'onde plasma courte.

Cependant, les électrons à accélérer ont, en général, une
vitesse différente de la vitesse de phase de l'onde plasma.
Après un certain parcours, un électron initialement dans la
partie accélératrice de l'onde va alors se retrouver dans la partie
décélératrice.
Ce parcours est appelé longueur de déphasage et noté
$L_{\varphi}$.
Pour un faisceau de particules ultra-relativistes, $L_{\varphi}$ vaut
$L_{\varphi}=\gamma_{p}^2\lambda_{p} = \gamma_{p}^3\lambda$, o\`u 
 $\gamma_{p}=\lambda_{p}/\lambda$ est le facteur de Lorentz de la
 vitesse de phase de l'onde plasma. 
L'emploi d'un plasma de petite longueur d'onde plasma, associée à
une densité électronique $n = \pi/(r_{_e} \lambda_{p}^2)$
élevée --- $r_{_e}$ est le rayon classique de l'électron ---
conduirait donc à une longueur d'accélération extr\^emement
courte.
Le gain d'énergie maximal est alors proche de 
$e L_{\varphi} E_{z,max}\propto \gamma_{p}^2 \propto 1/n$,
ce qui à nouveau incite à l'emploi de plasma peu dense.
%
%
Pour $\lambda=1\mu\text{m}$, $\lambda_{p}=100\mu\text{m}$, correspondant à
$n\approx 10^{17}\text{cm}^{-3}$, on obtient 
$E_{z,max}= \delta\times 32~\text{GV/m}$, $L_{\varphi}=1\text{m}$, 
et pour une onde plasma d'amplitude
$\delta =1$, un gain en énergie maximal par étage d'accélération
$eL_{\varphi}E_{z,max} = 32 ~\text{GeV}$.

~

L'excitation de l'onde plasma est obtenue par l'action de la force
pondéromotrice du laser, moyenne de la force de Lorentz sur un cycle
optique, sur les électrons du plasma. Un électron soumis à un champ
électromagnétique d'amplitude constante a une trajectoire fermée, en forme
de ``8'' pour une lumière polarisée linéairement, circulaire pour une
polarisation circulaire, et son énergie moyenne n'est pas modifiée.
Par contre, lorsque l'intensité incidente varie dans l'espace ou dans le
temps, (disons qu'elle diminue) on comprend que pour chaque cycle optique, la
force de Lorentz dans la seconde moitié de la trajectoire n'est pas suffisante
pour ramener l'électron à son point de ``départ''.
Ce déplacement, moyenné sur un cycle, est dû à une force, appelée
fort poétiquement pondéromotrice.
La force pondéromotrice est proportionnelle au gradient de
l'intensité, et opposée à celui-ci~: les particules chargées
du plasma fuient les zones de haute intensité laser.
Le déplacement des électrons du plasma sous l'action de la force
 pondéromotrice produit des sur-densités et des sous-densités
 électroniques locales qui oscillent ensuite sous forme d'onde
 plasma.
On obtient alors deux effets~:
\begin{itemize}
\item la variation temporelle de l'intensité d'une impulsion laser peut
 exciter une onde plasma longitudinale, que l'on peut employer pour
 l'accélération.
\item sa variation spatiale transverse excite une onde plasma
 transverse, ou radiale, dont le champ électrique associé
 défléchit les électrons du faisceau. 
Si le profil du laser est à symétrie cylindrique  l'amplitude du champ transverse est
 proportionnelle à la distance à l'axe au premier ordre, et l'onde
 plasma est équivalente à une lentille\cite{kyoto_capte}(\boul C)
\footnote{La boulette, $\bullet$, désigne
 les références qui peuvent \^etre trouvées en annexe de ce
 mémoire.
}.
Par contraste, une cavité RF n'exerce (au premier ordre) aucune
focalisation sur le faisceau.

Les conséquences importantes de la présence de ces champs
 transverses dans la conception d'un accélérateur laser-plasma
 seront examinées dans la section \ref{sec:multitev}.
\end{itemize}

En pratique deux méthodes ont été employées pour générer une
modulation temporelle adéquate de l'intensité laser~:
\begin{itemize}
\item 
 {\bf BWLA} :
dans la méthode du {\bf battement}, deux ondes laser de fréquences
 proches et d'amplitudes comparables interfèrent dans le plasma. Leur battement
 est équivalent à une modulation périodique de l'intensité d'une onde
 laser de fréquence égale à la moyenne des fréquences des deux ondes
 incidentes. Ceci produit donc une force pondéromotrice longitudinale
 modulée dans le temps. Lorsque la différence des fréquences des deux
 ondes laser est égale à la fréquence d'oscillation naturelle de l'onde
 plasma, on excite celle-ci de fa\c{c}on résonnante.
\item 
 {\bf LWFA} :
dans la méthode du {\bf sillage}, une impulsion courte unique excite
 l'onde plasma. Elle produit deux poussées successives, de signe contraire,
 sur les électrons du plasma. L'effet est donc optimal lorsque la durée
 d'impulsion du laser est de l'ordre d'une demi-période plasma
(Appendice \ref{sec:epw}).
\end{itemize}

\begin{table}
\caption{Différentes méthodes d'accélération évoquées dans ce mémoire, et acronymes employés
\label{tab:acro}}
\begin{center}\begin{tabular}{lll}
BWLA & beat wave laser acceleration &
\\
LWFA & laser wake-field   acceleration  &
EPW non déferlante
\\
SM-LWFA & self-modulated  laser wake-field   acceleration  &
EPW déferlante
\\
PWFA & plasma laser wake-field   acceleration & EPW créée par un paquet d'$e^-$
\\
EPW & electron plasma wave \\
\end{tabular}\end{center}
\end{table}

L'effort expérimental a d'abord porté sur le battement.
Des champs de l'ordre de quelques GV/m ont été produits, avec un gain
en énergie
de quelques MeV à quelques dizaines de MeV. 
Un diagnostic indépendant utilisant la diffusion Thomson cohérente
d'un faisceau sonde sur l'onde plasma fournit une mesure directe de
l'amplitude de l'onde plasma, ce qui permet d'étudier les
mécanismes de saturation de sa croissance.
 Différents groupes
ont utilisé des lasers de différentes longueurs d'onde, laser au CO$_2$
($\lambda=10~\mu\text{m}$, UCLA, Canada), ou laser au Néodyme
($\lambda=1~\mu\text{m}$, RAL, Ecole Polytechnique). Dans une seconde phase,
l'accélération d'un faisceau d'électrons injectés dans l'onde plasma a
été étudiée. Cette méthode du battement fait l'objet de la
sous-section suivante.

Puis le développement de l'amplification laser avec dérive de fréquence
(CPA)
 \cite{mourou}
 a permis de générer des impulsions laser à la fois courtes et
intenses, ce qui a permis d'étudier l'accélération par la méthode du
sillage. La première accélération de particules par cette méthode a
été obtenue à l'Ecole Polytechnique en 1997\cite{prl98}(\boul E).

\section{Expériences de première génération}\label{sec:prems}

Au début des années 80, on ne savait pas produire d'impulsion
laser à la fois courte et intense nécessaire à
l'accélération par sillage~: la méthode du battement d'ondes a
donc été explorée en premier.

\subsection{Accélération par battement d'ondes}
L'excitation d'une onde plasma par le battement de deux impulsions
laser est un processus résonnant, dans lequel l'accord de
fréquence doit \^etre d'autant plus précis que le nombre de cycles
est élevé. 
L'ajustement de la fréquence plasma étant réalisé par celui de
la densité électronique du plasma, la première étape a été
l'étude de la création d'un plasma homogène et stable.

Ensuite l'onde plasma produite a été observée par diffusion
Thomson cohérente d'un faisceau laser sonde sur l'onde plasma.
Dans le calcul de l'amplitude cohérente diffusée par les
électrons du plasma, il vient un terme en
$\cos(\omega_{s}t - \vec{k}_{s}\cdot \vec{r})$ d\^u à la sonde, 
multiplié par un terme en $\cos(\omega_{p}t - \vec{k}_{p}\cdot
\vec{r})$ d\^u à la répartition des électrons~:
 on voit appara\^{\i}tre des raies satellites à 
$\omega_{s} \pm \omega_{p}$.
L'intensité relative des raies satellites fournit une mesure de l'amplitude 
de l'onde plasma.
Ces expériences ont permis de démontrer la validité du principe
de l'excitation d'une onde plasma par battement d'ondes laser.
Elles ont aussi permis de déterminer les mécanismes de saturation
qui limitent la croissance de l'onde, et qui diffèrent en fonction
de la longueur d'onde employée, $\lambda=10\mu\text{m}$ (laser au
CO$_2$) ou $\lambda=1\mu\text{m}$ (laser au Néodyme).

Dans un troisième temps seulement, un faisceau d'électrons 
est injecté dans l'onde plasma, et le spectre de
ceux d'entre eux qui ont été accélérés est étudié.

\subsubsection{Battement d'ondes~: laser au CO$_2$
 ($\lambda=10\mu\text{m}$)} 

Plusieurs groupes ont utilisé un laser CO$_2$, avec des moyens et
des résultats assez semblables
\cite{Ebrahim,kitagawa,Claytonprl93}.
 Je décris brièvement les
travaux du groupe de UCLA, précurseur dans le domaine.
Ce groupe a développé tout d'abord un laser bi-fréquence, 
$\lambda_{1}=9.56~\mu\text{m}$
et 
$\lambda_{2}=10.59~\mu\text{m}$
fournissant une énergie de 16~J avec une durée d'impulsion
$\tau$~=~1.5~ns.
Le facteur de Lorentz de l'onde plasma correspondante est $\gamma_{p}
= \lambda/ \delta\lambda \approx 10$, et la densité à la
résonance $n = 1.1~10^{17}\text{cm}^{-3}$.
Le plasma est obtenu par une décharge électrique
dans l'hydrogène\cite{claytonprl85}.

Les diagnostics de diffusion Thomson cohérente d'un faisceau laser
sonde (rubis) injecté à petit angle ($7^\circ$), et du faisceau
principal lui m\^eme à l'avant ont permis d'observer très
clairement les satellites associés à l'onde
plasma\cite{claytonprl85}. L'intensité 
de ces satellites a permis d'estimer l'amplitude de l'onde plasma à
$\delta \approx 1-3~\%$, alors que le calcul prédit 8~\%.
Le champ électrique longitudinal correspondant est de 0.3 à 1 GV/m.
L'amplitude maximale est ici limitée par un effet de déphasage
relativiste~: lorsque  l'amplitude cro\^{\i}t, les électrons du plasma
deviennent relativistes, et leur mouvement devient anharmonique. La
fréquence plasma diminue, et l'accord de phase entre le battement
d'ondes et le plasma est rompu.

Ces résultats 
ont permis de valider 
la méthode du battement d'ondes pour l'excitation de l'onde plasma.
Par contre les mesures précises de la densité électronique locale
fournies par la diffusion Thomson sur les ondes plasma ont montré
que la décharge électrique utilisée pour l'ionisation ne produit
pas un plasma suffisamment homogène\cite{claytontr85}.

Le groupe s'est alors orienté vers l'emploi d'un ``$\theta$ {\em
 pinch}''
\cite{ClaytonthetaPinch}
~: un condensateur est déchargé dans une bobine de très
faible inductance entourant le gaz.
La variation temporelle du champ magnétique crée un champ
électrique qui ionise le gaz.
Au cours de la montée du champ, le front d'ionisation se propage de
la bobine vers son axe~: au maximum du champ on dispose d'un plasma
complètement ionisé et donc homogène.
Cette approche n'a pas permis de déboucher, parce que les conditions
de fonctionnement du $\theta$ {\em pinch}
qui fournissent un plasma homogène laissent un champ magnétique
piégé dans le plasma trop élevé, ce qui emp\^eche
l'injection efficace d'un faisceau de particules chargées.

Le groupe de UCLA a alors pu augmenter l'énergie fournie par le
laser CO$_2$ à 60~J en 300~ps (FWHM), permettant l'ionisation du gaz par effet
tunnel dans le champ électrique du laser\cite{ClaytonTunnelIon}.
Le laser est ici réglé sur les raies du CO$_2$
$\lambda_{1}=10.275~\mu\text{m}$ et
$\lambda_{2}=10.591~\mu\text{m}$, plus
proches, produisant une onde plasma plus rapide $\gamma_{p} \approx
33$, à une densité résonnante de $0.9\cdot 10^{16}\text{cm}^{-3}$.
Le faisceau d'électrons produit par un LINAC à 9.3~GHz, à une
énergie de 2.1~MeV, se présente sous la forme d'un macro-paquet de
1.5~ns FWHM, comprenant des micro-paquets de $\approx$~10~ps espacés
de 107~ps.
Un spectromètre magnétique analyse les électrons
accélérés, une barette de diodes en silicium les détecte.
L'imagerie dans une chambre à brouillard en champ magnétique a
permis d'identifier le signal comme provenant bien d'électrons
accélérés, et non pas d'électrons de bruit de fond déviés.
L'amplitude de l'onde plasma mesurée par diffusion Thomson 
est ici de 15-30\%, et les spectres d'électrons
accélérés s'étendent jusqu'à 28~MeV, ce qui pour une
longueur d'accélération de 1~cm, correspond à un champ de
2.8~GV/m, soit une amplitude de 28\%.
Ces résultats sont en accord avec la valeur théorique, 37~\%,
toujours limitée par le déphasage relativiste.
Une étude du devenir du faisceau non accéléré à 2~MeV a
fourni une information intéressante~: le faisceau est dévié de
fa\c{c}on collective jusqu'à plusieurs nanosecondes après
l'excitation de l'onde plasma\cite{ClaytonPhysPlasma}. 
Ceci est le signe de la persistence de champs intenses dans le plasma,
alors que la durée de vie de l'onde plasma mesurée par diffusion
Thomson est beaucoup plus courte, 50-100~ps.

%

\subsubsection{Battement d'ondes~: laser au Néodyme ($\lambda=1\mu\text{m}$)}

Le groupe de l'Ecole Polytechnique a choisi d'emprunter une voie
différente, l'emploi d'un laser au Néodyme
au Laboratoire pour l'Utilisation des Lasers Intenses (LULI).
Ici, la longueur d'onde plus courte permet d'obtenir un plasma
homogène par multi-photo-ionisation complète du gaz par le tout début
 de l'impulsion laser.

Le groupe a commencé par étudier la formation du plasma, et
l'homogénéité et la stabilité de la densité électronique
\cite{homogen}. 
Puis l'ajout d'un second oscillateur au Nd:YLF a permis l'étude de
l'excitation de l'onde plasma par le battement des deux ondes, et  l'étude des
mécanismes de saturation\cite{prl92}.
Les longueurs d'ondes des deux oscillateurs,
$\lambda_{1}=1.0530~\mu\text{m}$ et $\lambda_{2}=1.0642~\mu\text{m}$, sont
très proches, et leur battement produit une onde plasma fortement
relativiste, avec $\gamma_{p}=94.5$, et $\lambda_{p}=105~\mu\text{m}$.
L'énergie de chaque impulsion en sortie de cha\^{\i}ne est d'environ
10~J, avec une durée de 100~ps (FWHM).
La densité électronique résonnante est  $n=1.1\cdot
10^{17}\text{cm}^{-3}$, correspondant à 2~mbar de deutérium.

L'amplitude de l'onde plasma est mesurée à partir de 
 la diffusion Thomson d'un faisceau à 0.53~$\mu$m, obtenu
par doublage de fréquence d'une partie de l'un des faisceaux incidents.
L'observation sur l'axe à l'avant est impossible parce qu'un signal
important est généré dans le gaz neutre précédant et suivant
le plasma, par mélange à quatre ondes.
On observe donc le rayonnement diffusé à un angle de $10^\circ$ \cite{karda}.
Les résultats obtenus indiquent que l'amplitude de l'onde plasma est
de 1 à 5~\%, alors que
la limite due au déphasage relativiste est de 15~\%.
On observe de plus que le rayonnement diffusé n'est présent qu'au
tout début de l'impulsion laser \cite{karda},
 $55\pm 20$~ps avant le maximum. 
La durée de vie de l'onde plasma, estimée à partir d'une mesure
résolue en temps, est de l'ordre de 30~ps seulement.
De plus, il se trouve que la diffusion Thomson directe sur les ondes
plasma relativistes ne devrait \^etre observable que dans une plage
angulaire restreinte, de $\pm 2^\circ$, autour de l'axe commun de
propagation.
On observe en fait ici des ondes plasma ``filles'', produites par
couplage entre l'onde plasma relativiste et des ondes acoustiques
ioniques\cite{physplasm}.
Ce couplage appelé instabilité modulationnelle transfère 
l'énergie de l'onde plasma électronique relativiste à d'autres
composantes de vecteur d'onde quelconque, c'est-à-dire le plus
souvent pas longitudinales, et pas relativistes~: inutiles.

Le calcul du taux de croissance de l'instabilité modulationnelle
indique que l'onde sature extr\^emement rapidement, à une amplitude
de 1.5~\%, environ 60~ps avant le maximum du flux laser. 
La disparition rapide et totale du signal Thomson
indique qu'ensuite l'onde relativiste est détruite.
Les calculs théoriques sont en accord avec nos résultats~: le
mécanisme est bien compris.
Cette saturation de l'onde plasma par instabilité modulationnelle nous
co\^ute un facteur 10 sur son amplitude.

\paragraph{Accélération d'électrons} ~

L'observation de l'onde plasma par diffusion Thomson a permis
d'engager la phase suivante d'accélération d'un faisceau
d'électrons injectés.

Le faisceau
d'électrons est fourni par un accélérateur de Van de Graaff
opéré par le Laboratoire des Solides Irradiés (LSI), à une
énergie de 3~MeV, un courant de 200~$\mu$A, et une émittance de 
0.06~mm~mrad\cite{nim95}(\boul A).

Un schéma du dispositif expérimental
\footnote{Une version modifiée, avec injection de l'impulsion laser sous vide, 
est en figure  \ref{leschema}.}
 peut \^etre trouvé en figure
1 de la référence \cite{nim95}(\boul A).
(La figure 1 de ce texte, relative à l'expérience de sillage, est semblable.)
Les électrons pénètrent dans l'enceinte expérimentale par une fen\^etre
en aluminium de 1.5~$\mu$m, puis sont focalisés au m\^eme point que
le faisceau laser et de fa\c{c}on colinéaire à celui-ci, par un
aimant triplement focalisant --- stigmatique et achromatique\cite{achrom}.
La position et la taille du faisceau d'électrons est monitorée en
imageant le rayonnement de transition dans le domaine optique (RTO) émis à
la traversée d'une feuille métallique.
Le m\^eme système image la lumière laser diffusée sur la
feuille, ce qui fournit l'alignement relatif avec une précision de
l'ordre de 10~$\mu$m.
Le spectre des électrons accélérés est analysé par un
spectromètre magnétique et mesuré par un hodoscope de
scintillateurs-photomultiplicateurs.

Un soin tout particulier a été consacré à l'injection du
faisceau d'électrons (45~$\mu$m RMS dans 2~mbar de deutérium) dans
le petit plasma ($60 \pm 20 \mu$m FWHM).
La lutte contre le bruit de fond a aussi été menée avec vigueur,
la durée du signal -- quelques picosecondes -- étant grandement
inférieure à la durée d'intégration de l'appareillage -- 7 ns.
Un système de collimation dédié a été
utilisé, et le spectromètre stigmatique à l'énergie du
faisceau (3~MeV) permet l'éjection du faisceau non accéléré
vers un absorbeur de Z faible.

Nous avons observé l'accélération d'électrons avec un gain
pouvant atteindre 1.4~MeV\cite{prl95}(\boul B).
L'accélération n'est observée
que lorsque les conditions expérimentales suivantes sont
respectées~:
\begin{itemize}
\item présence des deux ondes laser avec un recouvrement temporel
 suffisant \cite{karda}.
\item injection du faisceau d'électrons~(aucun piégeage des
 électrons du plasma n'est observé)
\item pression proche de la pression résonnante, avec une
 largeur à mi hauteur d'environ 5\%  \cite{karda}.
\end{itemize}

Le dernier point identifie clairement le battement comme le
mécanisme en jeu.
La variation de l'énergie d'injection du faisceau d'électrons nous
a permis de caractériser le mécanisme limitant le gain d'énergie
des électrons à la traversée de l'onde plasma.
Lorsqu'un électron est injecté en phase avec l'onde --- au maximum
du champ --- et reste en phase pendant toute la traversée du plasma,
le gain d'énergie est simplement~:
\begin{equation}
\Delta W = e \int{E_{z} \dd z}= \pi z_{0} e E_{z,max}
\end{equation} o\`u $z_{0}$ est la
longueur de Rayleigh du laser.
Ici, les électrons sont injectés avec une vitesse plus faible que
la vitesse de phase de l'onde ($\gamma_{e}=5.9$, $\gamma_{p}=94.5$),
et l'expression de la longueur de déphasage est 
$L_{\varphi} \approx \gamma_{p}\gamma_{e}^{2}\lambda\approx
4.2~\text{mm}$, de l'ordre de la longueur de Rayleigh.
A partir des gains d'énergie d'électrons injectés 
à 2.5, 3.0, 3.3 MeV \cite{karda},
 on peut estimer indépendamment 
les valeurs de l'amplitude de l'onde plasma, et de la longueur
d'accélération $\delta \approx 2.4\%$ et $L=2 z_{0} \approx
2.8~\text{mm}$.
Le champ accélérateur maximal correspondant est de 0.7~GV/m.
La variation de $ \Delta W$ en fonction de la focale de la lentille de focalisation
du laser confirme ces résultats \cite{karda}.

Notons qu'en absence de ce déphasage, nous aurions obtenu $\Delta W
= \pi z_{0} e E_{z,max} \approx 3.1~\text{MeV}$~: le déphasage des
électrons par rapport à l'onde plasma nous a co\^uté un facteur 2.

\subsubsection{Battement d'ondes~: quels enseignements ?}

Les expériences de première génération utilisant 
une onde plasma excitée par battement d'ondes laser ont permis de
commencer l'exploration de l'accélération de particules dans un plasma excité par un laser. 
Des résultats importants ont été obtenus, avec en particulier
la détermination des mécanismes conduisant à la saturation de
l'onde.

\begin{itemize}
\item
Dans les expériences utilisant un laser au CO$_2$, le mécanisme
principal est le déphasage relativiste décrit plus haut.
\item
Avec un laser au Néodyme, de longueur d'onde plus courte, la densité
électronique
est plus élevée, ce qui permet d'atteindre un gradient
accélérateur plus grand à amplitude de l'onde identique; cependant 
le taux de croissance de l'instabilité modulationnelle,
proportionel à la densité, est alors plus élevé et c'est
cet effet qui limite ici la croissance de l'onde, à des valeurs de
quelques pour-cents.
\end{itemize}

Les gradients obtenus sont de l'ordre du GV/m.
La durée de vie de l'onde plasma est toujours inférieure à la
durée de la pompe laser, 50-100~ps/300~ps dans le cas du CO$_2$,
30~ps/100~ps dans le cas du Néodyme, ce qui traduit une mauvaise
efficacité dans l'utilisation de l'énergie incidente.

\begin{center}\begin{tabular}{cc|cc}
& & CO$_2$ & Néodyme\\
& & \protect\cite{Claytonprl93} & \protect\cite{prl95}(\boul B)\\ \hline
$\lambda$ & $\mu$m & 10 & 1.06 \\
$\gamma_p$ & & 33 & 94.5 \\
 $\lambda_p$ & $\mu$m & 360 & 100 \\
$E$ & GV/m & 2.8 & 0.7 \\
$\delta$ & \% & 28 & 2.4 \\
$\pi z_{0}$ & mm & 25 & 4.4\\ 
$\Delta W$ & MeV & 28 & 1.4 \\
\end{tabular}\end{center}

L'une des fa\c{c}ons de dépasser ce problème est de développer
un battement d'ondes court, voire réduit à quelques battements
seulement. 
On bat alors de vitesse les instabilités susceptibles de limiter la
croissance de l'onde, et on s'affranchit du délicat problème du
réglage fin de l'homogénéité, et de la stabilité de la
densité électronique.

L'autre voie est l'excitation par sillage d'une impulsion laser
courte, et fait l'objet de la sous-section suivante.

\subsection{Accélération par sillage laser}

L'accélération de particules par sillage laser a pu \^etre
étudiée après le développement de l'amplification laser avec
dérive de fréquence\cite{mourou}.

\begin{itemize}
\item
Un oscillateur produit tout d'abord une impulsion laser ultra-courte.
\item
Celle-ci est étirée dans un dispositif à dispersion temporelle
non nulle.
L'impulsion étirée conserve son émittance longitudinale faible,
limitée de Fourier~: on
a simplement établi une relation linéaire temps-fréquence.
\item
L'impulsion peut alors \^etre amplifiée dans une cha\^{\i}ne de puissance
tout en gardant une intensité inférieure au seuil de déformation
par effet non-linéaire dans les barreaux amplificateurs.
\item
Enfin l'impulsion est comprimée temporellement dans un dispositif
dont la dispersion temporelle est opposée à celle de l'étireur.
\end{itemize}

La géométrie de l'impulsion laser est détaillée en appendice \ref{sec:las}, et figure 
\ref{fig:waist} : les profils transverse et longitudinal sont gaussiens, et nous verrons l'intérêt d'un rapport d'aspect proche de l'unité : l'impulsion est une ``boule'' de lumière.

\begin{figure}[hbt]\begin{center}
\mbox{\epsfig{figure=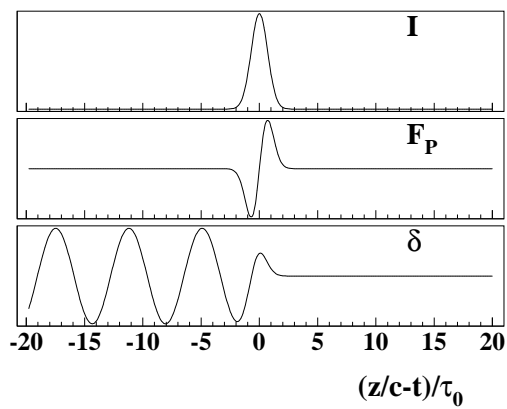,width=8cm}}
\caption{Profil d'intensité du laser $I$, force pondéromotrice $F_P$, et amplitude de l'onde plasma
$\delta$, en fonction de $(z-ct)/\tau_0$.\label{fig:epw}}
\end{center}\end{figure}

La force pondéromotrice de l'impulsion laser unique donne aux
électrons du plasma deux poussées successives de sens contraire~: vers
l'avant --- dans le sens de propagation --- lorsque l'intensité
cro\^{\i}t, puis vers l'arrière lorsqu'elle décroit
(figure \ref{fig:epw}). Ceci est détaillé dans l'appendice \ref{sec:epw}).
L'excitation de
l'onde plasma est optimale lorsque la durée d'impulsion laser 
est de l'ordre d'une demi-longueur d'onde plasma (fig. \ref{sillage})

Du point de vue de l'accélération de particules, les méthodes
d'accélération par sillage et par battement sont très proches~:
\begin{itemize}
\item
le laser crée un plasma complètement ionisé;
\item
il excite une onde plasma dans ce plasma;
\item
on envoie un faisceau de particules chargées se faire accélérer
dans l'onde plasma.
\end{itemize}

En pratique, la méthode du sillage présente plusieurs avantages
importants~: 
\begin{itemize}
\item Tout d'abord l'ajustement nécessaire de la densité du plasma est beaucoup 
moins fin que dans le cas du battement, o\`u l'excitation est résonnante (fig. \ref{sillage}).
\item De plus, l'excitation de l'onde plasma est un processus rapide,
 dans lequel les mécanismes de saturation décrits plus haut
 n'apparaissent pas.
\item Enfin, du point de vue de la réalisation, on s'affranchit du
 problème de la synchronisation des deux impulsions de la méthode
 du battement.
\end{itemize}

Comme dans le cas du battement, la création d'une onde plasma par
sillage laser a d'abord été étudiée par des moyens
entièrement optiques. J'en présente les résultats ci dessous.

Ensuite, je présente les résultats de deux expériences
d'accélération par sillage laser réalisées par une
collaboration Japonaise.
Les tentatives de publication de ces résultats \cite{KEK93,KEK97}
ont échoué, 
et j'essaierai d'expliquer pourquoi.
Le second papier a été finalement présenté à une conférence
\cite{KandoKyoto},
et nous avons d\^u publier un article de réfutation\cite{denisprel}(\boul F)).

Je présente ensuite l'expérience d'accélération par sillage
laser de l'Ecole Polytechnique.

\subsubsection{Sillage laser~: mesures optiques}

On se souvient que la création d'une onde plasma par battement d'ondes a
été d'abord étudiée par diffusion Thomson cohérente d'un
faisceau laser sonde sur l'onde plasma.
En pratique le faisceau pompe est utilisé le plus souvent, ou une
réplique de celui-ci éventuellement doublée en fréquence.

Cette méthode peut difficilement \^etre employée ici ;
l'impulsion laser ultra courte a une grande largeur
spectrale~: les satellites sont alors noyés dans les pieds de
l'impulsion elle m\^eme.
En effet, la largeur spectrale est au moins égale à la limite de
Fourier~: $\sigma_{\omega} \ge 1/(2\sigma_{t})$, alors que le
décalage des satellites Thomson est $\Delta\omega=\omega_{p}$, soit
à la quasi-résonance, $\Delta\omega=2/\tau_{0}$, du m\^eme ordre
de grandeur que la largeur de raie $\sigma_{\omega}$.
Les satellites sont alors indistinguables dans les pieds de la raie
centrale.

Pour cette raison, une autre méthode a été utilisée~:
l'interférométrie dans l'espace temps-fréquence\cite{interfero}.
Une impulsion pompe crée le plasma et l'onde plasma, qui est
parcourue par une paire d'impulsions lasers sonde.
Les impulsions sonde traversent le plasma dont l'indice de
réfraction $\eta$ dépend de la densité électronique~:
$ \eta = \sqrt{1-n/n_{c}}$,
o\`u la densité critique du plasma $n_{c}$ ne dépend que de la
longueur d'onde du rayonnement~: 
$n_{c} =  \omega^{2}/(4\pi r_e c^2)$.
Chaque impulsion sonde chevauche l'onde plasma avec une phase qui
reste constante lorsque la vitesse de groupe de l'impulsion sonde et
la vitesse de phase de l'onde plasma sont proches\footnote{La vitesse
 de phase de l'onde plasma est égale à la vitesse de groupe de
 l'impulsion pompe. Si les longueurs d'onde pompe et sonde étaient
 identiques, les impulsions sonde et l'onde plasma progresseraient
en phase.}.
Chaque impulsion traverse un plasma de densité électronique,
et donc d'indice de réfraction, qui dépend  de cette phase, et
de l'amplitude de la modulation de la densité électronique,
c'est-à-dire de celle de l'onde plasma.
Les impulsions subissent alors un décalage de phase relatif
$\delta\Phi$ qui dépend du délai qui les sépare.
Elles sont collectées et envoyées dans un spectromètre o\`u elles
interfèrent.
Le décalage en longueur d'onde observé est directement
proportionnel à $\delta\Phi$.
Un balayage du délai permet de reconstruire la dépendance
temporelle de l'onde plasma 
\cite{revuerapha}. 

L'expérience a utilisé le laser au saphir dopé au titane
(Ti:Saphir) du Laboratoire d'Optique Appliquée sur le site de
l'Ecole Polytechnique \cite{rapha, newrapha}.
Le laser de pompe fournit des impulsions de 40~mJ à une cadence de
10~Hz et une longueur d'onde de $0.8~\mu\text{m}$, avec une durée
d'impulsion de 130~fs (FWHM).
Une partie de l'impulsion est utilisée pour créer les impulsions
sonde après doublage de fréquence.

Ces expériences ont permis d'observer la création d'onde plasma
par sillage laser, jusqu'à une amplitude proche de 100\%, et un
gradient de 10~GV/m.
L'amplitude de l'onde plasma et sa fréquence sont en accord avec la
théorie linéaire du sillage laser détaillée en appendice.
En particulier la courbe de quasi-résonance a été observée
avec précision 
 \cite{newrapha}.

Cette expérience a fourni les premières mesures de durée de vie de
l'EPW. 
Notons que ces résultats ont été acquis en mode ``{\bf radial}'' : le
faisceau laser est fortement focalisé, de façon à utiliser au mieux la
 faible énergie disponible.
Des électrons éjectés du plasma dans le gaz neutre par le fort champ électrique
transverse de l'EPW, reviennent ensuite sur l'axe, ayant perdu la
relation de phase avec l'EPW, et ils la détruisent.
%

Des mesures de durée de vie de l'EPW en mode {\bf longitudinal} seront les
bienvenues.

\subsubsection{Expériences d'accélération par sillage
 laser au Japon}

Après la description des mesures de l'onde plasma par
interférométrie, j'en viens aux expériences d'accélération
proprement dites.

\paragraph{Accélération par sillage laser au Japon~: première expérience} ~

Une première tentative de démonstration d'accélération de
particules par sillage laser a eu lieu au Japon au début des
années 90 \cite{KEK93}. Le faisceau d'électrons est généré
par l'impact d'une fraction du faisceau laser sur une cible solide.
Un aimant collecte ces électrons et les dévie vers le point de
plasma.
Puis un second aimant effectue l'analyse en impulsion.
Un hodoscope de scintillateurs assure la détection.

Lorsque l'on n'injecte pas le laser de pompe, et donc en l'absence de
plasma, on observe simplement le spectre des électrons injectés, qui
s'étend entre 0 et 2~MeV, avec un maximum vers 1~MeV.
Lorsque le laser de pompe crée un plasma, le spectre présente une
queue jusque vers 8~MeV interprétée comme l'effet d'une
accélération (fig. 4 de la référence \cite{KEK93}).
La non-publication de la référence \cite{KEK93} peut \^etre due
à plusieurs causes. Du point de vue de l'accélération de
particules, 
on peut noter l'absence d'une discussion de la séparation du signal
correspondant à des électrons accélérés, du bruit de fond
d\^u par exemple à des électrons déviés.

Notons aussi l'absence d'optique de collection des électrons en
sortie de plasma (fig. 2 de la référence \cite{KEK93}) ;
en fait l'aimant à faces parallèles utilisé comme
spectromètre est fortement {\bf défocalisant} dans le 
plan vertical ce qui doit exploser le faisceau sur le plancher et le
plafond de l'enceinte.
Par la suite ce groupe a modifié son appareillage pour une seconde
tentative. 

\paragraph{Accélération par sillage laser au Japon~: seconde expérience} ~

Dans cette seconde expérience \cite{KEK97,KandoKyoto}, le faisceau d'électrons est produit
par un LINAC à une énergie de 17~MeV.
Le plasma est produit dans de l'hélium à une pression de quelques
millibars, par un laser Ti:Saphir à une longueur d'onde de
$0.79~\mu\text{m}$, avec une durée d'impulsion de 90~fs (FWHM) et
une énergie de 200~mJ.
Le spectromètre analysant le spectre des électrons
accélérés est modifié, avec l'ajout d'un doublet de
quadrip\^oles assurant la focalisation.

Des électrons accélérés sont observés avec un gain 
d'énergie supérieur à 100~MeV \cite{KEK97,KandoKyoto}, alors que la théorie linéaire 
 du sillage laser  prédit un gain maximal
 inférieur à 10~MeV.
Les auteurs interprètent leur résultat comme étant d\^u au
guidage du faisceau laser par le plasma, appuyant l'argument sur la
valeur élevée de la longueur de Rayleigh observée, proche de
1~cm. 

En fait, les différents mécanismes de guidage envisagés ne
peuvent rendre compte de l'effet \cite{denisprel}(\boul F).
Par contraste, la {\bf défocalisation} du faisceau sur la lentille
constituée par l'interface gaz/plasma \cite{malka} permet de
comprendre l'élargissement et l'allongement du {\em waist}
\cite{denisprel}(\boul F). 
L'intensité laser est alors beaucoup plus faible que ce qu'elle
aurait été dans le vide, ce qui doit \^etre pris en compte dans le
calcul de l'amplitude de l'onde plasma, et le gain d'énergie des
électrons est encore plus faible.

Quant aux électrons accélérés, on sait depuis les travaux du
groupe de UCLA \cite{ClaytonPhysPlasma}
qu'une onde plasma laisse derrière elle des champs magnétiques et/ou
des champs électriques transverses qui défléchissent le faisceau
d'électrons incident.
L'interaction des électrons défléchis sur les parois de
l'enceinte crée un bruit de fond associé à l'onde plasma, et qui
doit donc \^etre étudié de fa\c{c}on spécifique \cite{denisprel}(\boul F).
En fait, cette composante du bruit de fond a été identifiée
lors de l'expérience de l'Ecole Polytechnique décrite ci dessous,
comme un signal de faible amplitude dans les canaux de haute énergie.

\subsubsection{Expérience à l'Ecole Polytechnique}
Après l'étude de l'accélération par battement, notre groupe
s'est orienté vers l'exploration de l'accélération par sillage laser.
Nous avons utilisé le laser TW du LULI \cite{laserLULITW}, qui
fournit une impulsion de 400~fs (FWHM) de quelques joules, avec un
taux de répétition de un tir toutes les 20 minutes à l'énergie
maximale\footnote{
L'objectif principal dans la conception de ce laser a été
d'obtenir une énergie laser élevée, ce qui a conduit à
l'emploi d'amplificateurs en verre dopé au Néodyme. 
La largeur spectrale obtenue est alors plus faible que pour une cha\^{\i}ne
complète en Ti:Saphir, et la vitesse de refroidissement plus faible
du verre limite le taux de répétition.}.

\begin{figure}[tbh] 
\begin{center}
 \mbox{\begin{turn}{90} \epsfig{figure=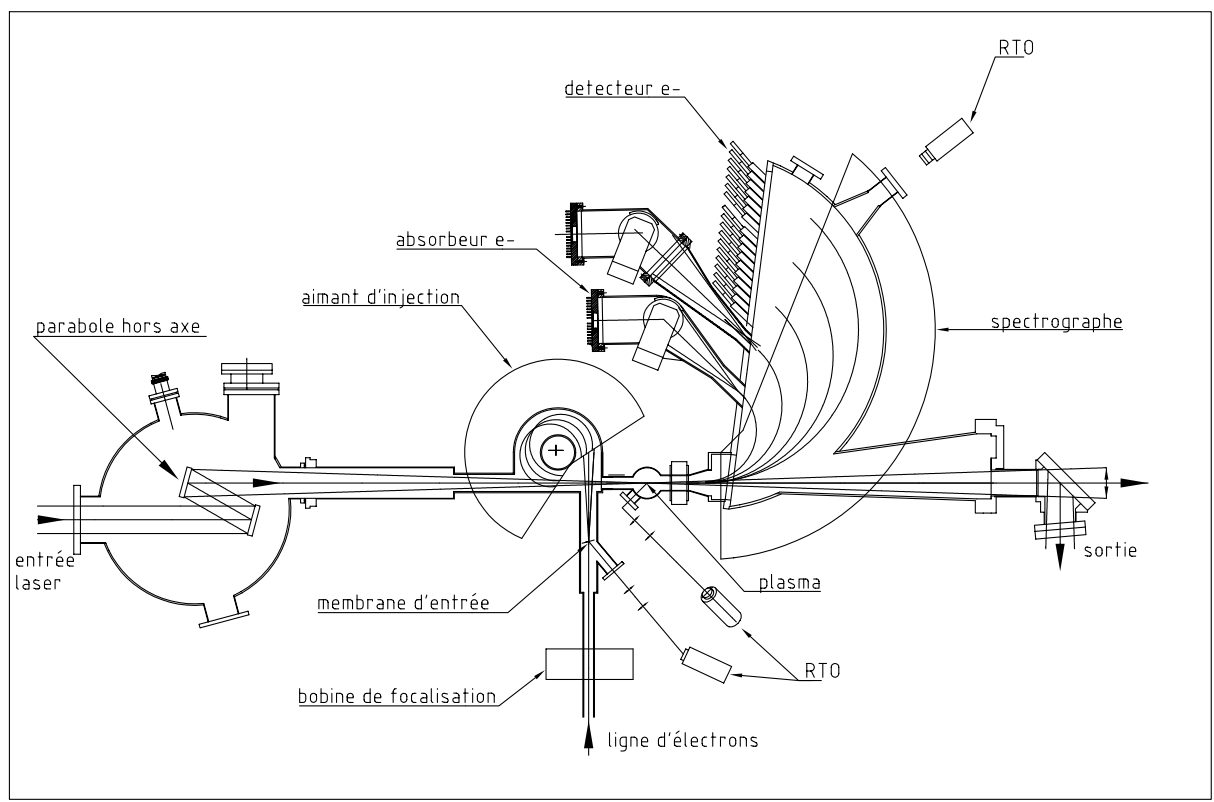,height=\linewidth}
\end{turn}}
\caption{Schéma de l'expérience d'accélération de particules
 par sillage laser à l'Ecole Polytechnique.
Voir le texte et la référence \protect\cite{nim95}\boul A). 
\label{leschema}}
\end{center}
\end{figure}

Notre dispositif expérimental (fig. \ref{leschema}) est pour
l'essentiel celui employé lors de l'expérience d'accélération
par battement \cite{nim95}(\boul A).
La compression temporelle est effectuée sous vide après le
transport jusqu'à l'enceinte expérimentale.
Le faisceau est focalisé par une parabole hors axe.

Nous avons observé des électrons accélérés avec un gain
maximal d'énergie de 1.6~MeV \cite{prl98}(\boul E).
Le signal est présent uniquement lorsque les conditions suivantes
sont satisfaites~:
\begin{itemize}
\item une impulsion laser ultra courte est injectée,
\item un faisceau d'électrons est injecté,
\item un recouvrement spatial et temporel correct du faisceau et du
 plasma est réalisé.
\end{itemize}
En particulier, nous n'avons observé aucun signal d\^u à des
électrons du plasma qui auraient été piégés par l'onde
plasma et accélérés.

\begin{figure}[tbh] 
\begin{center}
 \epsfig{figure=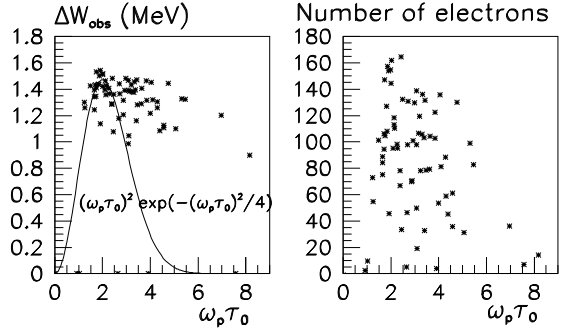,width=0.47\linewidth} 
\caption{
Expérience d'accélération de particules
 par sillage laser à l'Ecole Polytechnique.
Gain maximal et nombre d'électrons observés en fonction de  $\omega_{p}\tau_{0}$.
\label{fig:resultat:densite}}
\end{center}
\end{figure}

Le gain maximal d'énergie $\Delta W$ des électrons lors d'un tir présente
un maximum autour de $\omega_{p}\tau_{0} \approx 2$  conformément à la théorie
linéaire du sillage.
Néanmoins, la valeur maximale observée, 1.6~MeV, est inférieure
à la valeur calculée, égale à 10~MeV.
De plus, la valeur de $\Delta W$ observée dépend beaucoup moins de
l'accord laser/plasma, c'est-à-dire de $\omega_{p}\tau_{0}$ que ce
que prédit la théorie linéaire du sillage 
(Fig. \ref{fig:resultat:densite}) (\cite{prl98}(\boul E)),

\begin{figure}[tbh] 
\begin{center}
 \epsfig{figure=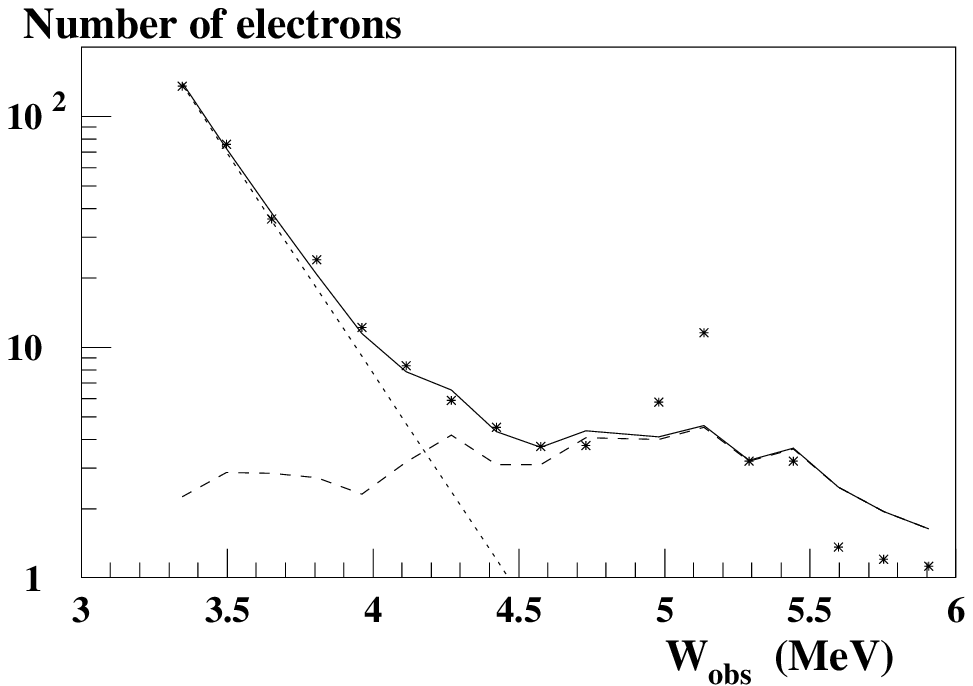,width=0.4\linewidth} 
 \epsfig{figure=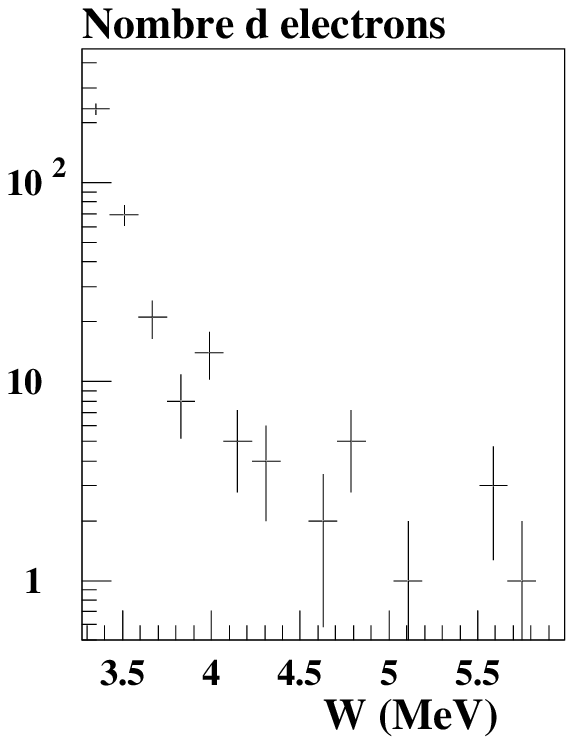,width=0.22\linewidth} 
 \epsfig{figure=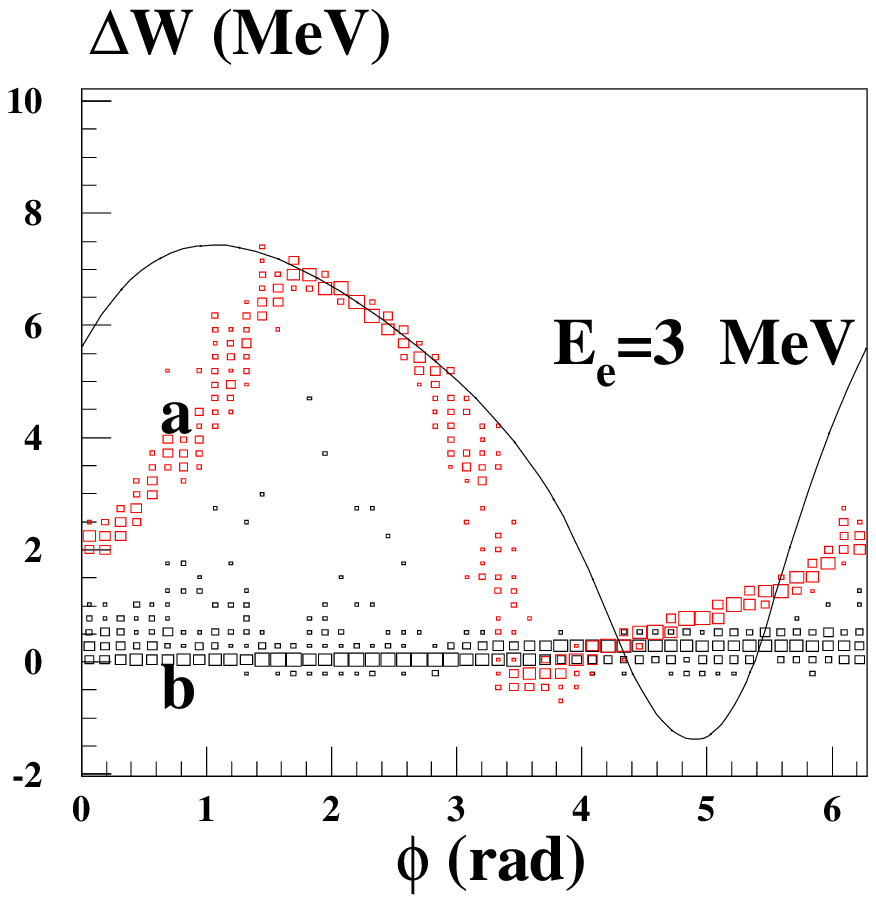,width=0.3\linewidth} 

\caption{
Expérience d'accélération de particules
 par sillage laser à l'Ecole Polytechnique.
Gauche : spectre expérimental \cite{prl98}.
Centre  : Spectre simulé.
Droit : 
Simulation 3D : ligne faisceau sur axe;
(a) Rouge :  émittance faible.
(b) Noir :  émittance réelle.
\label{fig:resultat:sillage}}
\end{center}
\end{figure}

Une simulation tridimensionelle de l'accélération des électrons
dans l'onde plasma reproduit et permet de comprendre les résultats
expérimentaux
(Fig. \ref{fig:resultat:sillage}),\cite{prl98}(\boul E)~:
\begin{itemize}
\item Pour une injection sur l'axe (émittance nulle), le gain
 d'énergie ne dépend que de la phase de l'électron injecté par
 rapport à l'onde plasma.
Il est égal au gain décrit par la théorie linéaire, soit ici
à peu près 10~MeV. 
\item Pour une émittance faible, $\epsilon = \sigma \sigma'$ avec $\sigma= 30~\text{nm}$
et $\sigma'=10~\mu\text{rad}$, soit une émittance de
$0.3~10^{-12}~\text{m}\times \text{rad}$, plus petite par 6 ordres de grandeur que
l'émittance réelle, les électrons parcourant la partie
défocalisante de l'onde plasma sont expulsés très t\^ot, alors
que les autres sont piégés par le champ focalisant et décrivent
une oscillation bétatronique dans l'onde plasma. L'accélération
maximale de ces derniers est  celle prédite par la théorie
linéaire. (a)
\item Par contre, pour une émittance réelle (b) ($\sigma= 30~\mu\text{m}$
et $\sigma'=10~\text{mrad}$), la plupart des
 électrons sont perdus avant la zone du {\em waist}, o\`u le champ
 accélérateur est le plus intense~: ceci produit le spectre
 exponentiel observé (Fig. \ref{fig:resultat:sillage}), gauche),
 \cite{prl98}(\boul E).
\end{itemize}

Ceci est à première vue étonnant, la taille du faisceau
d'électrons au waist $\sigma= 30~\mu\text{m}$ étant du m\^eme ordre
que la taille du plasma $w\approx 24~\mu\text{m}$.
L'effet peut \^etre compris par deux approches différentes, l'une
examinant seulement le comportement du faisceau dans le plan
transverse, l'autre incluant l'axe longitudinal

\paragraph{Oscillations bétatroniques} ~

L'effet est compris plus en détail dans le cadre du modèle
simpliste développé en référence \cite{kyoto_capte}(\boul C).
Dans ce modèle, on sépare l'espace entourant le {\em waist} en
trois zones distinctes~: tout d'abord un espace de dérive, dans
lequel les électrons progressent en ligne droite; puis la zone du
{\em waist}, dans laquelle ils sont piégés et subissent une
oscillation bétatronique; puis un espace de dérive en sortie.

 
\begin{figure}[hbt] \begin{center}
\setlength{\unitlength}{0.5mm}
 \begin{picture}(80,55)(0,0)
 \put(0,0){\makebox(80,40) {\epsfig{file=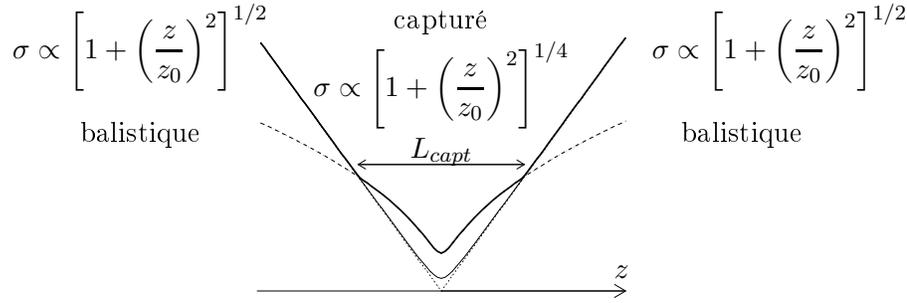,
width=100\unitlength}}}
 \put(88,-10){\makebox(0,0)[b]{$z$}}
 \put(-40,25){\makebox(0,0)[b]{balistique}}
 \put(40,55){\makebox(0,0)[b]{capturé}}
 \put(120,25){\makebox(0,0)[b]{balistique}}
 \put(40,30){\makebox(0,0)[b]
{$\displaystyle \sigma\propto 
\left[1+\left(\frac{z}{z_{0}}\right)^2\right]^{1/4}$}}
 \put(-40,40){\makebox(0,0)[b]
{$\displaystyle \sigma\propto 
\left[1+\left(\frac{z}{z_{0}}\right)^2\right]^{1/2}$}}
 \put(130,40){\makebox(0,0)[b]
{$\displaystyle \sigma\propto 
\left[1+\left(\frac{z}{z_{0}}\right)^2\right]^{1/2}$}}
 \put(40,20){\makebox(0,0)[b]{$L_{capt}$}}
 \end{picture}

\vspace{0.5cm}
\caption{Schéma de l'enveloppe du faisceau d'électrons (ligne en gras)
Vol balistique (fine).
Faisceau capturé (tiré).
\label{db:ledessin}}
\end{center} \end{figure}

Lorsque le faisceau piégé approche du {\em waist}, sa taille
décroit comme $\sigma = \sqrt{\beta\epsilon} \propto
[1+(z/z_{0})^2]^{1/4}$, alors que dans le vide il décroit plus vite,
comme $\sigma = w/2 \propto [1+(z/z_{0})^2]^{1/2}$
(fig. \ref{db:ledessin})
 \cite{balte}. 
A cause de ce piégeage des électrons dans le champ transverse de
l'onde plasma, le faisceau d'électrons est alors plus gros au {\em
 waist} que ce qu'il aurait été dans le vide~: la plupart d'entre
eux passent simplement à coté de la zone centrale de champ intense
au {\em waist}.

\paragraph{Couplage synchro-bétatronique} ~

Un autre effet intervient ici, mettant en jeu à la fois
l'évolution des électrons du faisceau par rapport à l'onde
plasma dans le plan transverse et sur l'axe longitudinal.
De ce point de vue, on peut parler de couplage synchro-bétatronique.

Le faisceau d'électrons est piégé transversalement dans l'onde
plasma sur une longueur de l'ordre de 20~mm \cite{balte},
supérieure à la longueur de déphasage 
$L_{\varphi}= \gamma_0^2\lambda_p \approx 8~\text{mm}$ où $\gamma_0$
est le facteur de Lorentz d'injection des électrons.
 
Les électrons explorent donc en fait tout une plage de phases par
rapport à l'onde plasma, et en particulier traversent à un moment
ou à un autre une région défocalisante de l'onde, et sont
expulsés avant d'avoir pu  se faire accélérer au {\em waist}.

\begin{figure}[tbh] 
\begin{center}
 \epsfig{figure=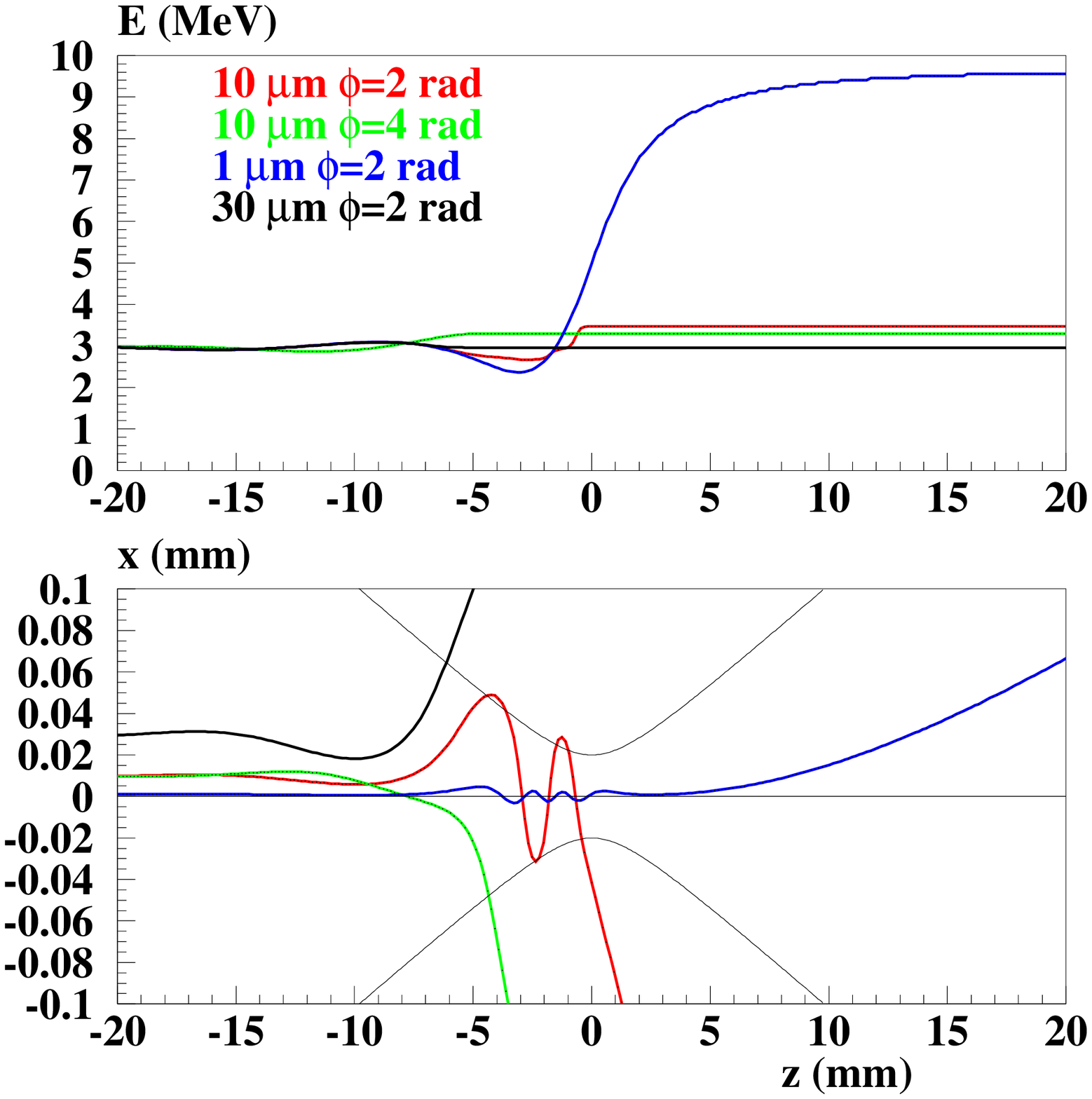,width=0.5\linewidth} 
 \epsfig{figure=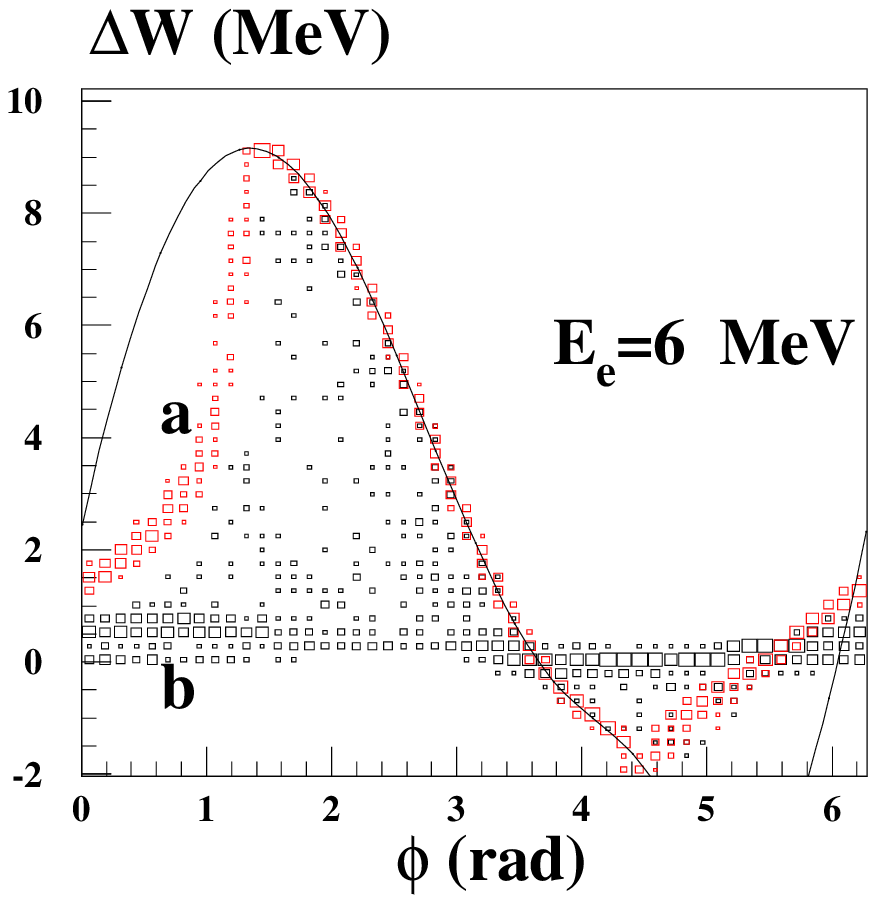,width=0.4\linewidth} 
\caption{
Trajectoires d'électrons dans l'onde plasma (a).
Simulation de l'injection 
à une énergie double de celle de l'expérience (b).
 Ligne : faisceau sur axe.
(a) Rouge :  émittance faible.
(b) Noir :  émittance réelle.
\label{fig:resultat:trace}}
\end{center}
\end{figure}

Un tracé de rayon confirme l'ordre de grandeur de la longueur 
sur laquelle se produit l'effet,
(figure \ref{fig:resultat:trace}a).
Une simulation avec des électrons injectés à une énergie supérieure (6
MeV, figure \ref{fig:resultat:trace}b), aurait permis d'observer des électrons
accélérés jusqu'au maximum.

J'en viens maintenant à un autre mode de création d'une onde
plasma, le sillage laser auto-modulé.


\subsection{SM-LWFA : Sillage auto-modulé}
Un autre mécanisme d'excitation d'une onde plasma, le sillage
auto-modulé, a été aussi étudié, avec des résultats
impressionnants, et en particulier des champs longitudinaux de
plusieurs centaines de GV/m.
Dans ce processus, une impulsion longue ($\tau \gg 1/\omega_{p}$)
génère une onde plasma extr\^emement intense dans un plasma dense
($n \approx 10^{19}~\text{cm}^{-3}$).

On peut se demander comment l'onde plasma est excitée hors
résonance.
Supposons qu'une onde plasma existe déja dans le plasma.
Alors, la densité électronique est suffisamment élevée pour
que l'onde plasma crée une modulation longitudinale 
de l'indice de réfraction 
(cette modulation est identique à celle obtenue par la
méthode du battement).
L'impulsion laser chevauchant ce milieu d'indice de réfraction
modulé va voir à son tour son intensité modulée~: c'est un
début de {\em bunching} (de groupage ?) du paquet de photons, avec
un pas égal à $\lambda_{p}$.

La force pondéromotrice de l'intensité laser modulée amplifie
alors l'onde plasma~: on est m\^eme exactement à la résonance !
Ce mécanisme se développe à partir d'une instabilité, on dit qu'il ``démarre sur le bruit''
et il
est donc {\bf impropre} à l'accélération de particules :
\begin{itemize}
\item
D'une part la phase de l'onde par rapport à l'impulsion laser
incidente est imprédictible.
\item
D'autre part l'onde plasma est faiblement relativiste dans ce plasma
très dense ($\gamma_{p} = \sqrt{n_{c}/n} < 10$)~: la longueur de
déphasage pour des particules ultra-relativistes est de l'ordre de
quelques centaines de microns, ce qui est inutilisable pour
l'accélération de particules de haute énergie.
\end{itemize}
Néanmoins cette création d'onde plasma par sillage auto-modulé
est un laboratoire o\`u nous pouvons tester notre compréhension des
ondes plasma intenses.
La durée de l'impulsion laser étant plus longue que
$1/\omega_{p}$, la raie laser est suffisament étroite pour pouvoir
utiliser  la diffusion Thomson.

Le sillage laser auto-modulé a été observé lors d'une
expérience menée au Rutherford Appelton Laboratory (RAL) avec le
laser Vulcan, qui fournit une impulsion de 20~J, en 1~ps
(FWHM)\cite{ModenaSMLWFA}. 
L'intensité laser  atteint $5~10^{18}\text{W/cm}^{2}$, dans un
jet de gaz d'hydrogène ou d'hélium de 4~mm de diamètre, avec
une densité électronique pouvant atteindre $1.5~10^{19}\text{cm}^{-3}$. 

L'interaction entre l'onde plasma extr\^emement intense, non
linéaire, et le laser produit des raies satellites d'ordres
successifs très intenses (figure 7 de la référence
\cite{astikarda}, pour $n=0.54~10^{19}\text{cm}^{-3}$).

Pour des valeurs plus élevées de la densité électronique,
l'onde plasma déferle, et sa cohérence est brisée (m\^eme
courbe, $n=1.5~10^{19}\text{cm}^{-3}$).
On observe alors la production d'une bouffée d'électrons
piégés par l'onde plasma et accélérés violemment vers
l'avant, (figure 3 de la référence \cite{astikarda}),
jusqu'à une énergie de 44~MeV, limite de détection de
l'appareillage. 
Le champ électrique longitudinal correspondant est estimé à
100~GV/m.
Une expérience plus récente effectuée par le m\^eme groupe, avec
un spectromètre ayant  une plage de détection plus étendue,
a détecté des électrons jusqu'à une énergie de
100~MeV\cite{gordon}.

De nombreuses expériences menées par d'autres groupes ont
confirmé ces résultats.

\subsection{Injection optique}

\begin{figure}[th] \begin{center}

~

\setlength{\unitlength}{1mm}
 \begin{picture}(80,40)
 \put(0,00){\makebox(80,40) {\epsfig{file=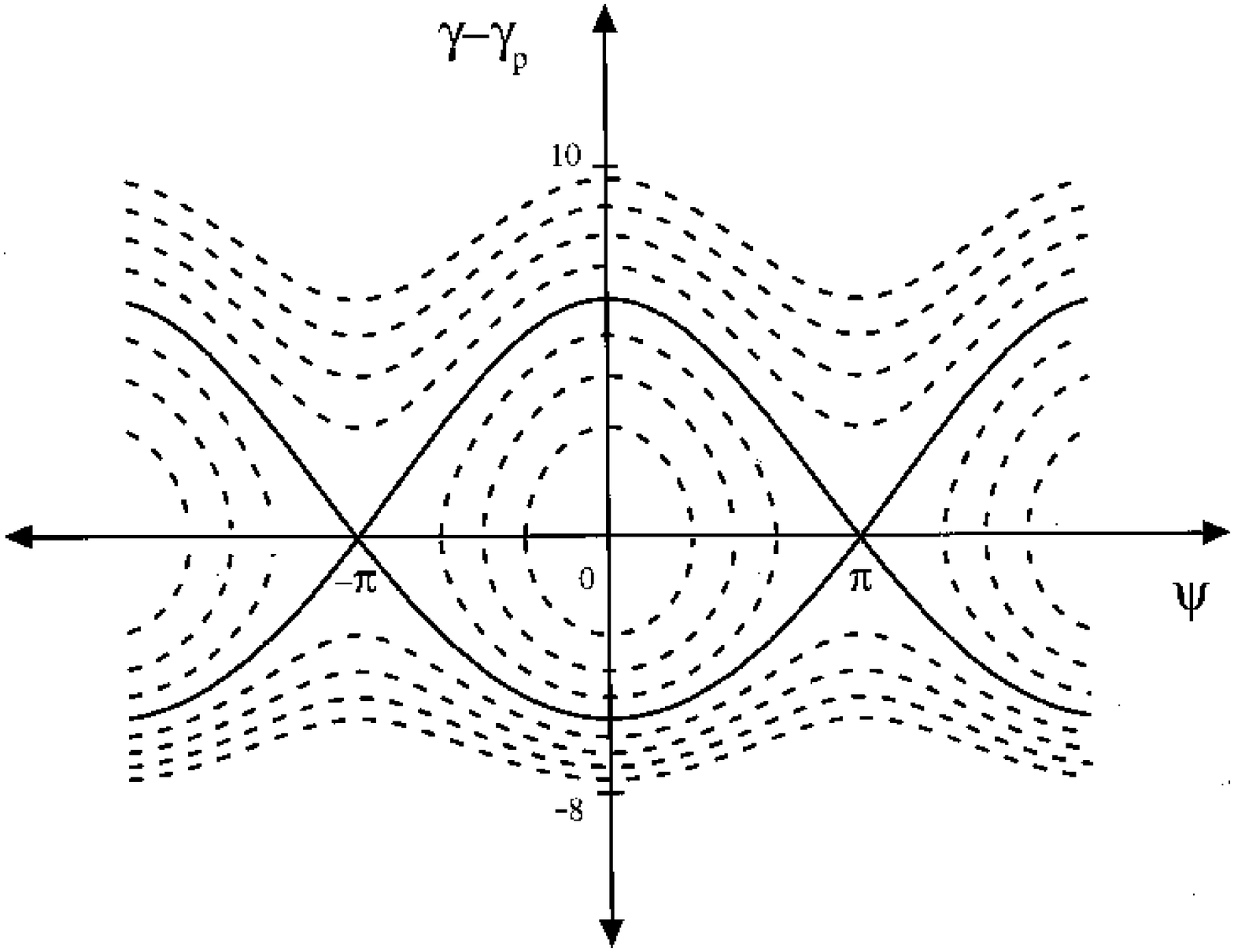,width=100\unitlength}}}
 \put(38,-1){\boul\bf(a)}
 \put(38,3.5){\boul\bf(b)}
 \put(52,15.){\boul {\bf(c)}}
 \put(38,29.5){\boul\bf(d)}
 \end{picture}

~

\caption{Trajectoire d'électrons dans le plan (phase, $\gamma-\gamma_p$)
($\gamma_p=20$) \protect\cite{esarey}.
{\bf(a)} Electron non piégeable,
{\bf(b)} électron piégeable,
{\bf(c)} point de compression maximale,
{\bf(d)} point de largeur en énergie minimale.
\label{fig:courbe:phase}}
\end{center} \end{figure}

Les électrons du plasma ne peuvent être piégés dans le puits de potentiel
d'une onde plasma non-déferlente : leur vitesse est trop inférieure à
la vitesse de phase de l'onde (fig. \ref {fig:courbe:phase}, point
{\bf(a)}).
Pour dépasser les limitations du sillage auto-modulé, plusieurs
groupes ont développé des schémas d'injection optique d'un petit
volume d'électrons dans une onde plasma sub-déferlante.
Une impulsion est communiquée au paquet qui est transféré sur une
trajectoire piégée (fig. \ref {fig:courbe:phase}, point {\bf(b)}).

Seul ce paquet est accéléré, et sa phase est déterminée par celle de
l'impulsion d'injection.

~

~

~
 
\paragraph{LILAC} (Laser Injected Laser Accelerator) 
L'idée originale \cite{inj:Um:PRL,inj:Um:PRE}  était 
de créer une EPW par un premier faisceau de pompe.
Les électrons du plasma décrivent une courbe fermée dans l'espace
(phase, énergie) : aucun n'est piégé par l'onde.
Une impulsion courte additionnelle est injectée
perpendiculairement à l'axe de l'onde plasma.
La force pondéromotrice de l'impulsion laser injecte un petit volume
d'électrons dans l'EPW.

\paragraph{CPI} (Colliding Pulse Injection)

Une autre méthode \cite{inj:Leemans:PRL} utilise un couple d'impulsions
contre-propagatives, outre l'impulsion pompe.  Leur battement crée une
force pondéromotrice qui injecte les électrons dans l'onde.

Dans une  version plus simple, une seule impulsion entre en collision avec
 l'impusion pompe \cite{inj:Leemans:PRE}. La force pondéromotrice du
 battement entre l'impulsion pompe et l'impulsion contre-propagative
 produit l'injection.
Ces méthodes permettent de prélever un petit volume des électrons du
plasma et de l'injecter avec une phase contrôlée dans l'onde plasma.
L'émittance et la largeur relative en énergie sont excellentes, la
longueur du paquet de quelques $\Uu{m}$, et la charge produite
d'environ 10 pC, pour une large gamme de densité électronique (table
\ref{tab:inj}).

\paragraph{Injection par deux impulsions co-propagatives}

\begin{figure}[tbh] 
\begin{center}
 \mbox{\epsfig{figure=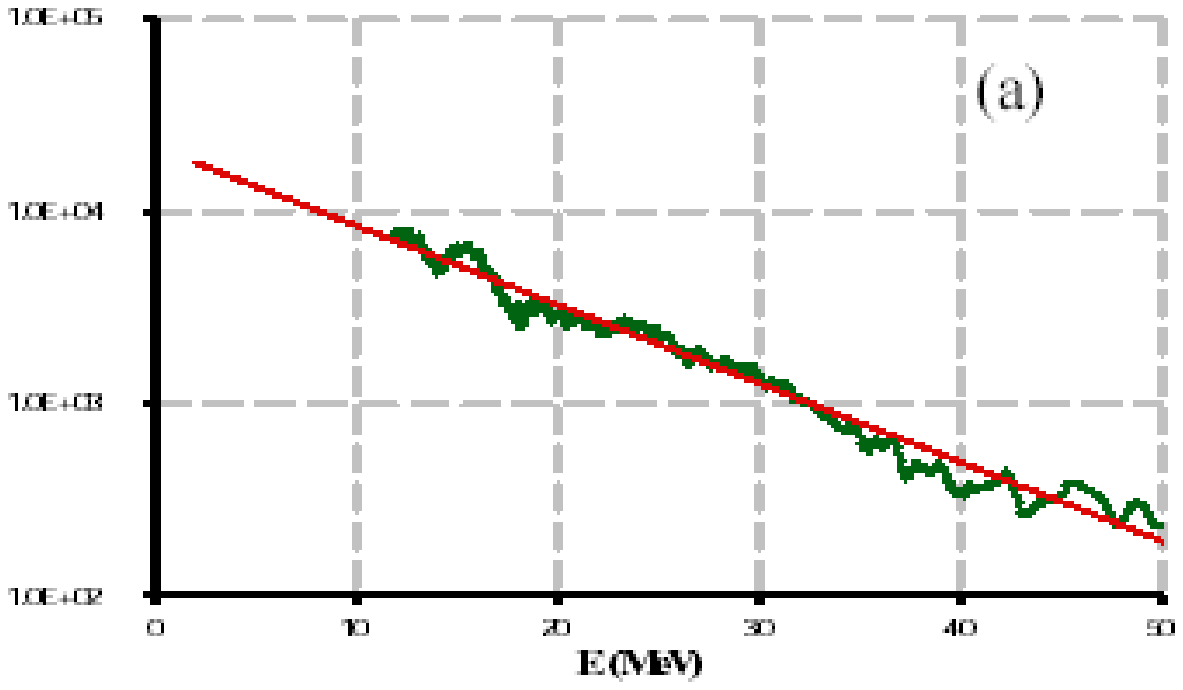,width=0.49\linewidth}}
 \mbox{\epsfig{figure=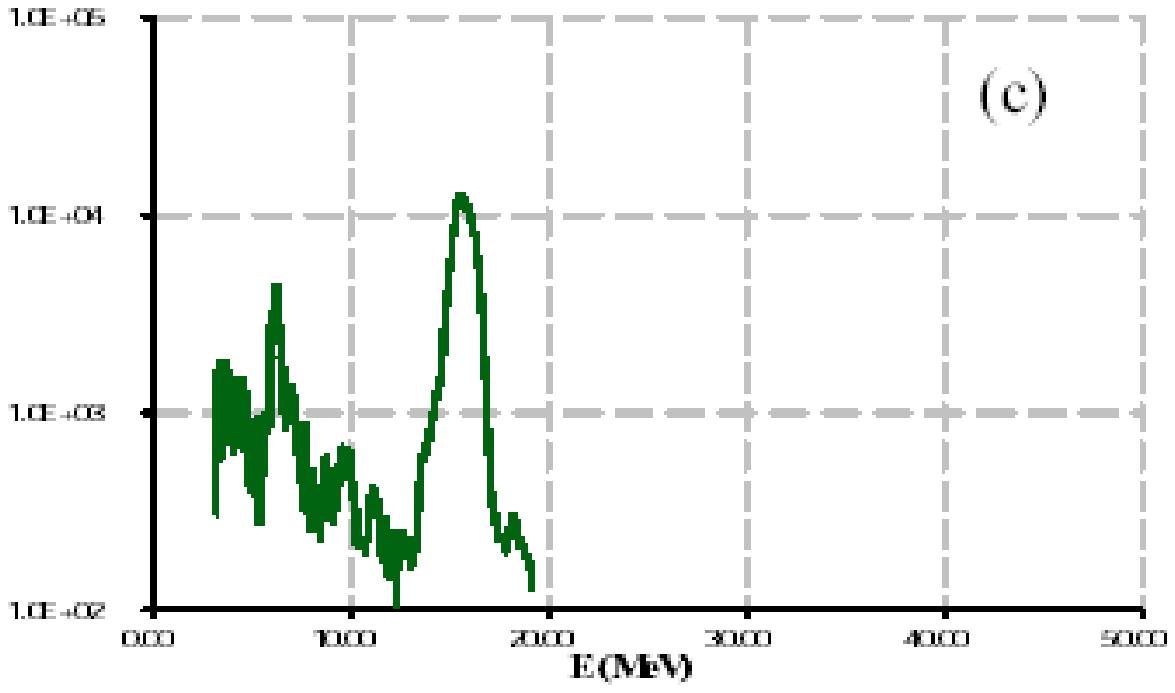,width=0.49\linewidth}}
\caption{Injection d'électrons dans une onde plasma au R.A.L. \cite{inj:AlecRAL}
Gauche : une seule impulsion.
Droite : deux impulsions.
\label{fig:inj:col}}
\end{center}
\end{figure}

Il s'agit ici de résultats expérimentaux 
\cite{inj:AlecRAL}.
Une impulsion intense  focalisée fortement
 ($a=4$, f/3, 250~mJ, 40~fs) crée une onde plasma qui éventuellement déferle, et produit un spectre thermique habituel (figure \ref{fig:inj:col}, gauche).
L'addition d'une impulsion similaire, moins focalisée ($a=0.2$, f/20) assure l'injection.

\begin{table}
\caption{Simulations 
(LILAC, CPI, CPI2)
et observation (Co-prop)
d'injection laser d'un paquet d'électrons du plasma dans une onde plasma non-déferlante\label{tab:inj}.}
\begin{center}\begin{tabular}{llllll}
 &  & LILAC & CPI & CPI2 & Co-prop \\
 &  & \cite{inj:Um:PRE} &  \cite{inj:Leemans:PRL} &  \cite{inj:Leemans:PRE}& \cite{inj:AlecRAL}\\
champs normalisés & $a_0, a_1, (a_2)$ & 1.6, 2.0  & 1.0, 0.5, 0.5  & 1.0, 0.2 & 4.0, 0.2\\
densité & $ n(\UUc{m}{-3})$ & $10^{19}$ & $10^{17}$  & $7.~10^{17}$ &  $10^{19}$   \\
energie & MeV & 20 & 27 & 10 & 15     \\
largeur & (\%) & 20  & 0.3 & 4 & $<10$    \\
longueur du paquet & $\sigma_z$ ($\Uu{m}$) & 3 & 1 & 0.3  \\
charge & $q (\Up{C})$ & 10 & 12 & 10 & 1 \cite{Alec}    \\
émittance normalisée & $\epsilon_{\perp N} (\pi~ \Um{m}.\Um{rad})$ & 2 & (1D)  & 0.4 \\  

\end{tabular}\end{center}
\end{table}

\subsection{Production d'impulsions d'électrons mono-cinétiques}

Trois groupes expérimentaux ont démontré en 2004 la production
d'impulsions mono-énergétiques, pour un réglage particulier des
paramètres d'intensité laser, de densité électronique, et de longueur
totale d'accélération
(table \ref{tab:mono}).

\begin{figure}[tbh] 
\begin{center}
 \mbox{\epsfig{figure=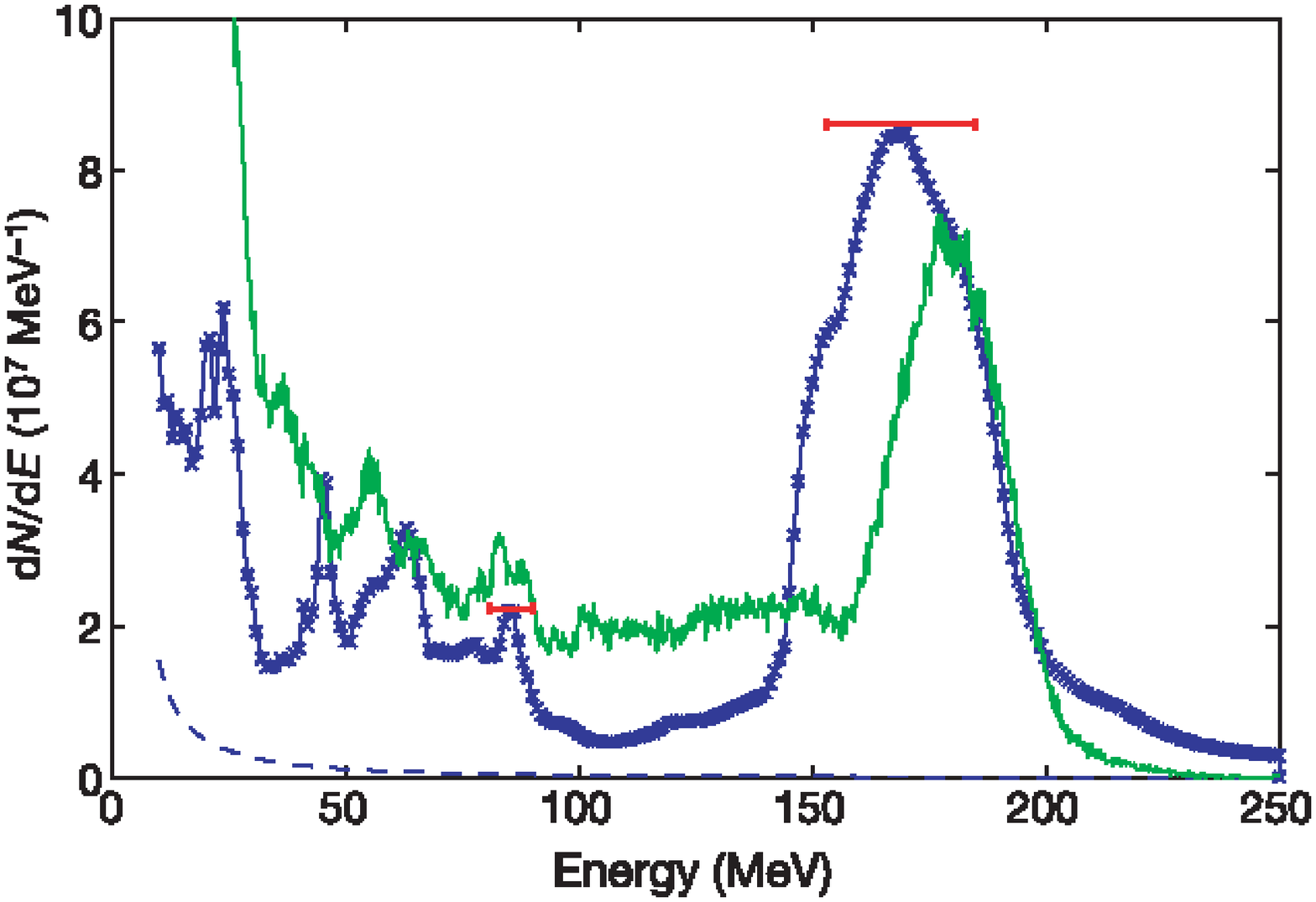,width=0.6\linewidth}}
\caption{Production d'impulsions d'électrons mono-cinétique :
Spectre observé (croix ``bleues'') et simulés (ligne ``verte'', 
 simulations PIC  3D) au Laboratoire d'Optique Appliquée (LOA).
 La barre ``rouge'' indique la résolution expérimentale
\protect\cite{LOAAAC04}.
\label{fig:LOAAAC04}}
\end{center}
\end{figure}

\begin{table}
\caption{Production de bouffée d'électrons monoénergétique par une impulsion
laser courte dans un plasma.
 (les quantités sous la ligne sont calculées par moi; j'ai supposé que
la longueur des paquets produits est $\approx$ 10 fs, conformément aux
simulations, dans le calcul du courant de crête.)
Les paramètres d'un des photo-injecteurs les plus brillants sur le
marché actuellement sont listés pour comparaison \cite{piot}.
$\Delta W$ désigne l'énergie mesurée, et  $e E_{z} L_\varphi$ la valeur maximale 
limitée par le déphasage.
Noter que la largeur en énergie mesurée est limitée par la résolution de l'appareillage.
\label{tab:mono}}
\begin{center}
\begin{tabular}{llll|l}
 & LBNL & LOA & RAL & ATF \\
 & \cite{LBLAAC04} & \cite{LOAAAC04} & \cite{RALAAC04} & \cite{piot}\\
\hline
 $E$ (J) & 0.5 & 1 & 0.5 & \\
 $\tau$ (fs) & 55 & 33 & 40 & 3000 \\
 $n$ ($e^-/\text{cm}^3$)&  $1.9 ~10^{19}$ &  $6. ~10^{18}$ &  $2. ~10^{19}$ \\
 $I$ ($\text{W/cm}^2$) &   $7. ~10^{18}$ &  $3. ~10^{18}$ &  $2.5. ~10^{18}$ \\
 $L$ (mm) & 1.7 & 3 & 2 \\
 $\Delta W$ (MeV) & 86 & 170 & 78 & 70\\
$\delta E/E$ HWHM (\%) & 2 & 12 & 3 & 0.03 \\
divergence (mrad) & 3 & 10 & (87) \\
diamètre laser ($\Uu{m}$)& 7 & 18 & 25  \\
$q$ (pC) & 300 & 500 & 22 & 1000 \\
\hline
$z_0$  (mm) & 0.2 & 1.3 & 2.5 \\
$\lambda_p$ ($\Uu{m}$) & 7.7 & 13.7 & 7.5 \\
$L_\varphi$ (mm) & 0.7 & 4.0 & 0.7 \\

$\epsilon_N$  RMS (mm.mrad) & 0.74 & 13 & 70 & 0.8 \\
$I$ (kA) & 30 & 50 & 2.2 & 0.17 \\
$B$ (A/(m.rad)$^2$) & $5.5~10^{16}$ &  $3.0 ~10^{14}$ &   $4.5 ~10^{11}$ &   $2.5 ~10^{14}$\\
~ \\
$\tau_r$ (fs) résonante pour $n$ & 13 & 24 & 13 \\
$E_{z} (\UG{V/m})$ & 109 & 78 & 54 \\
$e E_{z} L_\varphi (\UM{eV})$ & 76 & 312 & 38 \\
$a$ & 1.8 & 1.2 & 1.1 \\
$\delta$ & 1.1 & 0.46 & 0.38 \\ 
\end{tabular}\end{center}
\end{table}


Les trois expériences utilisent un plasma dense $ n\approx 10^{19}e^-/\text{cm}^3$.
L'emploi d'un jet de gaz permet la  focalisation du laser à son waist, à l'entrée du plasma.
La largeur du jet définit aussi la longueur maximale d'accélération $L$.

Les impulsions laser utilisées sont en régime légèrement relativiste
(le champ électrique normalisé du laser $a$ (eq. (\ref{eq:def:a})) est
proche de 1), et les ondes plasma sont donc produites à la limite de
la zone linéaire ($\delta$ proche de 1).


\begin{itemize}
\item
La différence principale entre ces résultats concerne l'utilisation de
guidage par un plasma préformé par LBNL,
sans lequel la production d'électrons n'est pas observée.
Ceci se comprend au vu de leur longueur  de Rayleigh extrêmement courte.
\item
Notons que 
le groupe du RAL a observé par imagerie transverse que la longueur
d'interaction effective est proche de 0.5 mm, beaucoup plus courte que
la longueur totale du plasma $L$, et du même ordre de grandeur que la
longueur de déphasage.
\end{itemize}

Des simulations \cite{LBLAAC04} indiquent que l'on observe le pic du
Jacobien de la trajectoire des électrons dans le plan (phase, énergie)
(figure
\ref{fig:courbe:phase}) : la fin de l'accélération en sortie de zone
d'interaction survient alors précisément lorsque le paquet d'électrons
se trouve au point {\bf(d)} de la figure \ref{fig:courbe:phase}.

L'ordre de grandeur du gain en énergie maximal limité par le déphasage
 $e E_{z} L_\varphi$, est en accord avec le gain $\Delta W$ observé expérimentalement
(table  \ref{tab:mono})
(à un facteur 2), ce qui conforte l'hypothèse.

Le groupe du LOA interprète le résultat comme la démonstration d'un
régime de ``bulle'', d'après un modèle développé par Pukhov {\em et
al.} \cite{pukhov}.
Les électrons sont éjectés hors de l'axe, une demi-longueur d'onde
après le passage de l'impulsion laser, puis reviennent sur l'axe une
longueur d'onde aprés, établissant un fort champ longitudinal.
L'applicabilité de ce régime, démontré par simulation pour des
impulsions fortement relativistes ($a = 12$)  \cite{pukhov} aux expériences
mentionnées ci-dessus, utilisant des impulsions faiblement relativistes
($a \sim 1$) 
a été vérifié ultérieurement \cite{Pukhov:2004ee}.

De façon similaire au sillage auto-modulé d'une impulsion longue
décrit plus haut, un double effet se produit, de l'impulsion laser
sur le plasma, et du plasma sur l'impulsion :

\begin{itemize}
\item
Les électrons à l'avant de l'impulsion sont comprimés par l'impulsion laser
(effet chasse-neige)
\item le front montant de l'impulsion laser se raidit sur ce fort gradient de densité électronique
(la tête de l'impulsion étant soumise à une  densité plus élévée, et donc à un indice de réfraction plus faible, ralentit (décalage vers le rouge)).
La valeur de $a$ après compression est alors plus élévée que la valeur initiale.
\end{itemize}

Dans ce régime de bulle
aussi, l'énergie maximale atteinte par le faisceau d'électrons au pic
mono-énergétique est déterminée par le déphasage \cite{pukhov}.  Elle
est donnée par une loi d'échelle \cite{Gordienko} (fixant le champ
normalisé du laser $a=10$):
\begin{equation}
\Delta W [\UG{eV}] \approx 0.1 ( E [\U{J}] )^{2/3}
\end{equation}

\begin{figure}[tbh] 
\begin{center}
 \mbox{\epsfig{figure=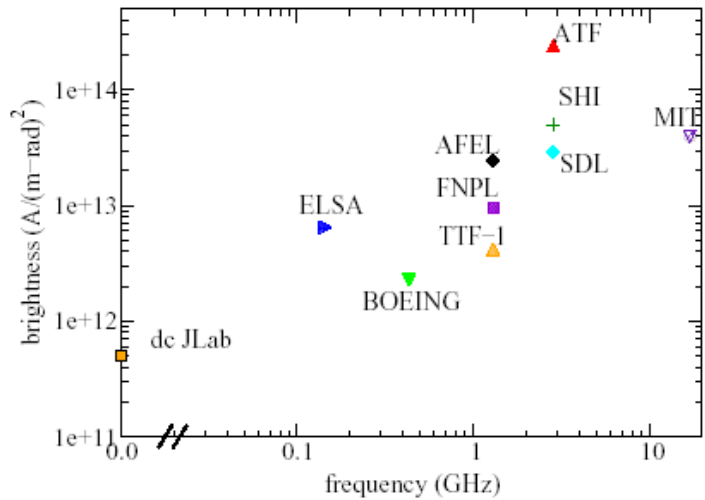,width=0.6\linewidth}}
\caption{Brillance des photo-injecteurs actuels en fonction de la fréquence de la RF
\protect\cite{piot}.
La brillance est le courant maximal divisé par le produit des émittances transverses normalisées.
\label{fig:phot-injectors}}
\end{center}
\end{figure}

Ces bouffées d'électrons ultra-courtes ont une brillance $B$
supérieure de plusieurs ordres de grandeur à celle des faisceaux
produits par les meilleurs photo-injecteurs, qui sont dans la gamme
$10^{12}$ -- $10^{15}$ A/(m.rad)$^2$ (figure \ref{fig:phot-injectors},
table \ref{tab:mono}).

\clearpage

\subsection{Expériences de première génération~: quels
 enseignements ?}

Cette décennie de développements nous a apporté de multiples
enseignements~: 
\paragraph{Création de l'onde plasma} ~

En ce qui concerne la création de l'onde plasma, il est apparu que
l'excitation de l'onde plasma par battement d'ondes est un processus
complexe, soumis à des phénomènes qui limitent la croissance de
l'onde~: le déphasage relativiste, qui désaccorde l'oscillateur
harmonique que constitue le plasma de la pompe laser; et quand la
densité électronique est suffisament élevée, l'instabilité
modulationnelle qui brise la cohérence de l'onde plasma, tout en
transférant l'énergie disponible à des ondes acoustiques
ioniques et à des ondes plasma (électroniques) non longitudinales
et non relativistes, inutiles en fait.

Ces problèmes peuvent \^etre dépassés par le développement de
battement d'ondes court, avec des impulsions de quelques picosecondes
seulement, qui ne comptent alors que quelques battements
\cite{claytonkyoto,claytonbalte}.

L'autre solution est l'emploi du sillage laser.
L'étude de l'excitation d'ondes plasma radiales par
interférométrie dans l'espace temps-fréquence a montré que
ce processus était bien compris, efficace, et qu'il permet de
créer une onde plasma jusqu'à la limite du domaine linéaire
(soit $\delta \approx 1$).

\paragraph{Accélération de particules} ~

Les premiers résultats obtenus en {\bf battement} ont permis de
démontrer le principe de la méthode.
Pour la première fois, des particules chargées ont été
accélérées avec un gradient supérieur à 1~GV/m.


L'expérience effectuée sur le site de l'Ecole a pu mettre en
oeuvre pour la première fois l'accélération par {\bf  sillage laser}. 
Les conditions expérimentales restant dans le domaine
linéaire, nous espérions pouvoir caractériser finement ce
processus, et en particulier reproduire la variation linéaire du
gain maximal d'énergie des électrons en fonction de l'énergie
laser, ainsi que sa dépendance en fonction de l'accord pompe/plasma,
c'est-à-dire en fonction de $\omega_{p}\tau$.
L'effet inattendu de la capture des électrons dans le champ
focalisant de l'onde plasma, très en amont du waist, associée au
glissement de phase amenant les électrons trop tôt dans une région
défocalisante de l'onde plasma, domine en fait la forme du spectre.

Plusieurs méthodes permettent d'éviter cet effet. L'emploi d'un
faisceau d'électrons d'énergie plus élevée, moins courbés
par le champ électrique transverse est une solution.
L'emploi d'un jet de gaz, limitant l'expansion longitudinale de l'onde
plasma en est une autre.
Ceci emp\^eche la capture des électrons loin du {\em waist}, alors
que la taille du faisceau est encore élevée.

\paragraph{Production de paquets d'électrons} ~

De nombreuses expériences de production d'électrons à partir d'un plasma ont été menées.

\begin{itemize}
\item
Dans un premier temps,  {\bf l'auto-modulation de phase} de l'onde puis son
déferlement ont produit  un paquet d'électrons collimés mais de
phase incertaine et de spectre énergétique large,
\item ensuite des procédés d'injection optique ont été proposés, qui 
permettent de fixer la phase,
\item enfin, un réglage fin des paramètres expérimentaux a permis la production
d'un pic {\bf  mono-énergétique}.
\end{itemize}

Ces derniers résultats sont obtenus dans un domaine de paramètres éloigné de
celui qui nous intéresse ici (ex : haute / basse densité).
Il ne fait pas de doute que ces nouvelles sources d'électrons ultra
courtes ($\approx $10 fs) et ultra-brillantes (table \ref{tab:mono})
trouveront leur usage, en particulier comme injecteurs.

~

Les différentes expériences d'accélération effectuées
jusqu'à ce jour ont focalisé leur attention sur le champ
accélérateur créé. 
Mais la spécification d'une ``bonne cavité'' ne se limite pas à
la valeur du champ. 
Plusieurs autres paramètres sont d'une grande importance, par exemple
la conservation des émittances longitudinale et transverse, et
l'efficacité du transfert de l'énergie de la prise électrique vers le
faisceau.

Dans la section suivante, j'explore brièvement ce que pourrait
\^etre une expérience de seconde génération, à la fois sous
l'angle de la performance ``pure'' du gain d'énergie, et des
méthodes de canalisation qui permettront peut-\^etre de transférer
une partie significative de l'énergie laser à l'onde plasma, puis
au faisceau à accélérer.

\clearpage

\section{Vers un développement de deuxième génération}\label{sec:deux}

Dans cette section, j'essaie d'utiliser les enseignements des
expériences de première génération pour déterminer ce que
pourrait \^etre une expérience de deuxième génération.

Je commence par le dimensionnement d'une expérience
d'accélération par sillage laser avec un faisceau laser
diffractant normalement, c'est-à-dire sans guidage.
(basé sur \boul \cite{kyoto_one} F), dans un premier temps.

Je décris ensuite l'intér\^et de guider, ou canaliser le faisceau
laser sur une longueur supérieure à la longueur de Rayleigh, à
la fois du point de vue de l'efficacité du transfert d'énergie du
laser vers l'onde plasma et de celui d'obtenir une grande longueur
d'accélération.
Je décris les expériences réalisées à ce jour pour explorer
les différentes méthodes de guidage, ainsi que leur intér\^et et
leurs limites pour l'accélération de particules dans un plasma excité par un laser.

\subsection{Dimensionnement d'une expérience d'accélération par sillage laser}\label{subsec:dim}

J'explore ici l'espace des paramètres d'une cavité à plasma,
définissant une possible expérience d'accélération par sillage
laser de deuxième génération.

\subsubsection{Détermination des limites}
Je commence par lister les différentes limitations inhérentes au
sillage laser. 

\paragraph{Valeur de l'amplitude de l'onde plasma} ~

\begin{figure} \bc
 \mbox{\epsfig{figure=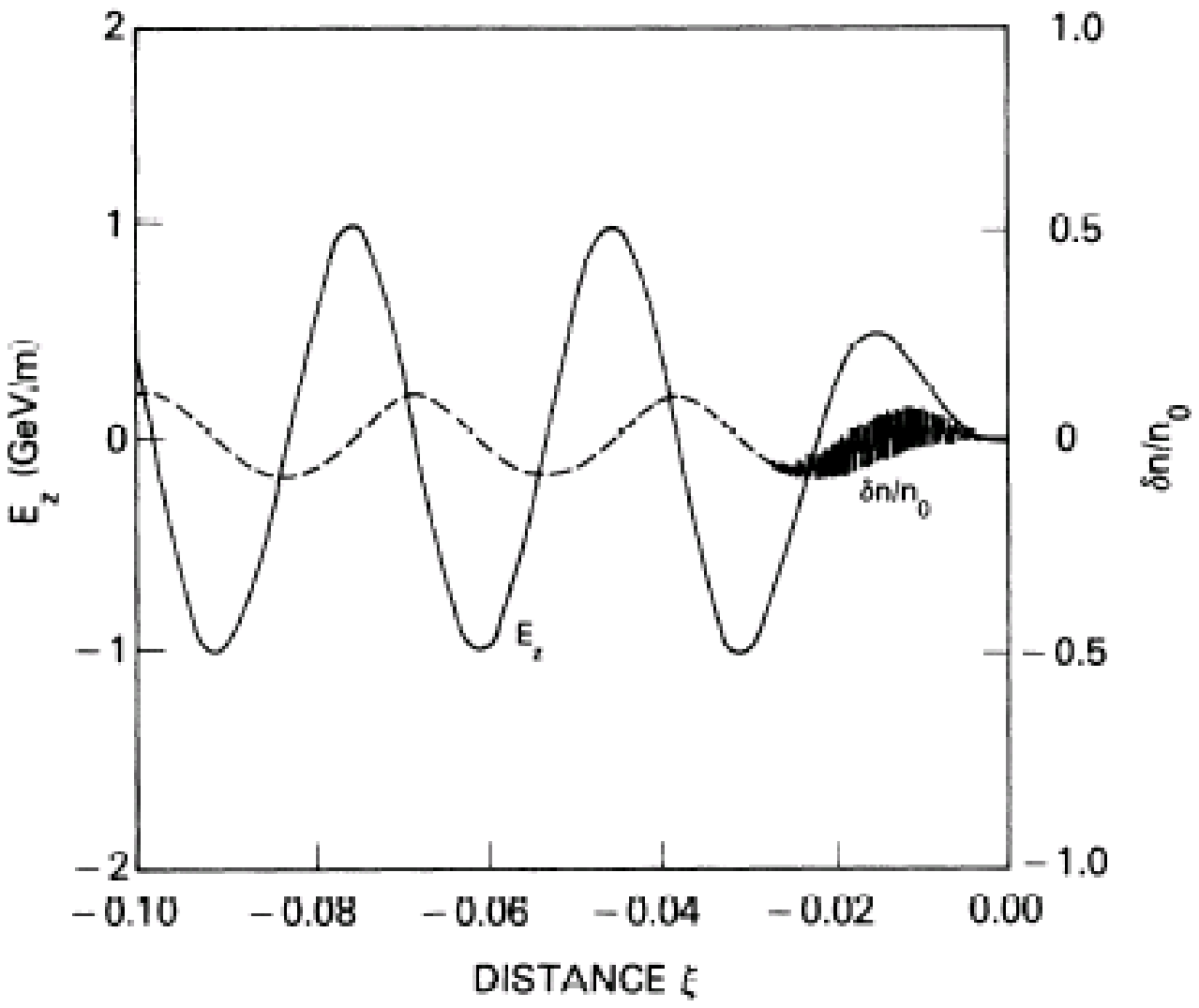,width=6cm}}
 \mbox{\epsfig{figure=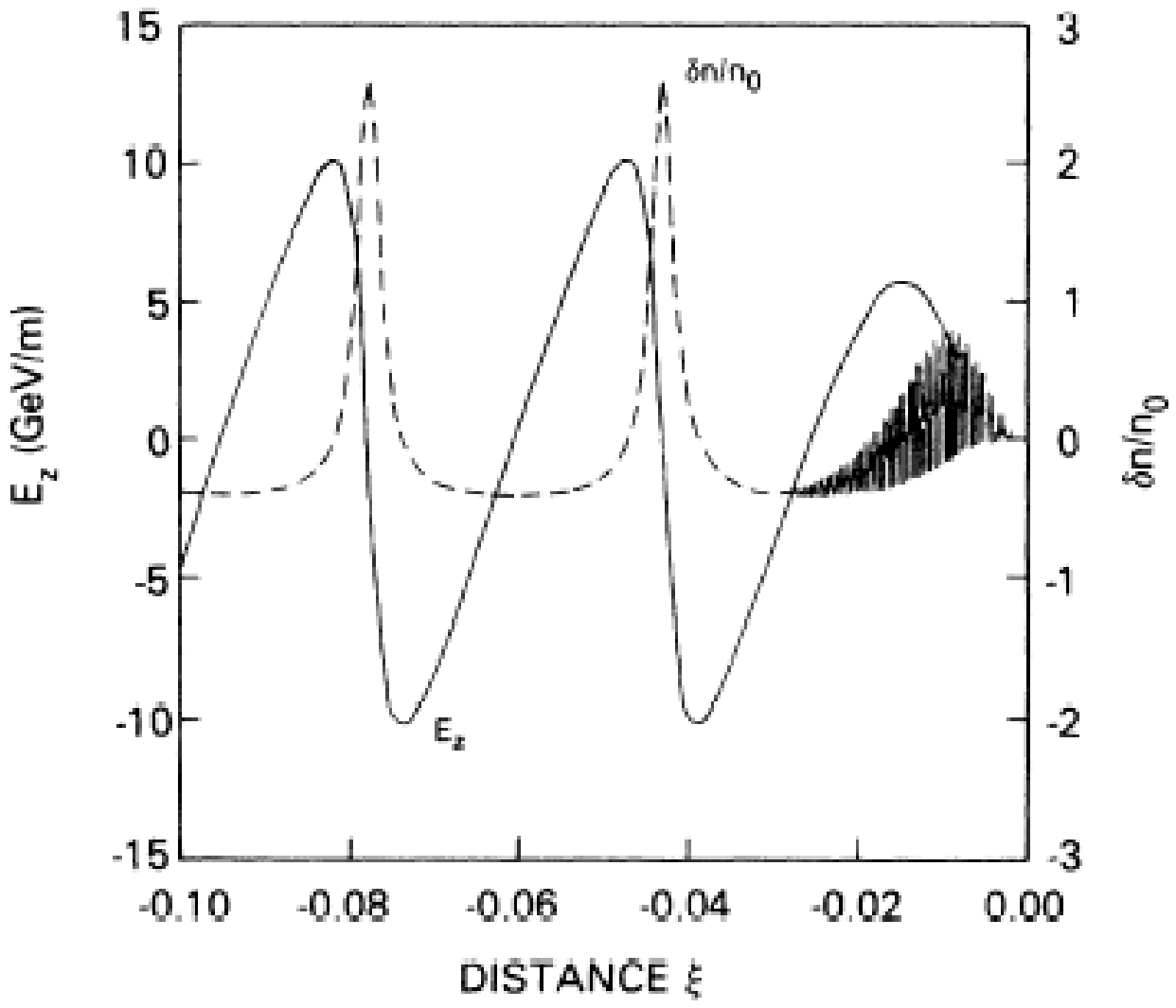,width=6cm}}
\caption{Onde plasma linéaire (gauche,
$a=0.5$), et non-linéaire (droite, 
$a=2.0$)
 \protect\cite{a2}.
\label{nonlineaire}}
 \ec \end{figure}

L'expression du champ accélérateur de l'onde plasma est 
(eq. \ref{eq:Ez:delta}) :
\begin{equation}
E_{z,max} = \frac{2\pi mc^{2}}{e\lambda_{p}}\delta=E_{0}\delta
\end{equation}

o\`u $E_{0}$ est le champ électrique de déferlement dans
l'approximation d'un plasma froid.
On veut  utiliser une onde plasma de grande amplitude.
Mais lorsque $\delta$ se rapproche de 1, l'onde plasma ne pouvant plus
\^etre traitée comme une petite perturbation de la densité
électronique, le traitement linéaire n'est plus valide.
Une résolution
numérique montre que l'onde plasma obtenue n'est plus
sinuso\"{\i}dale (figure \ref{nonlineaire}).

Dans le cadre de la théorie linéaire du sillage laser, on peut
établir une relation entre l'amplitude de l'onde plasma 
et la valeur du potentiel vecteur normalisé $a=eA/mc^{2}$ (eq. \ref{eq:a0:delta}):
\begin{equation}         
  \delta= \frac{\sqrt{\pi}}{2\mbox{\rm e}}a^{2}
\end{equation} 

L'onde plasma atteint donc le régime non linéaire ($\delta \approx 1$) 
lorsque le plasma devient relativiste ($a \approx 1.75$).
Ces ondes plasma non linéaires peuvent paraître intéressantes, car
dans la partie de plus grand gradient, le champ accélérateur est
extr\^emement élevé :
$\displaystyle {E}_{z}= a^{2} \left[1+
  a^{2}/4\right]^{1/2}\frac{\sqrt{\pi}}{2\mbox{\rm e}} E_{0} $.

Mais les ondes plasma non linéaires peuvent difficilement être employées pour un accélérateur, 
en particulier à cause de la variation de la longueur d'onde avec l'intensité 
(pour une EPW relativiste, on la notera $\lambda_{p,NL}$, et elle vaut 
asymptotiquement 
 $\lambda_{p,NL}=\lambda_p (2/\pi)(E_{max}/E_0)$ pour $E_{max}\gg E_0$).
Cette augmentation de la longueur d'onde est à peine visible sur la
figure \ref{nonlineaire}.
On prend donc :
       
\begin{equation}         
  \framebox{$\delta < 1$}
\end{equation} 

\paragraph{Cavitation due à l'onde plasma radiale} : importance de la taille transverse de l'EPW ~

Le champ électrique transverse de l'onde plasma peut influer fortement
sur le comportement transverse du faisceau, mais ce n'est pas ce qui
nous préoccupe ici.
Un autre effet apparaît lorsque l'amplitude plasma transverse
$\delta_{\perp}$ est supérieure à 1 : l'approximation de base du
traitement linéaire du sillage laser, c'est-à-dire le traitement
des ondes longitudinale et transverse comme deux perturbations
indépendantes de la densité électronique n'est plus valable.

Une simulation de la croissance de l'onde plasma par un code
particulaire \cite{DorchiesPhysPlasmas} 
 montre que
lorsque l'amplitude {\bf transverse} est élevée, $\delta_{\perp}>2$, le
développement de l'onde {\bf longitudinale} est affecté, avant m\^eme
que celle-ci ne devienne non linéaire 
 : il n'y a simplement plus
d'électrons dans la région proche de l'axe pour maintenir l'onde
plasma longitudinale.
On prend donc :

\begin{equation}         
  \framebox{$\delta_{\perp} < 1$}
\end{equation} 

\paragraph{Puissance maximale} ~

L'expression du   gain maximal (eq. \ref{defgainw})
\begin{equation}  
  \Delta \widehat{W} = \frac{r_{e} \lambda}{c^{2} \tau^{2}}
  \frac{16\ln{2}}{\mbox{\rm e}M^{2}} E 
\end{equation}
incite à l'utilisation d'impulsions courtes.
Cependant, au
dessus d'une puissance critique, $P_{crit}$, le faisceau subit une
auto-focalisation relativiste, ce qui conduit
rapidement au déferlement et à la destruction de l'onde.
\footnote{Dans le cas d'impulsions ultracourtes, $\omega_P \tau \ll 1$, la non-linéarité pondéromotrice (la densité électronique diminue dans les zones de champ élevé) compense
cette non-linéarité relativiste \cite{a2}.} 
La puissance critique d'autofocalisation relativiste est \cite{esarey}
:
\begin{equation}                  
P_{crit}= 2 \frac{m c^{3}}{r_{e}}
\left( \frac{\omega}{\omega_{p}}\right)^{2}
\end{equation}

soit $P_{crit}[\UG{W}]=17~\gamma_{p}^{2}$. 
On prend donc :
\begin{equation}         
  \framebox{$P < P_{crit}$}
\end{equation} 


\subsubsection{Optimisation}

On recherche alors un optimum de $\Delta W$ sous la contrainte
  $P=P_{crit}$, soit :
\begin{equation}                  
  2 \frac{m c^{3}}{r_{e}} \left( \frac{\omega}{\omega_{p}}\right)^{2}
  = \frac{E}{\tau}2\sqrt{\frac{\ln{2}}{\pi}}
\end{equation}

soit : 
\begin{equation}                  
E= \frac{\tau^{3} m c^{5}}{\lambda^{2} r_{e}}
\frac{\pi^{5/2}}{4 \ln{2}^{3/2}}
\end{equation}

ce qui donne : 
\begin{equation} 
  \Delta \widehat{W}= \frac{\tau}{\lambda} m c^{3}
  \frac{4\pi^{5/2}}{\mbox{\rm e}M^{2}\sqrt{\ln{2}}}
\end{equation}

On est alors conduit à {\bf augmenter} $\tau$, et donc $E$, et si
$\delta=1$ et $I$ sont fixés, on augmente la taille du waist $w_{0}$
et la longueur d'accélération $2z_{0}$ {\em ad infinitum}.

 On est alors conduit à fixer une limite arbitraire à cette
  escalade.  Constatant que deux approches différentes ont conduit
  les deux lasers actuels du LULI (350fs, 1.053$\mu$m, 10J) et du LOA
  (35fs, 0.8$\mu$m, 1J) à une m\^eme puissance maximale, je vais ici
  fixer la puissance laser.

On détermine alors la configuration par :

\begin{equation}         
  \framebox{
$\delta=1$
,~ ~
$P=P_{crit}$
,~ ~
$P=100~\UT{W}$
,~ ~
$\lambda=0.8\Uu{m}$
}
\end{equation} 

On déroule  la petite mécanique à l'envers, et on obtient les résultats listés en table 
\ref{tab:opt}.
%
%
La contrainte
  $\framebox{$\delta_{\perp} < 1$}$ est alors automatiquement respectée :
\begin{equation}         
 \frac{\delta_{\perp}}{\delta}=  \frac{E_{r}}{E_{z}} \approx 
 \frac{2}{\pi} < 1
\end{equation}

\begin{table}   
\caption{Valeurs des paramètres d'une ``cavité à plasma'', après
l'optimisation décrite dans le texte, en fonction de la puissance
crète de l'impulsion laser ($\delta=1$ et $P=P_{crit}$, $\lambda=0.8\Uu{m}$). \label{tab:opt}}
\begin{center} 
\begin{tabular}{lll} 
$P$ & $P$ & 100 TW 
\\ [2mm]
$\di \tau = \frac{\lambda}{\pi c}
\left[2\ln{2} \frac{P r_{e}}{mc^{3}}\right]^{\frac{1}{2}}$
& 
$\tau[\Uf{s}]=13.4~\lambda[\Uu{m}] \left(P[\UT{W}]\right)^{\frac{1}{2}}$
& 107 fs \\ [2mm]
$\di E= \frac{\lambda}{c} \left[\frac{r_{e}}{2\pi mc^{3}}\right]^{\frac{1}{2}}
P^{\frac{3}{2}}$
&
$E[\U{J}]=0.0142~\lambda[\Uu{m}] \left(P[\UT{W}]\right)^{\frac{3}{2}}$
& 11.4~J
\\ [2mm]
$\di \Delta W=\frac{8\pi^{2}}{\e M^{2}}
 \left[\frac{r_{e}mc P}{2\pi}\right]^{\frac{1}{2}}$
&
$\Delta W[\UM{eV}]=63.5\left(P[\UT{W}]\right)^{\frac{1}{2}}$
&
{\bf 635 MeV}
\\ [2mm]
$\di I_{max}= \e \sqrt{\pi} \frac{mc^{3}}{\lambda^{2} r_{e}}$
&
$I_{max}[\U{W}\UUc{m}{-2}]= 4.19~10^{18}/(\lambda[\Uu{m}])^{2}$
&
$ 6.55~10^{18}\U{W}\UUc{m}{-2}$
\\ [2mm]
$\di \gamma_{p}= \left[\frac{P r_{e}}{2mc^{3}}\right]^{\frac{1}{2}}$
&
$\gamma_{p}= 7.58  \left(P[\UT{W}]\right)^{\frac{1}{2}}$
&
75.8
\\ [2mm]
$\di \lambda_{p}= \lambda\left[\frac{P r_{e}}{2mc^{3}}\right]^{\frac{1}{2}}$
&
$\lambda_{p}= 7.58~\lambda[\Uu{m}]\left(P[\UT{W}]\right)^{\frac{1}{2}}$
&
$60.6~\Uu{m}$
\\ [2mm]
$\di w_{0}=  \lambda\left[\frac{2}{\pi^{\frac{3}{2}} \e}
\frac{P r_{e}}{mc^{3}}\right]^{\frac{1}{2}}$
&
$w_{0}[\Uu{m}]= 3.90~\lambda[\Uu{m}]\left(P[\UT{W}]\right)^{\frac{1}{2}}$
&
$31.2~\Uu{m}$
\\ [4mm]
$\di z_{0}=  \frac{2}{\e{\sqrt{\pi} M^{2}}}\frac{r_{e}}{mc^{3}}P\lambda$
&
$z_{0}[\Uu{m}]=  47.7~\lambda[\Uu{m}]P[\UT{W}]$
&
3.82 mm
\\ [2mm]
$\di (E_{z})_{max}=\sqrt{2\pi}\frac{mc^{2}}{r_{e}}\frac{1}{\lambda\sqrt{P}}
\left[\frac{c}{\epsilon_{0}}\right]^{\frac{1}{2}}$
&
$\di (E_{z})_{max}[\U{V/m}]=\frac{4.24~10^{11}}
{\lambda[\Uu{m}]\left(P[\UT{W}]\right)^{\frac{1}{2}}}$
&
53.~GV/m
\\ [4mm]
$\di n= \frac{2\pi mc^{3}}{r_{e}^{2}\lambda^{2}P}$
&
$\di n[\UUc{m}{-3}]=
\frac{1.94~10^{19}}{\left(\lambda[\Uu{m}]\right)^{2}P[\UT{W}]}$
&
$3.03~10^{17}\UUc{m}{-3}$
\\ [2mm]
 $\di \frac{\delta_{\perp}}{\delta}=  \frac{E_{r}}{E_{z}} \approx
 \frac{2}{\pi}$
&
&
0.64
\\[2mm]
$\di L_\varphi=\lambda \left[\frac{P r_{e}}{2mc^{3}}\right]^{\frac{3}{2}}$
&
$  L_\varphi[\Uc{m}]  = 0.0436 \lambda[\Uu{m}] \left(P[\UT{W}]\right)^{\frac{3}{2}}$
&
$34.8 \Uc{m}$
\\ [2mm]
$\di L_d= \lambda
\left[2\ln{2} \frac{P r_{e}}{mc^{3}}\right]^{\frac{3}{2}}
\frac{\text{e}}{\pi^{3/2}2\ln{2}}$
&
$ L_d[\Uc{m}]  = 0.0 \lambda[\Uu{m}] \left(P[\UT{W}]\right)^{\frac{3}{2}}$
&
$56. \Uc{m}$
\\ [2mm]
$\di N_{max}$
&
$  N_{max}=3.2~10^7 \lambda[\Uu{m}] \left(P[\UT{W}]\right)^{\frac{1}{2}}$
&
$2.6~10^{8}$
\\ [2mm]
$\di \Delta W_g \equiv E_{z,max}L_\varphi/2
 = \frac{P}{4} \left[\frac{\pi Z  r_{e}}{mc^{3}}\right]^{\frac{1}{2}}$
&
$\Delta W_g [\UG{eV}]= 0.092 \left(P[\UT{W}]\right)$
& 
{\bf $9.2 \UG{eV}$}
\\ [2mm]
$\di f\#= \left[
\frac{2\sqrt{\pi}}{\e}
\frac{P r_{e}}{mc^{3}}
\right]^{\frac{1}{2}}$
&
$\di  f\#=  \frac{\left(P[\UT{W}]\right)^{\frac{1}{2}}}{12.2}$
&
122.
 \\ [2mm]
\end{tabular}
\end{center}
\end{table}

\paragraph{Extension transverse et utilisation de l'énergie disponible} ~

Un faisceau d'électrons de petite taille transverse se propageant sur
l'axe
crée un champ de sillage sur une
distance transverse de l'ordre de $\lambda_{p}/2\pi$ \cite{Kats:Wilk}.
Ce même paquet est donc 
 capable d'extraire de l'énergie de l'onde plasma sur cette
distance transverse.
Une onde plasma plus large conduirait donc à une diminution de l'efficacité du transfert.
Notons que cette optimisation fournit une loi d'échelle :
\begin{equation} 
w_{0}/(\lambda_{p}/2\pi)=
4 \sqrt{\sqrt{\pi}/\text{e}} \approx 3.2
\end{equation}
Le paquet d'électrons n'utilise que  l'énergie  du coeur de l'onde plasma.

\paragraph{Transfert de l'énergie du laser vers l'onde plasma} ~

L'expérience montre que le laser est très peu perturbé à la
traversée de ce plasma très peu dense, quelques millibars d'hydrogène.
L'efficacité du transfert  de
l'énergie du laser vers l'onde plasma est  faible.
Une impulsion laser de durée $\tau$, de section $S$, de champ
électrique  ${\cal E}$ transporte une énergie
 $(c\tau S){\cal  E}^{2}/\epsilon_{0}$. 
Cette impulsion laser crée une onde plasma de m\^eme section, et
de longueur $L$. 
On obtient de m\^eme l'énergie de l'onde plasma : $(L
S)E_{z}^{2}/\epsilon_{0}$.
Le rapport $R$ des deux quantités est :
\be
R = \frac{L}{c\tau} \left(\frac{E_{z}}{\cal  E}\right)^{2}
\ee
Ce qui donne ici, avec $L=\pi z_{0}$, $R=2\%$, effectivement très faible.

On définit la {\bf longueur de déplétion} $L_{d}$ comme la longueur
sur laquelle le laser transfère toute son énergie à l'onde plasma,
soit $R=1$ et 
\be
L_{d} = c\tau \left(\frac{\cal  E}{E_{z}}\right)^{2}
\ee

A noter qu'après optimisation, le rapport de la longueur de
déplétion et de la longueur de déphasage relativiste $L_\varphi =
\lambda {\gamma_P}^3$, est constant :

\begin{equation} 
\frac{L_d}{L_\varphi} = 
\frac{4 \text{e} \sqrt{\ln{2}}}{\pi^{3/2}} \approx 1.35
\end{equation}

Le déphasage est donc le premier facteur limitant
(d'autant plus que pour éviter la zone défocalisante, on doit se limiter à $L_\varphi/2$).
Le gain d'énergie maximal en mode guidé $\Delta W_g$
sera donc de l'ordre de  $E_{z,max}L_\varphi/2$, qui est proportionnel à $P$.

\paragraph{Charge électrique maximale} ~

Lorsqu'un paquet d'électrons de charge élevée est injecté, la tête du
paquet bénéficie de la totalité du champ accélérateur, mais le champ de sillage du paquet atténue le champ de l'onde : la queue du paquet est moins accélérée.
La charge maximale est celle qui crée un champ de sillage ``égal et
opposé'' au champ de l'onde plasma. Le nombre de particules dans un
tel paquet est $N_{max}$ \cite{Kats:Wilk}:
\begin{equation} 
N_{max} 
\equiv 
\frac{ {\lambda_P}^2     e E_{z,max}}{16 \pi^3 r_e  mc^2}
\label{eq:Nmax}
\end{equation}

La dernière particule de ce paquet voit alors un champ nul, et le gain énergie 
des particules varie de 0 à 100 \% .
Un décalage de phase permet de compenser la baisse du champ résiduel,
dans une certaine mesure, méthode employée dans les accélérateurs
``métalliques''.
La même méthode peut être employée ici \cite{HighQualBeamChiou}.

Une simulation effectuée dans des conditions assez proches de celles de la
table \ref{tab:opt} montre \cite{ReitsmaTrines} que la largeur relative
en énergie augmente en début d'accélération, diminue, puis atteint un
minimum à 2.6\%; avec des paquets de longueur égale à $\lambda_p/6$, et
un beam loading de 80\% (c'est-à-dire un nombre d'électrons $N= 0.8
N_{max}$).
Ceci correspond ici (table \ref{tab:opt}) à $N = 2~10^8$ électrons, soit 30pC.

\begin{figure} \bc
 \mbox{\epsfig{figure=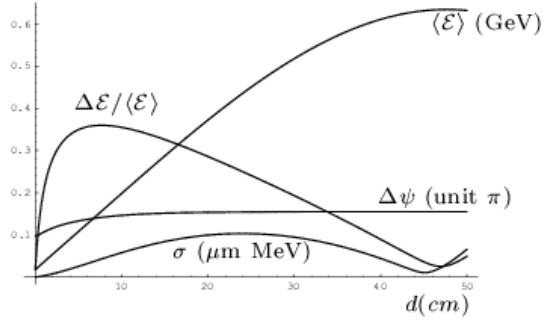,width=8cm}}
\caption{Energie moyenne $\langle {\cal E}\rangle$,
largeur relative en énergie,  $\Delta{\cal E}/ \langle {\cal E}\rangle$,
et émittance longitudinale $\sigma$ en fonction de la distance d'accélération
pour une longueur de paquet de $\lambda_p/6$, 
et une fraction de beam loading $\eta = 0.8$  
 \protect\cite{ReitsmaTrines}.
\label{fig:albert}}
 \ec \end{figure}


\paragraph{Optique}


Dans un accélérateur utilisant des cavités à sillage laser dans un
plasma, l'encombrement dû à l'optique de focalisation de l'impulsion
laser contribuera fortement à la longueur totale.
 La focale est limit\'ee par la tenue des
mat\'eriaux (r\'eseaux, miroirs après compression) \`a l'\'energie laser,
$F_{l} \approx 1\U{J}\UUc{m}{-2}$
\cite{Lenzner}, ce qui donne des miroirs de rayons :
\begin{equation}
  w_{i}^{2}= \frac{P\tau}{F_{l}} \frac{1}{2\sqrt{\pi\ln{2}}},
\end{equation}
 
soit $w_{i}=1.9~\Uc{m}$, et $f=2.3\U{m}$
(car l'ouverture de l'optique est fixée, $w_0$ étant fixé).
  La taille de l'optique de
focalisation affecte le taux de remplissage de
l'accélérateur en sections accélératrices : on perd  plus d'un ordre
de grandeur sur le gradient accélérateur moyen.

Cette focale 
 est proportionelle à $w_0 \sqrt{P\tau}$, soit à $P^{5/4}$, alors que
 le gain maximal par cavité avec guidage $\Delta W_g \equiv e E_z
 L_\varphi/2$ est proportionnel à $P^{1/2}$; le gradient effectif
 moyen ($4 \UG{V/m}$ pour $P=100\UT{W}$) décroît donc comme $P^{-3/4}$
 : une augmentation de la puissance disponible n'est pas souhaitable
 dans ce modèle.

\subsection{Guidage du Laser}

De nombreux mécanismes ont été envisagés pour réaliser le
guidage d'une impulsion laser sur une longueur  supérieure à
la longueur de Rayleigh.
Je les présente succintement ci-dessous, examinant en particulier 
la compatibilité avec le domaine favorable pour l'accélération de particules.
Deux types de méthodes ont été envisagées :
\begin{itemize}
\item guidage dans un milieu d'indice de réfraction inhomogène,
 ``fibre optique à plasma''.
\item et guidage aux parois d'un tube capillaire.
\end{itemize}

\subsubsection{Guidage dans un  milieu d'indice de réfraction inhomogène}

Comme dans une fibre optique, un excès d'indice de réfraction 
sur l'axe de propagation produit une focalisation de l'impulsion
laser. En particulier, si le profil d'indice est parabolique, 
$\eta = \eta_{0} + \Delta \eta (r/w_{0})^2$,
l'équation de Maxwell paraxiale donne 
l'évolution de la taille d'un faisceau gaussien comme :

\begin{equation}
 \frac{\partial ^{2} w }{\partial z^{2}} =
\frac{ w_{0}^{4}}{z_{0}^{2} w^{3}} 
\left( 1 -  \frac{\Delta \eta}{\Delta \eta_{c}} \left(\frac{w}{w_{0}}\right)^{4}
\right)
\end{equation}

o\`u $ \Delta \eta_{c}$ est une valeur ``critique'' de la variation de
l'indice de réfraction, pour laquelle une impulsion injectée avec un
diamètre $w_0$ est en équilibre :
$
\Delta \eta_{c} = \frac{\lambda ^{2}}{2 \pi^{2} \eta_{0} w_{0}^{2}}
$.
La taille du faisceau est adaptée au canal lorsque $w$ est conservé lors de la propagation, soit :
\begin{equation}
w= w_{0} \left(  {\frac{\Delta \eta_{c}}{\Delta \eta}} \right)^{1/4}
\end{equation}

Un faisceau injecté avec une taille différente verra sa taille osciller
lors de la propagation.

\subsubsection{Auto-focalisation relativiste dans un plasma}\label{sec:auto-foc}

La première méthode développée a été l'auto-focalisation d'une
impulsion intense dans un plasma de densité suffisante.
L'indice de réfraction du plasma est $\eta =\sqrt{1-(\omega_p/\omega)^2}$.
A haute intensité laser, $a>1$, les électrons du plasma deviennent
relativistes, de facteur de Lorentz $\gamma$,
et $\omega_p^{2} \sim \omega_{p0}^{2}/\gamma$ : le maximum d'intensité
laser sur l'axe de propagation produit donc un maximum de l'indice effectif, et
donc une focalisation ``relativiste''.
L'auto-focalisation relativiste a été démontrée au laboratoire de Los
Alamos \cite{Borisov} et confirmée depuis auprès du laser P102 du
CEA \cite{Monot}.
Comme déja mentioné, la diffraction est compensée exactement pour une puissance égale à :
\begin{equation}
P_{crit} = \gamma_p^2 2 mc^3/r_e,
\end{equation}
soit $ \approx 17 \text{GW}\gamma_p^2$.
Les seuils d'auto-focalisation relativiste et de déferlement, suivant
une auto-modulation de l'impulsion, sont proches : les nombreuses
expériences suivantes se sont orientées vers la production de bouffées
d'électrons du plasma, piégés et accélérés.

Ce mode de focalisation est {\bf inutilisable} pour l'accélération.
En effet, l'indice de réfraction est modifié au bout d'un temps
qui est de l'ordre de $1/\omega_p\sim \tau$, c'est-à-dire, à la
quasi-résonnance, après le passage d'une grande partie de
l'impulsion.  Effectivement, des simulations \cite{guidingSprangle}
montrent que seule la queue  de l'impulsion est guidée, et que la t\^ete
diffracte.

Par contre c'est un moyen efficace de guidage d'impulsions longues
($\tau \gg 1/\omega_p$), qui sont impropres à la génération d'ondes plasma 
linéaires propres à l'accélération contrôlée de particules.

\begin{table}
\caption{Méthodes de guidage d'impulsion laser courte.
\label{tab:guid}}
\begin{center}
\begin{tabular}{lllllllll}
              &  capillaire   &  capillaire  &  décharge  &       z-pinch  &  axicon      &  axicon     & ionisation-chauffage        \\  
              &            &             &  capillaire  &              &                  &  + décharge      \\ 
              & Kitagawa       &   Cros      &  Spence      &  Hosokai     &   Milchberg  &  Gaul        & Geddes  \\
& \cite{Kitagawa} &\cite{CrosPhyScript2004}&\cite{Spence} & \cite{Hosokai} &  \cite{DurfeePRL,DurfeePRE}    & \cite{Gaul}   & \cite{LBLAAC04} \\
gaz           &           &             &   H$_2$           &    He        &   Ar, N$_2$         &  He       &   H$_2$       \\
$n(\UUc{m}{-3})$&  $6~10^{16}$ & 0 ; $5~10^{16}$  & $2.7~10^{18}$ & $2~10^{17}$  & $3~10^{18}$    & $4~10^{18}$  &  $3~10^{19}$   \\
$I(\U{W}\UUc{m}{-2})$ & $3~10^{18}$ & 1 -- 2 $10^{17}$&   $10^{16}$  &  $10^{17}$   &  $10^{14}$      & $1.3~10^{17}$ & 1 -- 2.5 $10^{18}$ \\
$L$ (cm)      & 1              &     12      &      4       &    2         &       1      &     1.5     & 0.25  \\
$T$ (\%)      & 40             &    60       &      80      &     64       &    70        &    50       & 35  \\
$w_0 (\Uu{m})$&     25         &    20       & 37.5         &  20          &       25     & 14          & 7  \\
$E$ (J)       &  10            &       4     &   0.23       &  0.2         &    0.04      &    0.03     & 0.5  \\
$\tau(\Uf{s})$& 500            &    350      &     120      &   90         &    100 ps    &     80      & 55  \\
              &                &             &              &              &              &               \\
\end{tabular}
\end{center}
\end{table}

\subsubsection{Guidage d'une impulsion dans un plasma préformé}

Ici on établit un profil radial de densité électronique, et donc d'indice de réfraction, 
avant d'injecter l'impulsion courte à guider.
Dans le cas d'impulsions courtes, ou de l'ordre de $1/\omega_p$, la compensation exacte des instabilités Raman et modulationnelles permet la propagation guidée sur de grandes longueurs 
\cite{SprangleStableGuiding}.

Lorsque le profil de densité est parabolique, de profondeur $\Delta
n$,  la structure peut
guider un mode gaussien  de rayon donné $w_{0}$ tel que :
\be
\Delta n = \frac{1}{\pi r_{e} w_{0}^{2}}
\ee


La valeur du creux de densité électronique nécessaire au guidage
d'une impulsion $w=31~\mu\text{m}$ est $\Delta n  =
1.2~10^{17}\text{cm}^{-3}$, soit un tiers de la  densité
électronique résonnante, ce qui est tout à fait raisonnable.

Le guidage d'une impulsion laser par un plasma préformé a été
développé lors d'une série d'expériences réalisées à l'université du
Maryland \cite{DurfeePRL,DurfeePRE}.

Une impulsion laser est focalisée par un axicon, de fa\c{c}on à
créer un filament de $1~\mu\text{m}$ de rayon et de 1~cm de
longueur.
Le plasma chaud et de pression élevée s'étend radialement tout en
ionisant le gaz neutre qui l'entoure. Un minimum de densité électronique
 se développe. 
Une seconde impulsion sonde est injectée dans le canal créé.

Plusieurs méthodes sont envisagées pour créer le choc établissant la
distribution de  densité :

\begin{itemize}
\item la création d'un plasma par une première impulsion laser,
 en configuration axicon, comme cité ci-dessus, 

\begin{itemize}
\item  l'ionisation complète du plasma par l'impulsion laser longue peut être un problème
(si l'ionisation n'est pas complète, le faisceau laser court à guider va la terminer, et la réfraction
au dioptre entre les deux zones empêche une injection efficace).
L'emploi d'une décharge électrique ensemence le gaz en électrons ce qui favorise l'ionisation par le laser de chauffage (\cite{Gaul}) 
\end{itemize}
\item 
l'utilisation directe d'une décharge électrique dans un capillaire contenant un gaz, 
(\cite{Spence})

\item  l'utilisation d'une impulsion courte (15 mJ, 60 fs) qui ionise le gaz, suivie d'une impulsion longue (150 mJ, 250 ps) qui le chauffe et produit la détente \cite{Volfbeyn,LBLAAC04}.
L'intérêt ici est que l'impulsion longue, contrairement à une
impulsion courte, peut être guidée par autofocalisation relativiste.
C'est donc elle qui détermine la géométrie du canal.

\end{itemize}

Dans tous les cas mentionnés ci-dessus, une transmission efficace a
été observée, sur quelques centimètres, avec une atténuation de
quelques $\UU{m}{-1}$.
Mais ces méthodes reposent sur l'expansion hydrodynamique du plasma,
efficace seulement dans une plage de quelques $10^{18}$ à quelques
$10^{19}\UUc{m}{-3}$.

L'emploi d'une décharge en configuration $Z-pinch$ permet d'atteindre
des densités électroniques plus faibles \cite{Hosokai}, mais
l'inhomogénéité de la densité électronique peut être problématique
\cite{HosokaiNIM}.

\subsubsection{Guidage d'une impulsion dans un capillaire}

Ici on utilise un diélectrique creux, un capillaire, dans lequel une
onde mono-mode peut se propager avec de faibles pertes.
Le mode fondamental $\text{EH}_{11}$ a un profil d'intensité proche de
celui d'un faisceau gaussien.
Le couplage d'un faisceau gaussien à l'entrée du capillaire avec ce
mode est maximal lorsque le rayon du mode $w_0$ et celui du capillaire
$a$ sont tels que :
\begin{equation}
 w_{0} /a = 0.645,
\end{equation}

\begin{figure}[hbt] 
\caption{Profil d'intensité d'un mode ET$_{11}$, et du mode gaussien adapté. 
\label{fig:GaussBessel}}
\begin{center}
 \mbox{\epsfig{figure=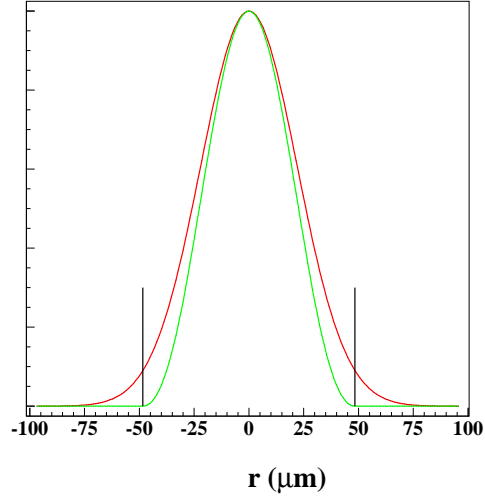,width=0.46\linewidth}}
\end{center} 
\end{figure}

Le couplage vaut alors 98\%.
La tenue du matériau constituant le capillaire dépend de 
 l'intensité au mur à l'intérieur du tube :
celle-ci, normalisée à l'intensité sur l'axe, vaut :
\begin{equation}
I(r=a) = 
I(r=0)
\left(
\frac{1.2 \lambda}{2\pi a}
\right)^2
\frac{\epsilon}{\sqrt{\epsilon-1}}
\end{equation}

où $\epsilon$ est la constante diélectrique du matériau constituant le tube
(2.25 pour du verre).
La longueur d'atténuation $\Lambda$ pour ce mode vaut 
\begin{equation}
\Lambda =
\frac{2 a^3}{u^2}
\left(
\frac{2\pi}{\lambda}
\right)^2
\frac{\sqrt{\epsilon-1}}{\epsilon+1}
\end{equation}

Pour les valeurs obtenues en table \ref{tab:opt}, ceci donne 
\footnote{$u$ est la position du premier zéro de la fonction de Bessel $J_0$, $u\approx 2.4$}
$a=48 \Uu{m}$,  $I(r=a)/I(r=0)=2.0~10^{-5}$,
soit  $I(r=a)=1.3~10^{14}    \U{W}/\UU{cm}{2}$,
(soit à peu près le seuil de dégradation des optiques en silice à
incidence normale en impulsion courte), et $\Lambda= 82 \Uc{m}$.
Cette approche a été validée  expérimentalement \cite{CrosPhyScript2004}.
Une impulsion de 350 fs d'une intensité de $10^{17}  \U{W}/\UU{cm}{2}$
a été guidée sur une longueur de 12 cm, soit 50 $Z_R$.

\begin{figure}[hbt] 
\caption{Guidage d'une impulsion laser dans un capillaire :
intensité transmise en fonction de la distance de guidage
\protect\cite{CrosPhyScript2004}.
\label{fig:cros}}
\begin{center}
 \mbox{\epsfig{figure=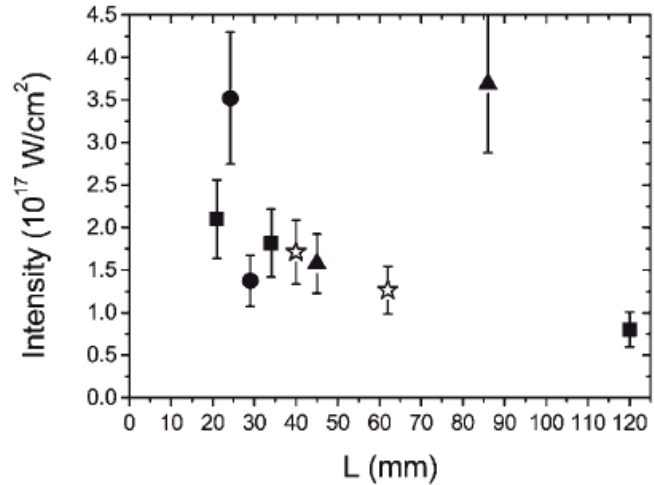,width=9cm}}
\end{center} 
\end{figure}

L'emploi de cette technique nécessite un faisceau laser ayant :

\begin{itemize}
\item 
un excellent contraste temporel, de façon à éviter que le pied de
l'impulsion crée sur les parois un plasma, dont l'expansion rapide
bouchererait le tube avant l'arrivée du coeur de l'impulsion;
dans l'expérience mentionnée plus haut, un doublage de fréquence a
permis d'obtenir le contraste nécessaire.  L'emploi de {\bf miroir à plasma}, 
coupant la pré-impulsion, est aussi envisagé.
Leur emploi dans la plage de temps qui nous intéresse ($\approx$ 100
fs) a été caratérisé expérimentalement et par des simulations
 \cite{Doumy}.
L'utilisation de plusieurs miroirs en série est possible,
avec 
une amélioration du contraste de 2 ordres de grandeur par miroir.

L'emploi d'un dernier miroir à plasma permettrait en outre d'améliorer
le taux de remplissage de l''accélérateur \cite{BiJanChan}.

\item 
un excellent contraste spatial (transverse), de façon à ce que la
fraction de l'intensité laser incidente sur la face avant ne détruise
pas le capillaire, et permette l'utilisation multi-tirs.
\end{itemize}

Il a été démontré par des simulations \cite{Andreev:capil} que l'onde
plasma générée dans un tel guide est très proche de celle générée par
un mode gaussien, et se prête bien à l'accélération de particules.

\subsection{Conclusion}

Il devrait être possible d'accélérer un faisceau d'électrons d'une fraction de nC dans une
cavité à plasma, de paramètres listés dans la table \ref{tab:opt} avec
un gain de quelques GeV sur une longueur d'une dizaine de centimètres.

L'incertitude principale concerne la méthode de guidage employée :
\begin{itemize}
\item 
Le  guidage par capillaire est la méthode de choix
 à ces basses densités, 
démontré jusqu'à 
$3~10^{18}\U{W}\UUc{m}{-2}$  \cite{Kitagawa}.
Les contrastes spatial et  temporel 
seront le facteur limitant expérimentalement en utilisation multi-tirs.
\item 
L'emploi d'un $z-pinch$ est aussi peut-être possible, si les champs magnétiques résiduels sont suffisament faibles.
\end{itemize}

Les premières expériences de guidage de plusieurs Joules, à des
intensités relativistes, sont toutes récentes
(table \ref{tab:guid})
\cite{CrosPhyScript2004,Kitagawa,LBLAAC04}. Le futur proche devrait
permettre de tester expérimentalement les limites de chaque technique.

\section{Vers un accélérateur laser-plasma multi-TeV ?}\label{sec:multitev}

\begin{table}
\caption{Paramètres d'accélérateurs en projet. \label{tab:acc:projet}
}
\begin{center}
\begin{tabular}{lccccc}
      &     &   Sband   &   CLIC   &   LWFA \\
      &     & \cite{sband} & \cite{clic} & \cite{Xie} \\

Fréquence RF (GHz)   &     &  3   &  30    \\
$\lambda_p$ (mm)     &     &      &       &  0.1  \\
gradient   &  $E_z$ (GV/m)  & 0.017 &  0.15 &   10 \\
Energie dans le centre de masse & $\sqrt{s}$ (TeV)   &  0.5 &  3    &  5   \\
Luminosité &  ${\cal L} (10^{35} \UUc{m}{-2}\UU{s}{-1}$) & 5   &  0.8  &   1 \\
Longueur (km) (final focus)   &     & 33 (1.5) & 33.2 (5.2) & 2.46  \\
Taux de répétition (Hz)    &     & 50   & 100   & 156k  \\
Nombre de particules par paquet & $N (10^{10}$) &     1.1  & 0.4   & 0.016 \\
Nombre de paquets par train  &   &  333 & 154   & 1 \\
Puissance faisceau (MW) & $P_b$  & 14.5  & 14.8   & 20  \\
Puissance consommée (MW) &   & 140  & 230   & \\
Efficacité (\%) &   &  10.4 &  9.3  &    \\
Emittance normalisée&  $\gamma\epsilon_x, \gamma\epsilon_y$ mm mrad  & 5, 0.25  & 0.68, 0.01   & 0.025    \\
Taille faisceau &  $\sigma$ nm  & 335, 16  & 60., 0.7   &  0.56\\
Fonction bétatronique & $\beta_x, \beta_y (\Um{m})$ & 11, 0.45 & 6, 0.07 & 0.06 \\  
Longueur paquet &  $\sigma_z (\Uu{m})$   & 300  & 35   &  1 & \\
Largeur  en énergie en fin de Linac &  $\delta E/E$  (\%)  & 0.35  & 0.35 \\
Facteur de disruption & D &    0.32 7.1  & 0.07  6.3 & 0.29  \\
Paramètre de beamstrahlung & $\Upsilon$   & (0.46)  &  5  &  631 \\
Perte moyenne en énergie au croisement &  $\delta_B$  (\%) & 2.8 &  21.  & 20.\\
Nombre de $\gamma$ par $e$  & $n_\gamma$  &   & 1.53   & 0.72  \\
Facteur de pincement &  $H_D$  & 1.5  & 1.7   &    \\
 &   &   &    &    \\
 &   &   &    &    \\
 &   &   &    &    \\
 &   &   &    &    \\
\end{tabular}
\end{center}
\end{table}

Aucun projet abouti d'accélérateur utilisant des cavités à plasma
n'est disponible à ce jour.
Je présente en table \ref{tab:acc:projet} une tentative  d'un
tel design \cite{Xie}.
Nous avons jusqu'ici focalisé notre attention sur l'obtention d'un
gradient accélérateur élevé, de façon à limiter la longueur d'un
accélérateur multi-TeV.
Une spécification importante est la luminosité ${\cal L}$ :

\begin{equation}
 {\cal L} = \frac{f N^2}{4\pi \sigma_x \sigma_y}
\end{equation}

où l'on fait entrer en collision $N_b$ paquets, chacun contenant $N$ particules, avec un taux de répétition $f_0$ et un taux de croisement des
paquets individuels $f=N_b f_0$.

On écrit  ${\cal L}$ en fonction de la puissance moyenne transportée par les faisceaux, $P_b$ :
\begin{equation}
P_b = N E_b f,
\label{eq:puis}
\end{equation}

où $E_b$ est l'énergie de chacun des faisceaux.
On obtient :
\begin{equation}
 {\cal L} = \frac{P_b}{4\pi E_b} \frac{N}{\sigma_x \sigma_y}
\label{eq:lum:puis}
\end{equation}

Etant limité par la puissance disponible, on est conduit
à minimiser la surface de la zone de collision, et maximiser 
le nombre de particules par paquet.

Un premier effet limitant est l'interaction faisceau-faisceau.
Il est quantifié par le facteur de disruption $D$, rapport de la longueur d'un paquet par la focale de la lentille créée par le champ électromagnétique de l'autre paquet :
\begin{equation}
D_{(x,y)} =  \frac{2 r_e N \sigma_z}
{\gamma \sigma_{(x,y)} (\sigma_x +\sigma_y)} 
\end{equation}

La diminution de la taille des faisceaux due à cette focalisation
(pour des collisions de particules de charges opposées, $e^+e^-$),
appelée pincement, 
induit une augmentation de  {\cal L} d'un facteur noté $H_D$.
Dans la suite j'omets ce facteur, qui est trouvé proche de 1 lorsque
les conséquences du beamstrahlung sont raisonnables.

Un second effet délétère est le ``beamstrahlung'', rayonnement émis par une particule dans le champ du paquet opposé, créateur de bruit de fond au point de croisement, et élargissant le spectre en $\sqrt{s}$ de luminosité différentielle.
Le beamstrahlung est quantifié par le paramètre $\Upsilon$

\begin{equation}
\Upsilon =   \frac{2\hbar \omega_c}{3E_b} \approx 
 \frac{5 r_e^2 \gamma}{6\alpha \sigma_z} 
\frac{N}{\sigma_x +\sigma_y} 
\end{equation}
où $\hbar \omega_c$ est l'énergie critique du spectre bremstrahlung des photons.
Le nombre moyen de photons émis par $e$ est :

\begin{equation}
n_\gamma \approx  \frac{12 \alpha^2 \sigma_z}{5 r_e \gamma} 
 \frac{\Upsilon}{\sqrt{1+\Upsilon^{2/3}}}
\end{equation}

La perte relative d'énergie par beamstrahlung, $\delta_B$ est une
mesure de l'élargissement du spectre de $\dd {\cal L}/\dd \sqrt{s}$,
et vaut :

\begin{equation}
\delta_B \approx  \frac{5 \alpha^2 \sigma_z}{4r_e \gamma}
 \frac{\Upsilon^2}{(1+(1.5\Upsilon)^{2/3})^2}
\end{equation}

et  ${\cal L}$ peut s'écrire :

\begin{equation}
 {\cal L} = \frac{3 \alpha }{10\pi r_e^2}
\frac{\Upsilon   \sigma_z P_b }{E_b\gamma}
\left(\frac{1}{\sigma_x}+\frac{1}{\sigma_y}\right)
\end{equation}

\begin{itemize}
\item {\bf faisceaux plats}
Maximiser $\di \left(\frac{1}{\sigma_x}+\frac{1}{\sigma_y}\right)$
tout en limitant la valeur de $n_\gamma$ et de $\delta_B$, et donc de
$\Upsilon$, donc à $\sigma_x + \sigma_y$ fixé, conduit à utiliser des
faisceaux plats, avec par exemple $\sigma_x \gg \sigma_y$.
\item {\bf faisceaux ronds} La faisabilité d'une cavité à plasma
conservant les émittances transverses d'un faisceau plat est très
incertaine, ce qui conduit à envisager l'emploi de faisceaux ronds, de
largeur $\sigma \equiv \sigma_x = \sigma_y$.
\end{itemize}

Les deux projets d'accélérateur ``métal'' (\cite{sband,clic})
sont proches de la limite
($\Upsilon=1$) du régime de beamstrahlung faible ($\Upsilon<1$) alors
que les projets ``plasma'' sont dans le régime de beamstrahlung très
fort ($\Upsilon\gg 1$).

\begin{itemize}
\item 
Dans l'approximation de  {\bf beamstrahlung faible}, on obtient :

\begin{equation}
 {\cal L} =\frac{P_b n_\gamma }{4\pi E_b \alpha r_e \sigma}
\end{equation}

Le seul paramètre libre pour atteindre la luminosité nécessaire avec
une valeur de $n_\gamma$ supportable est $\sigma$.
Avec les valeurs de la table \ref{tab:acc:projet}
 \cite{Xie}, on obtient $\sigma \approx 1 $\AA$n_\gamma $, clairement hors de portée.
De plus, l'expression de ${\cal L}$ en fonction de $\delta_B$ est 
\begin{equation}
 {\cal L} =\frac{3}{5\pi E_b (2\sigma)}\sqrt{\frac{\delta_B \sigma_z}{5 r_e^3 \gamma}}
\end{equation}

Ceci favorise l'emploi de paquets  {\bf longs}.

\item 
On se résigne alors au régime de   {\bf beamstrahlung fort} et on obtient :



\begin{equation}
n_\gamma \approx  2\alpha \left(\frac{6\alpha r_e \sigma_z N^2}{5 \gamma (2\sigma)^2}\right)^{1/3}
\end{equation}

et $\delta_B \approx n_\gamma /3.3$ \cite{Dugan}.
La luminosité est alors

\begin{equation}
 {\cal L} =
\frac{P_b}{16\pi\alpha^2 E_b (2\sigma)}
\sqrt{\frac{5\gamma n_\gamma^3}{3 r_e \sigma_z}}
\label{eq:lumi:beamStrahlFort}
\end{equation}

Ici au contraire, on veut  {\bf diminuer} à la fois la  {\bf taille} $\sigma$ et la
  {\bf longueur} $\sigma_z$ des paquets.
\end{itemize}

La détermination du reste des paramètres est obtenue de la façon suivante :
\begin{itemize}
\item les paramètres dont on ne dispose pas librement,
$E_b, P_b, n_\gamma$, sont choisis.
\item la longueur du paquet $\sigma_z$ est ajustée à la limite de
faisabilité estimée $\approx 1\Uu{m}$.  
\item la taille du faisceau transverse $\sigma$ est alors déterminée par 
l'équation (\ref{eq:lumi:beamStrahlFort}).

\item cette taille s'exprimant $\sigma = \sqrt{\epsilon_{\perp}\beta^{*}}$, la valeur de la fonction bétatronique au croisement est
choisie la plus petite possible, pour relâcher autant que faire se
peut la contrainte sur l'émittance normalisée $\epsilon_{\perp N}$.
Le facteur limitant de la focalisibilité du faisceau est le rayonnement synchrotron dans l'optique de focalisation finale de l'accélérateur, 
imposant une taille minimale 
\cite{Oide,Irwin}
\begin{equation}
\sigma_m [\U{m}] = C \epsilon_{\perp N}^{5/7},
\end{equation}
 avec $C= 1.7 \times 10^{-4}$.
Ceci limite $\beta < \beta_M = \sigma^{3/5} \gamma C^{7/5}$
\item on en déduit la valeur de l'émittance,
\item puis la valeur de $N$ à partir de 
l'équation  (\ref{eq:lum:puis}),
\item enfin le taux de répétition $f$ à partir de 
l'équation  (\ref{eq:puis}).
\end{itemize}

J'examine rapidement ces résultats ci-dessous. Les paramètres de
l'accélérateur sont ceux de \cite{Xie} (table
\ref{tab:acc:projet}). Ceux de la cavité à plasma sont ceux de la
table \ref{tab:opt}.

\subsection{Beam loading}

Un des facteurs limitants de la qualité de l'accélération est le beam
loading : 
fraction de l'énergie de l'onde extraite par le faisceau.
Le nombre maximal $N_0$ de particules accélérables peut-être obtenu en
comparant l'énergie par unité de longueur transportée par l'onde
plasma (champ longitudinal uniquement), $u\equiv \epsilon_0 (E_{z}
w_0)^2/2$ soit $12~\U{J/m}$ avec les paramètres de la table
\ref{tab:opt}, avec l'énergie par unité de longueur transmise au
faisceau $N E_{z}$.  On obtient, pour $N_0 E_{z} =  u$ :

\begin{equation}
N_0  =  \epsilon_0 E_z w_0^2/2,
\end{equation}

soit au maximum 
1.4 nC, deux ordres de grandeur au dessus
des besoins de \cite{Xie}.
Notons que $N_0 \propto E_z w_0^2$ avec $E_z \propto 1/\lambda_p$ et
 $w_0 \propto \lambda_p$ : on obtient $N_0 \propto \lambda_p$,
 favorable à un plasma peu dense.

La limitation de la charge maximale  accélérable dans un paquet d'électrons
n'est donc pas le beam loading;
c'est l'atténuation du champ de l'onde plasma par le champ de sillage du 
paquet, discuté plus haut (eq. \ref{eq:Nmax} et table  \ref{tab:opt}).

\subsection{Longueur de paquets, émittance transverse}

La longueur de faisceau envisagée $\sigma_z = 1 \Uu{m}$ est probablement possible
avec la charge demandée (table \ref{tab:inj}).

L'émittance normalisée $\epsilon_{\perp N} = 0.025 \U{mm.mrad}$, sera
probablement difficile à obtenir directement à partir d'un injecteur à
plasma (tables \ref{tab:inj} et \ref{tab:mono}). 
De telle émittances ont été produites, dans le plan vertical, dans des anneaux d'amortissement \cite{Honda:2003tx}  (de tels anneaux sont nécessaires pour refroidir les paquets de positrons).
Néanmoins la synchronisation des systèmes RF d'anneaux de stockage
avec une précision de quelques femtosecondes, nécessaires au
fonctionnement d'un linac à plasma, est un défi non trivial.

Notons que la minimisation de l'élargissement en énergie dans une
 cavité à plasma  demande des paquets plus
 longs ($\approx \lambda_p/6$ \cite{ReitsmaTrines} soit $10\Uu{m}$
 (table \ref{tab:opt})).
L'emploi de paquets longs est vraiment rédhibitoire en mode
de beamstrahlung fort : $\sigma$ est proportionnel à
$1/\sqrt{\sigma_z}$ (eq. \ref{eq:lumi:beamStrahlFort}) : l'émittance à
la limite de Oide \cite{Oide}
 est alors proportionnelle à $\sigma_z^{-0.7}$.

On s'orientera donc vers l'emploi de trains de paquets, la phase
d'injection de chacun étant ajustée de façon à compenser l'atténuation
du champ créé par les précédents \cite{HighQualBeamChiou} : l'emploi
de paquets courts est à nouveau possible.

\subsection{Taux de répétition, durée de vie de l'onde plasma}

La difficulté d'accélérer des paquets d'électrons courts, pour un
collisionneur, et de charge élevée, dans une onde plasma, incite à les
fractionner et à charger l'onde d'un train de paquets.

En effet le taux de répétition $f = 156 \Uk{Hz}$, 4 ordres de grandeur
 au dessus des lasers femto-seconde multi-Joule actuellement
 disponibles, nous incite à maximiser le nombre de paquets par train
 $N_b$.
Les ondes plasma ont une durée de vie limitée. 
Le temps caractéristique de déplacement des ions est
de l'ordre de $\sqrt{m_p/m_e}2\pi/\omega_p = 10 \Up{s}$, supérieur à
celui des électrons par un facteur  $\sqrt{m_p/m_e} \approx 40$.

\begin{figure}[tbh] 
\begin{center}
 \epsfig{figure=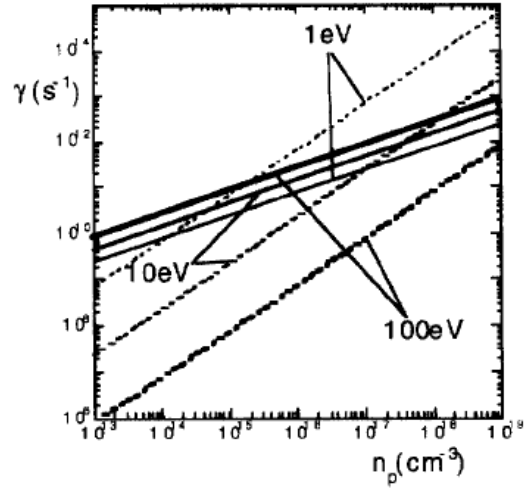,width=0.47\linewidth} 
\caption{Taux de décroissance d'une onde plasma dû à l'instabilité
modulationnelle (couplage avec des ondes ioniques)
(ligne continue), et collisions
entre particules du plasma (ligne pointillée)
en fonction de la densité et de la
température du plasma \protect\cite{Ogata}
\label{fig:Ogata}}
\end{center}
\end{figure}

Une étude  \cite{Ogata} du taux de décroissance d'une onde
plasma fournit des résultats similaires (figure \ref{fig:Ogata}) :
pour une densité $n = 3.~10^{17}\UUc{m}{-3}$, le taux de décroissance
devrait être proche de $10^{12}\UU{s}{-1}$, soit 
$\approx 30 \omega_p$.

On peut penser gagner au moins un ordre de grandeur sur $f$ par
l'utilisation de trains multi-paquets.

~


La façon d'éviter les collisions des ``mauvais'' couples de paquets,
avant ou après le point d'interaction, n'est néanmoins pas triviale, étant donné
leur très faible séparation ($\lambda_p \approx 60 \Uu{m}$).

\clearpage

\subsection{Taille de faisceau}

On a vu qu'une des particularités des cavités à plasma est la présence
de champs électriques transverses du même ordre de grandeur que le
champ longitudinal.
Ceci crée une série de complications additionnelles.
Les paquets ayant une longueur finie, les électrons ne voient pas
exactement le même champ, en fonction de leur phase $\zeta$ par rapport à
l'EPW.
(L'oscillation bétatronique peut-être représentée comme une rotation
dans l'espace des phases $x,x'$, où $x$ décrit l'une des coordonnées
transverses, et $x'$ est la notation paraxiale de $\dd x:/ \dd s$.)
A une position donnée $s$ le long de l'axe de propagation, le faisceau
est décrit par une ellipse : les ellipses de phases $\zeta$ différentes
tournent à des vitesses angulaires différentes.

Lorsque qu'un faisceau d'émittance $\epsilon$ est injecté avec une
taille $\sigma_0$ dans un canal de fonction bétatronique $\beta$ telle
que $\sigma_0 = \sqrt{\beta\epsilon}$ (condition d'accord
bétatronique), la propagation s'effectue à taille constante dans le
canal.
Dans le cas contraire, la taille du faisceau   $\sigma$  oscille autour de  $\sigma_0$.
Utilisant la définition de l'émittance normalisée $ \epsilon_N
=\epsilon \gamma$, et la valeur de $\beta$ (eq. \ref{eq:fn_beta}), on obtient  :

\be
\sigma_0 =\sqrt{\frac{\epsilon_N w_0}{2}\sqrt{\frac{1}{\delta\gamma\sin{\zeta}}}},
\ee

soit $\sigma_0 \approx 4 \Uu{m} (E(\UG{eV})/ \sin{\zeta})^{-1/4}$.
La taille du faisceau décroit  comme $\gamma^{-1/4}$ dans un
système à focalisation continue, et elle est de l'ordre de $0.6  \Uu{m}$ à 2.5~TeV.

\subsection{Filamentation; alignement}

Un électron soumis au champ transverse d'une onde plasma, dans sa
phase focalisante, subit une oscillation ``bétatronique'' autour de
l'axe de propagation, de longueur d'onde $2\pi \beta$, où (eq. \ref{eq:fn_beta}):

\begin{equation}
\beta = w_0 \sqrt{\di \frac{\gamma}{4\delta \sin{\zeta}}}
\end{equation}

soit $\beta = 0.7 \Um{m}  \sqrt{E(\UG{eV})/ \sin{\zeta}}$.

Le nombre d'oscillations au cours de la propagation, $\mu = L/(2\pi \beta)$, est donc 
$\mu = L/(\pi w_0)\sqrt{\delta\sin{\zeta}/\gamma}$, 
et la variation $\delta \mu$ de $\mu$ due à l'étalement longitudinal
du faisceau $\delta \zeta = 2\pi\sigma_z/\lambda_p$ vaut : 

\begin{equation}
\delta \mu = \frac{L}{2 w_0} \sqrt{\di \frac{\delta}{\gamma}} \frac{\cos{\zeta}}{\sqrt{ \sin{\zeta}}} \frac{\sigma_z}{\lambda_p}
\end{equation}

soit $\delta \mu/L \approx 6 \U{rad/m}  \frac{\cos{\zeta}}{\sqrt{ \sin{\zeta}}} / \sqrt{E(\UG{eV})}$.

Le terme de phase doit être moyenné le long de la propagation : on
n'en dispose pas en pratique car la phase d'injection du paquet est
ajustée en premier lieu de façon à minimiser la largeur en énergie.
Un mélange complet des phases bétatroniques, ``filamentation'', a donc
lieu dans une grande partie de l'accélérateur.
Lorsque le faisceau est injecté dans le canal avec une taille
 différente de la taille d'équilibre $\sigma_0$, ou en présence de
 désalignement $\delta_x$ de l'injection du faisceau, la filamentation
 induit une augmentation d'émittance $\delta \epsilon$
 \cite{AssmannYokoya}

\be
\frac{\delta \epsilon}{ \epsilon}= \left( \frac{\delta_x}{\sigma_0}\right)^2
\ee

En pratique un écart $\delta_x \approx \sigma_0/10$ est tolérable
\cite{AssmannYokoya}, et l'alignement dans la partie de haute énergie
du LINAC doit être réalisé et stabilisé avec une précision de
$60\Un{m}$.  
La stabilisation d'une onde plasma, et en particulier d'une impulsion
laser, à une telle précision, ne manque pas d'inquiéter.
Le guidage dans un capillaire est peut-être la solution.

Le paramètre qui contrôle les effets néfastes d'un désalignement sur
l'émittance du faisceau accéléré est la taille du faisceau laser
$w_0$.  Le résultat présenté ici $w_0 \approx \lambda_p/2$ (table
\ref{tab:opt}) pourrait être relâché. Mais un petit paquet d'électrons
n'est sensible au champ accélérateur de l'onde plasma que sur une
distance transverse de l'ordre de $\lambda_p/(2\pi)$ \cite{Kats:Wilk}. 
Une augmentation importante de la taille du faisceau dégraderait
l'efficacité de l'utilisation de l'énergie de l'onde par le faisceau.

~

Chiou {\em et al.} 
 \cite{hollow} ont proposé de s'affranchir des champs transverses en
réalisant un canal creux : un cylindre est entièrement vidé de ses
électrons. L'impulsion laser crée alors une onde plasma à l'extérieur
du canal, dont le champ longitudinal ne dépend plus du rayon $r$ à
l'intérieur du canal, et dont le champ transverse est nul.

\subsection{Diffusion multiple}

La diffusion multiple dans le plasma tout au long du linac peut
augmenter la taille transverse du faisceau, et donc l'émittance finale.
Le taux de croissance dû à la diffusion multiple dans le plasma est 
\cite{montague} :

\be
\frac{\dd \epsilon_N}{\dd s} =
 \frac{\gamma(s)\beta(s)}{2}
\frac{\dd \langle \theta^2 \rangle}{\dd s}
\ee

où 
\be
\frac{\dd \langle \theta^2 \rangle}{\dd s}
=
\pi n \left( \frac{4 r_e^2}{\gamma(s)^2} \ln{(\frac{\lambda_D}{r_p})}
 \right)^2,
\ee

où $r_p = 1 \Uf{m}$ est le rayon du proton, et $\lambda_D$ est la longueur de Debye
($\lambda_D= \lambda_p \sqrt{kT/mc^2}/(2\pi)$),  soit, en intégrant et en utilisant la dépendance
 $\beta \propto \sqrt{\gamma}$ : 

\be
\delta  \epsilon_N = \frac{4\pi n r_e^2}{\gamma'} \ln{(\lambda_D/r_p)} (\beta_f - \beta_i)
\ee

soit 

\be
\delta  \epsilon_N = \frac{2 n r_e^2 \lambda_p}{\delta} \ln{(\lambda_D/r_p)} (\beta_f - \beta_i)
\ee

avec $\lambda_D = 0.03 \Uu{m}$, $\beta_f = 35 \Um{m}$, on obtient
$\delta \epsilon_N = 2.5 \U{m.rad}$, ce qui est négligeable.


\subsection{Conclusion ?}

La route qui mène vers un accélérateur à cavités laser-plasma est
encore semée de nombreux défis. 

\begin{itemize}
\item 
Détermination d'un mode de guidage d'une impulsion laser, satisfaisant
de multiples contraintes : basse densité électronique, haute
intensité, capacité multi-tirs, stabilisable en transverse à $\approx
60\Un{m}$.

\item 
Développement de laser femtoseconde, multi-Joule, multi-kHz.

\item 
Optimisation du
transfert d'énergie du laser à l'onde plasma, et de celle-ci au
faisceau, de façon à ce que l'efficacité de l'utilisation de
l'énergie, de la prise électrique au faisceau, soit de l'ordre de 10 -- 20 \%.
Actuellement obtenir une telle efficacité pour chacun des trois étages indépendamment
est encore un objectif à atteindre ..

\item 
Développer une source de faible émittance transverse.
Il faut noter  que
la synchronisation femtoseconde d'un anneau d'amortissement avec un
linac à plasma est probablement difficile.
L'accélération de positrons dans un linac à plasma pose donc question.
Ceci pourrait inciter à l'emploi d'un collisionneur $\gamma$-$\gamma$, 
à partir de deux faisceaux d'électrons.
\end{itemize}

\clearpage
\section{Systèmes laser}

Dans un système à amplification à dérive de fréquence, l'énergie
maximale par impulsion est déterminée par
\begin{itemize}
\item
le seuil $F_d$ de tenue des optiques au flux de l'impulsion étirée, d'une
durée de l'ordre de la nanoseconde,
\item
le flux de saturation $F_s=h\nu/\sigma$ de l'amplificateur, où
$\sigma$ est la section efficace de l'interaction photon-molécule à la
fréquence de lasage $\nu$.
\end{itemize}

Pour des impulsions de quelques nanosecondes, $F_d$ est de l'ordre de
$20~ \U{J}\UUc{m}{-2}$, et 
$F_s$ est de l'ordre de
$1 ~\U{J}\UUc{m}{-2}$ pour du Ti:saphir
\footnote{Certains matériaux comme le Yb:glass, verre dopé à
l'ytterbium, ont un flux de saturation plus élevé, $40
\U{J}\UUc{m}{-2}$, mais la tenue au flux devient un problème.  Et le
refroidissement du verre est extrèmement lent ..}.
Des amplificateurs en Ti:saphir de $20 \Uc{m} \times 20 \Uc{m}$ ont
déja été produits \cite{zetta}, et l'obtention d'impulsions de
quelques dizaines ou même centaines de Joules est envisageable sans problème.
Notons que la tenue au flux des optiques soumises à l'impulsion recomprimée 
(réseaux, optique de focalisation)
est plus faible, 
inférieure au $\U{J}\UUc{m}{-2}$.

Le problème suivant concerne la possiblité d'utiliser un tel système
à haute fréquence : le {\bf refroidissement} est alors un facteur limitant.

Le laser du LOA utilise un amplificateur au  Ti:saphir refroidi
\footnote{A $-30^{\circ}$, le Ti:saphir la conductivité thermique du
cuivre.}, pompé par laser avec une bonne efficacité, avec
une impulsion de quelques Joules : à 10 Hz, le refroidissement est
déja limitant.

La conception de lasers à impulsion courte, multi-J et multi-kHz,
souhaités pour un accélérateur à plasma, est un défi intéressant.
Une modification de la géométrie de l'amplificateur (disques au lieu
de barreaux) est une possibilité d'amélioration.

\clearpage
\section{Vers un programme d'expérience de 2ème génération : suite}

L'irruption des résultats de production de bouffées d'éléctrons
mono-cinétiques en 2004 ouvre la porte vers un programme expérimental
``tout optique'', où un injecteur à plasma fournirait un paquet
d'électrons ``test'' à une cavité plasma.

\subsection{Injecteur} ~

 En ce qui concerne l'injecteur lui même, l'énergie de production déja
disponible (table \ref{tab:mono}, \cite{LBLAAC04,LOAAAC04,RALAAC04})
de l'ordre de 100~MeV, est tout à fait satisfaisante, ainsi que la
charge du paquet, 4~pC, par comparaison avec le projet à 5 TeV de
\cite{Xie}, $q = 16 \Up{C}$ (table \ref{tab:acc:projet}).

\begin{itemize}
\item {\bf Emittance } \\
Un axe de recherche important est la minimisation de l'émittance 
normalisée du paquet, actuellement de l'ordre du mm.mrad
(Table \ref{tab:mono}) alors que $0.025 \U{mm.mrad}$ sont souhaités
(table \ref{tab:acc:projet}).

\item {\bf Largeur en énergie} \\
De même, la largeur relative en énergie mesurée est  de quelques pour-cent.
Les simulations (effectuées dans des conditions différentes, 
avec $n \approx 10^{17}\UUc{m}{-3}$) fournissent des chiffres semblables 
\cite{ReitsmaTrines,HighQualBeamChiou,MonoenergeticAndreev}.

\item {\bf Stabilité} \\
Enfin les trois expériences ont observé une
forte instabilité de la production de pic mono-énergétique, et de
l'énergie moyenne.
Cette instabilité est probablement due au réglage fin nécessaire pour
que le paquet quitte l'onde plasma précisement lorsqu'il se trouve au
sommet de la courbe
(figure \ref{fig:courbe:phase}).
Ceci nécessite une grande stabilité des paramètres qui contrôlent
l'accélération, et en particulier l'énergie et la qualité optique 
(profils temporel et transverse)
de l'impulsion laser, 
 et la
densité électronique du plasma.

\item {\bf Phase de l'injection dans la cavité} \\
Enfin, le contrôle de la phase de
l'injection du faisceau produit par l'injecteur dans la cavité exclut
l'emploi d'un injecteur avec déferlement par auto-modulation : une
injection laser sera probablement nécessaire.
\end{itemize}

\subsection{Cavité} ~

En ce qui concerne l'accélération dans la cavité à plasma, c'est à
dire par sillage laser dans un plasma de basse densité, le seul
résultat disponible à ce jour est celui de l'expérience
 de l'Ecole Polytechnique \cite{prl98}.
On a vu que son interprétation en terme de sillage linéaire est
compliquée par les effets du champ transverse.

Il est important de caractériser la méthode du sillage laser, par
l'exploration de l'espace des paramètres de la table \ref{tab:opt}.
Ceci se fera préférentiellement avec un guidage de l'impulsion laser.

\begin{itemize}
\item {\bf Guidage}\\ les 2 méthodes que l'on peut envisager d'utiliser
dans ce régime de basse densité sont :
\begin{itemize}
\item guidage par capillaire. Le guidage d'une intensité de 
  $3. ~10^{18}  \U{W}\UUc{m}{-2}$ a été démontré  \cite{Kitagawa} :
une amélioration du contraste spatial et temporel de l'impulsion laser
 est nécessaire, de façon à pouvoir effectuer plusieurs tirs sans
 détruire le capillaire.
\item guidage dans un profil de densité obtenu par 
``Z-pinch'': ici $10^{17} \U{W}\UUc{m}{-2}$ ont été guidés, mais le
guidage d'impulsions relativistes devrait être possible.
Le problème principal est la persistance possible de champs
magnétiques dans la plasma qui empêcherait l'injection d'un paquet
d'électrons, effet observé dans le passé par le groupe de
UCLA \cite{ClaytonthetaPinch}.
\end{itemize}

En sortie du canal focalisant que constitue l'injecteur, le faisceau
n'etant plus guidé diverge : sa taille transverse évolue de façon
notable après un parcours de l'ordre de $\beta = \sigma/\sigma'$
typiquement de l'ordre de $ 1\Uc{m}$ il faudra prévoir une optique de
focalisation\footnote{probablement achromatique, en particulier en présence d'instabilité de l'injecteur.}.

\subsection{Programme scientifique}
Le  {\bf programme scientifique} pourrait comprendre : 
\begin{itemize}
\item 
Dans un premier temps avec des capillaires de moyenne longueur 
($\le 10~ \Uc{m}$), avec un effet négligeable du déphasage et de la déplétion,
une étude de l'accélération   en {\bf sillage linéaire}, et en particulier la courbe de résonance
(figure \ref{sillage})
et la dépendance $E_z \propto P$ (eq.  \ref{eq:EzDeI:tau}).
\item
Ensuite, avec des capillaires plus longs, on pourra 
 chercher à atteindre la {\bf longueur de déphasage}, observant le pic
du Jacobien de la même façon qu'avec un injecteur. 
On se souvient que la  {\bf déplétion} de l'impulsion laser se produit sur une longueur du même ordre que le déphasage.
\item 
Si le programme ci-dessus a pu être réalisé, étudier les limites
de la  {\bf qualité} du faisceau obtenu, en particulier sa largeur en énergie
(\cite{ReitsmaTrines}).
Avec un faisceau de si faible énergie ($\sim$ GeV) la carte
d'émittance sera complètement filamentée, et l'émittance normalisée en
sortie sera au mieux égale à l'émittance normalisée en entrée :
 seule la conservation pourra être vérifiée.
\item 
A chaque étape, la {\bf durée de vie} de l'onde plasma et sa cohérence en fonction du
temps pourront être mesurées soit par l'interférence de deux
impulsions sondes se propageant avec un retard ajustable \cite{rapha,
newrapha}, soit en ajustant le délai de l'injection du paquet
d'électrons.
\item 
Les limites de la {\bf linéarité} de l'onde plasma pourront être étudiées,
et en particulier l'augmentation avec $a$ de $\lambda_p, L_\varphi, L_d$.
\item 
Puis, augmentant la charge du paquet injecté, observer l'effet
du  {\bf beam loading}, directement par déformation du spectre des électrons, puis par modification de la longueur de déphasage 
\cite{ReitsmaTrines}.

\item .. Ensuite, l'accélération de  {\bf trains} de paquets, séparés
d'environ (un multiple de) $\lambda_p/c$, et l'ajustement fin de leur
décalage de phase de façon à minimiser la largeur en énergie de
l'ensemble, pourra être tentée.

\end{itemize}
\end{itemize}

\clearpage

\section{Liste des articles joints en Annexe}\label{sec:liste}

~

Ils sont repérés dans le mémoire par une boulette \boul.

\begin{itemize} \small
\item {\bf A} \cite{nim95} 
The Plasma Beat-Wave Acceleration Experiment at Ecole Polytechnique, 
F.Amiranoff {\it et al.}, 
Nucl. Instr. and Meth. A 363 (1995) 497. 

\item {\bf B} \cite{prl95}
Electron Acceleration in Nd-Laser Plasma Beat-Wave Experiments, 
 F.~Amiranoff {\em et al.}, Phys. Rev. Lett. {\bf 74},
 5220 (1995).



\item {\bf C} \cite{kyoto_capte}
Electron Capture in an Electron Plasma Wave,
D. Bernard,
 in Proceedings of the 13th Advanced ICFA Beam Dynamics Workshop and 1st ICFA Novel and Advanced Accelerator
 Workshop, Kyoto, Japan, 1997,
 Nucl. Instrum. Methods Phys. Res. A 410 (1998) 418.

\item {\bf D} \cite{kyoto_one}
One GeV Acceleration with Laser Wake-Field in the Linear Regime,
D. Bernard, 
 Nucl. Instrum. Methods Phys. Res. A 410 (1998) 520.

\item {\bf E} \cite{prl98} 
Observation of Laser Wakefield Acceleration of Electrons,
 F.~Amiranoff {\em et al.}, Phys. Rev. Lett. {\bf 81 }, (1998) 995.


\item {\bf F} \cite{denisprel} Alternative Interpretation of Nuclear Instruments and Methods A410 (1998) 357 (H. Dewa et al.),
 D. Bernard et al.,
 Nucl.Instrum.Meth.A432:227-231,1999.

\end{itemize}
\clearpage

\section{Tableau récapitulatif des notations utilisées dans le mémoire}\label{sec:tableau}
 
\begin{center}\begin{tabular}{ll}
$a$ & rayon d'un capillaire\\
$a$ & champ électrique du laser normalisé à la limite de \\
& déferlement dans un plasma froid\\
$\beta$ & fonction bétatronique dans l'onde plasma \\ 
$c$ & vitesse de la lumière \\
$\delta$ & perturbation relative de densité électronique du plasma 
\\
$\delta_{\perp}$ & perturbation relative transverse de densité
électronique du plasma \\
$\Delta W$ & gain (perte) de l'énergie d'un électron
 accéléré (décéléré) par l'onde plasma \\
$\Delta W_g$ &  valeur de $\Delta W$ limité par déphasage, en présence de guidage \\
$e$ & charge électrique de l'électron \\
e & $\exp(1)$\\
$\epsilon$ & émittance du faisceau d'électrons $\epsilon = \sigma \sigma'$\\
$\epsilon_{0}$ & permittivité du vide \\
 ${\cal E}$ & champ électrique du laser \\
$E$ & énergie de l'impulsion laser \\
$E_{0}$ & champ électrique longitudinal de l'onde plasma \\
        & au déferlement dans un plasma froid\\
$E_{z}$ & champ électrique longitudinal de l'onde plasma \\
$E_{r}$ & champ électrique radial de l'onde plasma \\
$\eta$ & indice de réfraction du plasma\\
$\eta$ & fraction de beam loading \\
$\gamma_{e}$ & facteur de Lorentz des électrons à l'injection \\
$\gamma_{p}$ & facteur de Lorentz associé à la vitesse de phase de l'onde plasma \\
$k$ & vecteur d'onde du laser \\
$k_{p}$ & vecteur d'onde de l'onde plasma \\
$\lambda_{p}$ & longueur d'onde de l'onde plasma \\
$\lambda$ & longueur d'onde du laser \\
$L_{\varphi}$ & longueur de déphasage électron/onde plasma \\
$L$ & $L=2 z_{0}$ \\
$I$ & intensité lumineuse \\
$m$ & masse de l'électron \\
$n$ & densité électronique du plasma \\
$n_{c}$ & densité électronique critique \\
$\omega_{p}$ & fréquence plasma \\
\end{tabular}\end{center}

\begin{center}\begin{tabular}{ll}
$P$ & puissance laser à son maximum \\
$P_{crit}$ & puissance laser critique d'auto-focalisation \\
$r_{e}$ & rayon classique de l'électron \\
$\sigma$ & taille du faisceau RMS \\
$\sigma'$ & divergence angulaire du faisceau RMS \\
$\tau$ & durée de l'impulsion laser FWHM \\
$\tau_{0}$ & durée de l'impulsion laser~: 
 $I\propto\exp{(-t^{2}/\tau_{0}^{2})}$ \\
$w$ & rayon du faisceau laser ``à $1/e^{2}$ en intensité'', soit
$w=2\sigma$\\
 $z_{0}$ & longueur de Rayleigh du laser~: c'est la position
sur l'axe de propagation $(Oz)$ \\
 & pour laquelle l'intensité laser vaut la moitié de sa valeur maximale \\ 
$Z$ & l'impédance du vide\\
\end{tabular}\end{center}

\clearpage

{\bf \Large Appendices}

~

\appendix

Je donne ici un extrait d'une note que j'ai écrite en préparation à
l'expérience de sillage à l'Ecole Polytechnique.
\cite{alp9702}.

~

On consid\`ere un d\'eroulement en deux temps : tout d'abord, une
impulsion courte ($\approx 0.35$ps) cr\'ee un plasma et excite une
onde plasma, ensuite l'onde ``vit'' longtemps (plusieurs ps), et peut
acc\'el\'erer des particules.  Seul le mode lin\'eaire est d\'ecrit
ici.

Les valeurs num\'eriques utilis\'es dans ce texte sont celles qui
\'etaient pr\'evues \`a l'\'epoque. 


\section{Description du faisceau laser}\label{sec:las}
\subsection{G\'eom\'etrie, densit\'e volumique d'\'energie}

L'impulsion laser a un profil transverse gaussien : c'est la solution de plus petit diamètre 
des équations de Maxwell paraxiales.
Le profil temporel et aussi gaussien.
La densit\'e volumique d'\'energie est :
\begin{equation}                                
 \framebox{$\displaystyle \rho = \frac{E}{(\pi/2)^{3/2} w^{2} w_{z}}
  \exp{\left(-\frac{2 r^{2}}{ w^{2}}-\frac{2(z-ct)^{2}}{ w_{z}^{2}}\right)} $}
\end{equation}
   
$E$ est l'\'energie totale de l'impulsion laser, et $ w_{z}$ d\'ecrit la
longueur du paquet : $ w_{z}= c\tau /\sqrt{2\ln{2}}$ o\`u $\tau$ est
la dur\'ee de l'impulsion \`a mi-hauteur. Pour
 $\tau=350\Uf{s}$, on a
$ w_{z}/c=297\Uf{s}$, soit $w_{z}=89.2\Uu{m}$.
La taille transverse du faisceau est d\'ecrite par $w$ :
\begin{equation}
 w =w_{0}\left[1+\left(\frac{z}{z_{0}}\right)^{2}\right]^{1/2}
\end{equation}

\begin{figure}[hbt]\begin{center}
\vspace{-5mm}
\setlength{\unitlength}{1.0mm}
\begin{picture}(80,40)(0,0)
 \put(0,0){\makebox(80,40) {\epsfig{file=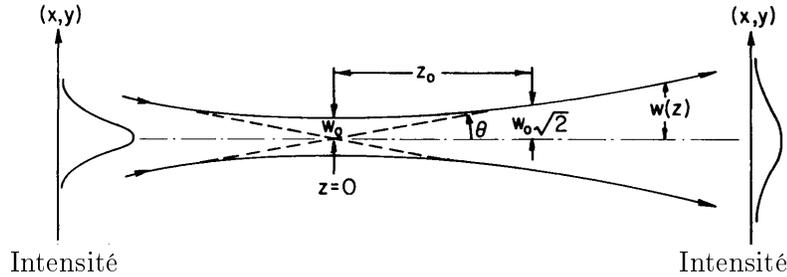,width=100\unitlength}}}
 \put(-6,0){\makebox(0,0)[b]{Intensit\'e}}
 \put(83,0){\makebox(0,0)[b]{Intensit\'e}}
\end{picture}
\vspace{5mm}
\caption{Variation de la taille $w$ du faisceau en fonction de $z$
pr\`es du waist.
 \label{fig:waist}}
\end{center}\end{figure}

C'est le rayon du faisceau ``\`a 1/e$^{2}$ en intensit\'e''. Notons que
$w$ est \'egal au double de l'\'ecart-type de la distribution en
intensit\'e du faisceau, $w=2\sigma$.

Avant focalisation, le faisceau est d\'ecrit par le m\^eme formalisme
gaussien. Le waist au foyer $w_{0}$ s'exprime \`a partir du waist
$w_{i}$ du faisceau incident par :
\begin{equation}                                
\framebox{$\displaystyle   
 w_{0} = M^{2}\frac{\lambda f}{\pi w_{i}} $}
\end{equation}

$f$ est la focale de l'optique de focalisation. Le facteur de
qualit\'e du faisceau $M^{2}$ est  appel\'e coefficient de
Siegman. Pour un faisceau limit\'e par diffraction, il vaut $M^{2}=1$.
Pour $\lambda = 1.053 \Uu{m}$, $f$=1.4m, $\phi_{i}=2 w_{i}= 80\Um{m}$,
et $M^{2}=1$, on obtient $w_{0}=11.7\Uu{m}$.
$z_{0}= M^{2}\pi (w_{0}/ M^{2})^{2}/\lambda$ est la longueur de Rayleigh (ici
$z_{0}=411\Uu{m}$), soit :
\begin{equation}                               
 \framebox{$\displaystyle z_{0}= M^{2} \frac{\lambda f^{2}}{\pi
   w_{i}^{2}} $}
\label{rayleigh}
\end{equation}

Sur l'axe, l'amplitude du champ varie comme $w_{0}/w$, c'est \`a dire
comme $ 1/\sqrt{1+(z/z_{0})^{2}}$. L'intensit\'e lumineuse est une
Lorentzienne en $z$ de demi-largeur \`a mi-hauteur \'egale \`a $z_{0}$.

A grand $z$, le comportement asymptotique $w \sim w_{0} z/z_{0}$
permet de d\'efinir une divergence angulaire $w'=w/z \sim
w_{0}/z_{0}$. On constate que la taille du waist $w_{0}$ et la
longueur de Rayleigh $z_{0}$ varient ensembles comme $M^{2}$. La
divergence angulaire ne d\'epend donc pas de $M^{2}$ : seules les
propri\'et\'es du faisceau au foyer sont affect\'ees par la valeur de
$M^{2}$. On obtient l'\'emittance
$\epsilon=\sigma_{0}\sigma'=w_{0}w'/4=M^{2}\lambda/4\pi$.

La densit\'e volumique d'\'energie maximale est :
\begin{equation}
 \rho_{max} = \frac{E}{(\pi/2)^{3/2} w_{0}^{2} w_{z}}
\end{equation}

soit 
\begin{equation}
 \rho_{max} = \frac{E}{c\tau}
\left(\frac{ w_{i}}{M^{2}\lambda f}\right)^{2} 4\sqrt{\pi\ln{2}}
\end{equation}

soit pour $E$=10~J, $\rho_{max} =414~\UM{J}\UU{cm}{-3}$.

\subsection{Intensit\'e, puissance}

L'intensit\'e est :
\begin{equation}
I= \rho c
\end{equation}

soit :
\begin{equation}                              
\framebox{$\displaystyle 
 I_{max} = \frac{E}{\tau}
\left(\frac{ w_{i}}{M^{2}\lambda f}\right)^{2} 4\sqrt{\pi\ln{2}} $}
\end{equation}

d'o\`u $I_{max} = 12.4~10^{18} \U{W}\UU{cm}{-2}$.
En int\'egrant dans le plan transverse, on obtient la puissance :
\begin{equation}
  P=\int I \dd S =  \sqrt{\frac{2}{\pi}} \frac{E c}{w_{z}}
  \exp{\left(-\frac{2(z-ct)^{2}}{w_{z}^{2}} \right)}
\end{equation}

soit $P=\pi w_{0}^{2} I/2$.  La puissance maximale est:
\begin{equation}                              
  \framebox{$\displaystyle P_{max}=  \sqrt{\frac{2}{\pi}}\frac{E c}{w_{z}} =
    \frac{E}{\tau} 2 \sqrt{\frac{\ln{2}}{\pi}}$}
\end{equation}

soit $P_{max}= 0.939 E/\tau$, et ici $P_{max}=26.8~\UT{W}$.

\subsection{Champ \'electrique}

On obtient la valeur du champ \'electrique ${\cal E}$ par :
\begin{equation}
I = \epsilon_{0} c {\cal E}^{2}/2
\end{equation}

soit :
\begin{equation} {\cal E}=\sqrt{2 I / \epsilon_{0} c}
\end{equation}

En utilisant $Z=\sqrt{\mu_{0}/ \epsilon_{0}}$ et $c=1/\sqrt{\mu_{0}
 \epsilon_{0}}$, on obtient :
\begin{equation} \epsilon_{0} c = \frac{1}{Z}
\end{equation}

$Z$ est l'imp\'edance du vide et vaut $Z=376.6\Omega$.
Il vient finalement : 
\begin{equation}                              
\framebox{$\displaystyle {\cal E}=\sqrt{2 Z I} $}
\end{equation}

et ${\cal E}_{max}=9.66\UT{V/m}$. On utilise souvent le champ
\'electrique normalis\'e\footnote{On veillera \`a ne pas confondre $w$,
mesure de l'\'etendue transverse du faisceau, et
 $\omega$,
 la pulsation.} :
\begin{equation}                              
\framebox{$\displaystyle a=\frac{e {\cal E}}{m \omega c} $}
\label{eq:def:a}
\end{equation}

soit ${\cal E}[\UT{V/m}]= 3.21 a/\lambda[\Uu{m}]$, et :
\begin{equation}                              
\framebox{$\displaystyle 
a^{2}=\frac{2 r_{e} \lambda^{2} I}{\pi m  c^{3}} $}
\label{champnorm}
\end{equation}

soit $a^{2}= 0.731~10^{-18} (\lambda[\Uu{m}])^{2}I[\U{W/\UU{cm}{2}}]$,
et ici $a^{2}$=10.1.  {\bf On est donc ici en r\'egime relativiste}.

On obtient aussi la relation entre le champ \'electrique et la
puissance maximale :
$P_{max}=\pi w_{0}^{2} I/2 = \pi {\cal E}_{max}^{2} w_{0}^{2}/4Z$.

\section{Excitation de l'onde plasma}\label{sec:epw}

La fr\'equence d'oscillation naturelle des \'electrons dans le plasma,
$\omega_{p}$, est donn\'ee en fonction de la densit\'e \'electronique $n$ :
\begin{equation}                              
\framebox{$\displaystyle 
\omega_{p} = \left(\frac{n e^{2}}{m \epsilon_{0}}\right)^{1/2}
= \left(\frac{4\pi \alpha n \hbar c}{m}\right)^{1/2}
= c (4\pi r_{e} n)^{1/2} $}
\end{equation}

$r_{e}$=2.818~fm est le rayon classique de l'\'electron.
La longueur d'onde plasma correspondante est :
\begin{equation}                              
\framebox{$\displaystyle 
 \lambda_{p} = \frac{2\pi c}{\omega_{p}}=\left(\frac{\pi}{r_{e} n}\right)^{1/2} $}
\end{equation}

Pour une densit\'e \'electronique $n=10^{17}\UUc{m}{-3}$, on obtient 
$ \lambda_{p}=105.5\Uu{m}$.
Lors du passage de l'impulsion laser, les \'electrons sont soumis
\`a la force pond\'eromotrice, moyenne temporelle de
la force de Lorentz :
\begin{equation}                             
\framebox{$\displaystyle 
\vec{F}_{P} = - \frac{e^{2}}{2 \epsilon_{0} m c\omega^{2}}\vec{\nabla}I $}
\end{equation}

~

On peut aussi l'exprimer en fonction de variables d\'ecrivant le  plasma :
\begin{equation}
\vec{F}_{P} = - \frac{1}{2 n c}\left(\frac{\omega}{\omega_{p}}\right)^{2}\vec{\nabla} I
= - \frac{1}{2 n_{c} c}\vec{\nabla} I
\end{equation}

$ n_{c}$ est la densit\'e \'electronique critique, pour laquelle 
$\omega_{p}=\omega$.

\subsection{Amplitude plasma}

Je suis ici le traitement de la r\'ef\'erence \cite{am}. La
perturbation de densit\'e est not\'ee $\delta n$. $\vec{E}$ d\'esigne
ici le champ \'electrique de l'onde plasma \'electronique, et $\Phi$
le potentiel associ\'e.
 $\Phi_{P}$ est le potentiel pondéromoteur de l'impulsion laser.
%
Les \'equations d\'ecrivant le plasma s'\'ecrivent :
\begin{equation}\mbox{\rm conservation du nombre de particules\hspace{1cm}}
 \frac{ \partial \delta n } {\partial t} +n\vec{\nabla}. \vec{v} =0
\label{conserve}
\end{equation}
\begin{equation}\mbox{\rm \'equation du mouvement\hspace{1cm}}
 m \frac{ \partial \vec{v} } {\partial t} = - e \vec{E} -
 \vec{\nabla} \Phi_{P}
\label{mouve}
\end{equation}
\begin{equation}\mbox{\rm \'equation de Poisson\hspace{1cm}}
\Delta \Phi = - \vec{\nabla} . \vec{E}= \delta n \frac{e}{\epsilon_{0}}
\label{poisson}
\end{equation}

On en tire :
\begin{equation}
 ( \frac{ \partial^{2} } {\partial t^{2}} +\omega_{p}^{2} ) \delta n =
 +\frac{n}{m} \Delta \Phi_{P}
\end{equation}
\begin{equation}
 ( \frac{ \partial^{2} } {\partial t^{2}} +\omega_{p}^{2} ) \vec{E} =
- \omega_{p}^{2} \vec{\nabla} \Phi_{P}/e
\end{equation}
\begin{equation}
 ( \frac{ \partial^{2} } {\partial t^{2}} +\omega_{p}^{2} ) \Phi = 
+ \omega_{p}^{2} \Phi_{P}/e
\label{dern}
\end{equation}

On peut r\'esoudre (\ref{dern}) et en d\'eduire la
perturbation de densit\'e et le champ \'electrique \`a l'aide des
relations $\delta n = \epsilon_{0} \Delta \Phi/e$ et
$\vec{E}= -\vec{\nabla} \Phi$.
On suppose $\Phi_{P}$ s\'eparable en $r$ et en $(z,t)$.
\begin{equation}
 \Phi_{P} = -e \Phi_{0}(r) f_{0}(z-c t)
\end{equation}

On a alors d'apr\`es (\ref{dern}): 
\begin{equation} \Phi= \Phi_{0}(r) f(z-c t)
\end{equation}

En posant $u= \omega_{p}(t-z/c)$, l'\'equation (\ref{dern}) devient :
\begin{equation}
 (\frac{\partial^{2}}{\partial t^{2}} + 1)f(u)= -f_{0}(u)
\end{equation}

dont la solution qui s'annule \`a $u = -\infty$ est :
\begin{equation}
 f(u)=-\int_{ -\infty}^{u} f_{0}(u') \sin(u-u') \dd u'
\end{equation}

soit encore :
\begin{equation}
 f(u)=
-\sin(u)\int_{ -\infty}^{u} f_{0}(u') \cos(u') \dd u'
+\cos(u)\int_{ -\infty}^{u} f_{0}(u') \sin(u') \dd u'
\label{lautre}
\end{equation}

Apr\`es passage de l'impulsion laser, c'est \`a dire pour $u$ grand,
on peut \'etendre la borne d'int\'egration dans (\ref{lautre}) \`a
$+\infty$. Si, de plus, $f_{0}$ est une fonction paire, il reste :
\begin{equation}
 f(u)=-\sin(u)\int_{-\infty}^{+\infty} f_{0}(u')\cos(u')\dd u'
\end{equation}

Dans notre cas, l'impulsion est gaussienne ($\Phi_{P}\propto I$) :
\begin{equation}
 f_{0}(u)= - \exp{(-u^{2}/u_{0}^{2})}
\end{equation}

On obtient alors : 
\begin{equation}
 f(u)=\sin(u)\int_{-\infty}^{+\infty} \exp{(-u'^{2}/u_{0}^{2})}
 \cos(u')\dd u'
\end{equation}

c'est \`a dire :
\begin{equation}
 f(u)= \sqrt{\pi} u_{0} \exp{(-u_{0}^{2}/4)}\sin(u)
\end{equation}

On se souvient que $\Phi_{P} = e^{2}I/2 \epsilon_{0} m c\omega^{2}$ a
\'et\'e param\'etris\'e ici $\Phi_{P} = -e\Phi_{0}(r) f_{0}(z-c t)$, soit
$\Phi_{P} = e\Phi_{max} \exp{(-u^{2}/u_{0}^{2})} \exp{(-2 r^{2}/w^{2})}$,
avec $\Phi_{max}=I_{max} e/(2\epsilon_{0} m c \omega^{2})$.
On obtient donc $\Phi$ en grandeurs dimensionn\'ees :
\begin{equation}
 \Phi(r,z,t)= \sqrt{\pi} \omega_{p} \tau_{0} \exp{(-\frac{
   \omega_{p}^{2} \tau_{0}^{2}} {4})} \exp{(-\frac{
  2 r^{2}}{w^{2}})} \sin{(\omega_{p} t -k_{p}z)} \Phi_{max}
\end{equation}

avec $\tau=\tau_{0} 2\sqrt{\ln{2}}$, et $k_{p}=\omega_{p}/c$, soit :
\begin{equation}
 \Phi(r,z,t)=\exp{(-\frac{2 r^{2}}{w^{2}})}\sin{(\omega_{p}t-k_{p}z)}
 \varphi
\end{equation}

avec :
\begin{equation}
 \varphi=\sqrt{\pi} \omega_{p} \tau_{0} \exp{(-\frac{ \omega_{p}^{2}
   \tau_{0}^{2}} {4})} \Phi_{max}
\end{equation}

%
On en d\'eduit le champ \'electrique de l'onde plasma :

\begin{center}\framebox{\parbox{9cm}{
\begin{eqnarray}
 E_{r}= \frac{4r}{w^{2}} \exp{(-\frac{2 r^{2}}{w^{2}})}
 \sin{(\omega_{p} t -k_{p}z)} \varphi\label{elongit}
\\
 E_{z}= k_{p} \exp{(-\frac{2 r^{2}}{w^{2}})} \cos{(\omega_{p} t -k_{p}z)}
 \varphi
\label{etransverse}
\end{eqnarray}
}}\end{center}

soit \`a un terme trigonom\'etrique pr\`es, et au maximum de $E_r$, $r = w/2$ :
\begin{equation}
 E_{r}/ E_{z} \approx \frac{4 r}{k_{p} w^{2}}
 \approx \frac{\lambda_{p}}{\pi w}
\end{equation}

Oubliant le terme de phase, le champ longitudinal s'\'ecrit  :
\begin{equation}
\framebox{$\displaystyle 
 E_{z}= k_{p} \exp{(-\frac{2 r^{2}}{w^{2}})}
\sqrt{\pi} \omega_{p} \tau_{0} \exp{(-\frac{\omega_{p}^{2} \tau_{0}^{2}} {4})}
\frac{I_{max} e}{2\epsilon_{0} m c \omega^{2}}
$}
\label{elongue}
\end{equation}

\begin{figure} \bc
 \mbox{\epsfig{figure=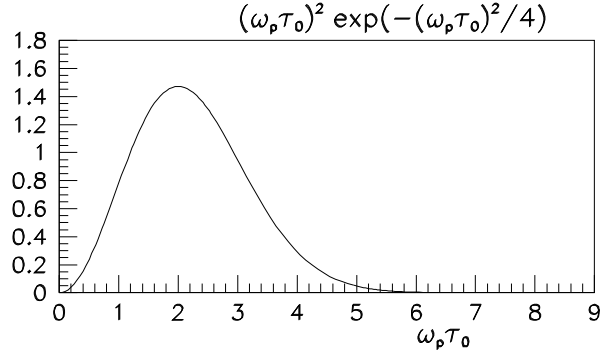,width=0.5\linewidth} }
\caption{Amplitude de l'onde plasma en fonction de $\omega_{p}\tau_{0}$.
\label{sillage}} \ec
\end{figure}

Ce champ est maximum {\bf \`a dur\'ee d'impulsion fix\'ee} lorsque
$\omega_{p}^{2}\exp{(-\omega_{p}^{2} \tau_{0}^{2}/4)}$ est maximum
(figure \ref{sillage}),
soit pour $\omega_{p}= \widehat{\omega}_{p}$ 
telle que\footnote{Le chapeau dénote les quantités à la résonnance.} :
\begin{equation}
 \widehat{\omega}_{p} \tau_{0}=2\mbox{\rm \hspace{1cm}soit\hspace{1cm}}
\framebox{$\displaystyle \widehat{\omega}_{p} \tau = 4\sqrt{\ln{2}} $}
\end{equation}

La densit\'e \'electronique correspondante est donn\'ee par :
\begin{equation}                            
\framebox{$\displaystyle 
  \widehat{n}= \frac{4\ln{2}}{\pi} \frac{1}{c^{2} \tau^{2} r_{e}}$}
\label{nhat}
\end{equation}

o\`u $\tau$ est la dur\'ee \`a mi-hauteur d\'efinie dans la premi\`ere
section. On obtient $\widehat{n} [\UUc{m}{-3}] = 3.5~10^{15}/(\tau
[\Up{s}])^{2}$, soit ici $\widehat{n}= 2.8~10^{16}\UUc{m}{-3}$.  La
pression correspondante est $\widehat{P}_{r}[\U{bar}] = V_{0}
\widehat{n}/( Z_{e} N_{a} )$, o\`u $V_{0}$ est le volume molaire,
$N_{a}$ le nombre d'Avogadro, et $Z_{e}$ le nombre d'\'electrons
fournis par chaque mol\'ecule : 
 $\widehat{P}_{r}=0.5\Um{bar}$ pour de l'H\'elium.
La longueur d'onde plasma correspondante est $\widehat{\lambda}_{p}$ :
\begin{equation}                       
\framebox{$\displaystyle 
 \widehat{\lambda}_{p}= c \tau\frac{\pi}{2\sqrt{\ln{2}}} $}
\end{equation}
 
soit $\widehat{\lambda}_{p}[\Uu{m}]= 566~\tau [\Up{s}]$, et ici 
$\widehat{\lambda}_{p}= 198 \Uu{m}$.
Le facteur relativiste de Lorentz $\gamma_{p}$ associ\'e \`a la
vitesse de phase de l'onde plasma est :
\begin{equation}
 \gamma_{p}= \frac{\omega}{\omega_{p}} =\sqrt{\frac{n_{c}}{n}}=
 \frac{\lambda_{p}}{\lambda}
\end{equation}

soit :
\begin{equation}                       
 \framebox{$\displaystyle \widehat{\gamma}_{p}=
  \frac{\pi}{2\sqrt{\ln{2}}} \frac{ c \tau}{\lambda} $}
\end{equation}
 
soit $\widehat{\gamma}_{p} = 566 \tau [\Up{s}]/\lambda[\Uu{m}]$,
et ici $ \widehat{\gamma}_{p} = 188$.

%

\subsection{Champ \'electrique longitudinal}

Reprenant l'expression (\ref{elongue}) du champ \'electrique
longitudinal $E_{z}$, on obtient :
\begin{equation}
E_{z}= \exp{(-\frac{2 r^{2}}{w^{2}})}
\left[\omega_{p}^{2} \tau_{0}^{2} 
\exp{(-\frac{\omega_{p}^{2}\tau_{0}^{2}}{4})}\right]
\frac{I_{max} \lambda^{2} r_{e}}{e\tau c^{2}} 
\sqrt{\frac{\ln{2}}{\pi}}
 \end{equation}

d'o\`u\footnote{On prendra garde \`a ne pas confondre $e$, unit\'e de
  charge \'electrique, et e=exp(1).} :
\begin{equation}
\widehat{E}_{z}= \exp{(-\frac{2 r^{2}}{w^{2}})}
\left[\frac{4}{\mbox{\rm e}}\right]
\frac{I_{max} \lambda^{2} r_{e}}{e\tau c^{2}} \sqrt{\frac{\ln{2}}{\pi}}
 \end{equation}

et :
\begin{equation}                  
 \framebox{$\displaystyle (\widehat{E}_{z})_{max}= \frac{I_{max}
   \lambda^{2} r_{e}}{e\tau c^{2}} \left[\frac{4}{\mbox{\rm e}}\right]
  \sqrt{\frac{\ln{2}}{\pi}} $}
 \end{equation}

soit :
\begin{equation}                  
 (\widehat{E}_{z})_{max} [\UG{V/m}] = 1.35~10^{-18}
 \frac{I_{max}[\U{W}\UUc{m}{-2}] (\lambda[\Uu{m}])^{2}}{\tau[\Up{s}]}
\label{eq:EzDeI:tau}
\end{equation}

Ici, on obtient $ (\widehat{E}_{z})_{max}$=53.2~GV/m.

\subsection{Gain d'\'energie maximal d'un \'electron}

On n\'eglige ici le d\'ephasage des \'electrons acc\'el\'er\'es par
rapport \`a l'onde plasma au cours de la propagation et on \'ecrit
l'expression de leur gain en \'energie par la simple int\'egrale du
champ \`a la travers\'ee du waist :
\begin{equation}  
\Delta W = e \int{E_{z} \dd z}
\label{gainw}
\end{equation}

Pour un faisceau de profil transverse gaussien, on a vu que le profil
longitudinal de l'intensit\'e laser, et donc aussi de $E_{z}$, est
lorentzien, de demi-''longueur'' \`a mi-hauteur \'egale \`a la
longueur de Rayleigh. On obtient alors :
\begin{equation}                  
 \framebox{$\displaystyle \Delta W = e \pi z_{0} (E_{z})_{max} $}
\label{gainwbis}
\end{equation}

soit : 
\begin{equation} 
  \Delta \widehat{W} = e \pi M^{2} \frac{\lambda f^{2}}{\pi w_{i}^{2}}
  \frac{E}{\tau} \left(\frac{ w_{i}}{M^{2}\lambda f}\right)^{2}
  4\sqrt{\pi\ln{2}} \frac{ \lambda^{2} r_{e}}{e\tau c^{2}}
  [\frac{4}{\mbox{\rm e}}] \sqrt{\frac{\ln{2}}{\pi}}
\end{equation}

On obtient finalement :
\begin{equation}                  
 \framebox{$\displaystyle \Delta \widehat{W} = \frac{r_{e} \lambda}{c^{2}
   \tau^{2}} \frac{16\ln{2}}{\mbox{\rm e}M^{2}} E $}
\label{defgainw}
\end{equation}

Les variables $f$ et $w_{i}$ qui sont reli\'ees \`a l'ouverture de
l'optique de focalisation n'apparaissent pas dans l'expression de $
\Delta W$ : lorsque la focale diminue, le waist est plus petit, et le
champ augmente mais la longueur d'acc\'el\'eration, proportionnelle
\`a la longueur de Rayleigh, diminue, et les deux effets se
compensent.  \c{C}a donne :
\begin{equation}                  
 \Delta \widehat{W} [\UM{eV}]= 0.798
 \frac{1}{M^{2}}
 \frac{\lambda[\Uu{m}]E[\U{J}]}{(\tau[\Up{s}])^{2}}
\label{defgainwbis}
\end{equation}

, soit
$\Delta\widehat{W}$=68.6~MeV. 
Deux
effets limitent la valeur de $\Delta\widehat{W}$, la non-lin\'earit\'e
de l'onde \`a haute \'energie, et le d\'ephasage des \'electrons
acc\'el\'er\'es par rapport \`a l'onde plasma.
La variation en $E/\tau^{2}$, soit $P/\tau$, de $\Delta\widehat{W}$
montre l'int\'er\^et  d'utiliser des {\bf impulsions  courtes}.
A partir de (\ref{gainwbis}) et utilisant 
$z_{0}= \pi w_{0}^{2}/\lambda M^{2}$  et  $P=\pi w_{0}^{2} I/2$, 
on voit parfois le gain exprim\'e sous la forme : 
\begin{equation}                        
  \framebox{$\displaystyle \Delta \widehat{W}= P \frac{1}{M^{2}}
    \frac{\lambda}{\lambda_{p}} \frac{r_{e}}{c} \frac{4
      \pi^{3/2}}{\mbox{\rm e}} $}
\end{equation}

soit :
\begin{equation} 
  \Delta \widehat{W} [\UM{eV}]= 481 \frac{1}{M^{2}}
  \frac{\lambda}{\lambda_{p}} P[\UT{W}]
\end{equation}

\subsection{Amplitude de l'onde plasma longitudinale}

On obtient l'amplitude de l'onde plasma \`a l'aide de la loi de
Poisson :
\begin{equation}        
  \delta n = \frac{\epsilon_{0}}{e} \Delta \Phi =
  \frac{\epsilon_{0}}{e} \left( \frac{ \partial^{2} } {\partial z^{2}}
    + \frac{1}{r} \frac{ \partial} {\partial r} \frac{1}{r} \frac{
      \partial} {\partial r} \right) \Phi
\end{equation}

d'o\`u : 
\begin{equation}        
  \delta n = -\frac{\epsilon_{0}}{e} \left[ k_{p}^{2}+ \frac{8}{w^{2}}
    \left( 1-\frac{2 r^{2}}{w^{2}} \right) \right] \Phi
\end{equation}

En particulier en $r=0$ :
\begin{equation}        
  \delta n_{(r=0)} = -\frac{\epsilon_{0} k_{p}^{2} \Phi}{e} \left(1+
    \frac{8}{ k_{p}^{2}w^{2}} \right)
\end{equation}

On peut \'ecrire : $\delta n= \delta n_{\parallel}+ \delta n_{\perp}$,
avec :
\begin{equation}        
  \delta n_{\parallel} = \frac{\epsilon_{0}}{e} \frac{ \partial^{2}
    \Phi} {\partial z^{2}}
\mbox{\hspace{1cm}et \hspace{1cm}}
  \delta n_{\perp} = \frac{\epsilon_{0}}{e} \Delta_{\perp}\Phi
\end{equation}

On a alors  en $r=0$ :
\begin{equation} 
 \delta n= \delta n_{\parallel}
 \left(1+ \frac{8}{ k_{p}^{2}w^{2}}
   \right)
\end{equation}

et : 
\begin{equation}
  |E_{z}|= \frac{e}{\epsilon_{0}k_{p}} |\delta n_{\parallel}|= k_{p}
  \varphi
\end{equation}  

Oubliant les valeurs absolues, on obtient l'expression de  la
perturbation de densit\'e \'electronique longitudinale : 
\begin{equation}
  \delta n_{\parallel} = \exp{(-\frac{2 r^{2}}{w^{2}})} \sqrt{\pi}
  \left[\omega_{p}^{3} \tau_{0}^{3}
  \exp{(-\frac{\omega_{p}^{2}\tau_{0}^{2}}{4})}\right] \frac{I_{max}
    \lambda^{2}}{2\pi^{2} m c^{5}\tau^{2}} \ln{2}
\end{equation} 

On d\'efinit la perturbation relative de densit\'e : 
\begin{equation}
  \delta_{\parallel}= \frac{ \delta n_{\parallel} }{n}
\end{equation} 

Il vient :
\begin{equation}
  E_{z} = \frac{m c \omega_{p}}{e} \delta_{\parallel}
\label{eq:Ez:delta}
\end{equation}  

soit : 
\begin{equation}
  \framebox{$\displaystyle \delta_{\parallel}= \frac{E_{z}}{E_{0}}$}
\end{equation}  

o\`u $E_{0}$ est la valeur du champ \'electrique longitudinal de
l'onde plasma au d\'eferlement, dans l'approximation non relativiste
et de plasma froid, $E_{0} = m c \omega_{p}/e$ (c'est justement la
valeur pour laquelle la perturbation de densit\'e est maximale, cf
\'eq. (\ref{poisson})).

On en tire l'expression du gain $\Delta W$ en fonction de la
perturbation relative de densit\'e~:
\begin{equation}
  \framebox{$\displaystyle \Delta W = 2\pi^{2} mc^{2} \frac{ \delta_{\parallel} z_{0}}
{\lambda_{p}} $}
\label{gainwdedelta}
\end{equation}

Enfin, en utilisant l'expression (\ref{nhat}) de $\widehat{n}$, il
vient :
\begin{equation}              
  \framebox{$\displaystyle \widehat{\delta}_{\parallel}=
    \frac{1}{\mbox{\rm e}\sqrt{\pi}} \frac{r_{e}}{m c^{3}} I_{max}
    \lambda^{2} $}
\end{equation} 

L'expression de $\widehat{\delta}_{\parallel}$ est ind\'ependante de
$\tau$ \`a $I_{max}$ donn\'e.  On a $ \widehat{\delta}_{\parallel}=
2.38~10^{-19} I_{max}[\U{W}\UUc{m}{-2}] (\lambda[\Uu{m}])^{2} $, et
ici $\widehat{\delta}_{\parallel}= 3.28$.

\subsection{Expressions utilisant  le champ \'electrique normalis\'e}

Utilisant l'expression (\ref{champnorm}) du champ \'electrique
normalis\'e $a$ du laser, on obtient :

Pour le champ \'electrique longitudinal de l'onde plasma :

\begin{equation}
E_{z}=
 \exp{(-\frac{2 r^{2}}{w^{2}})}
\left[\omega_{p}^{2} \tau_{0}^{2}
\exp{(-\frac{\omega_{p}^{2}\tau_{0}^{2}}{4})}\right]
\frac{m c a^{2}}{e\tau}
\frac{\sqrt{\pi\ln{2}}}{2}
\end{equation} 

soit :
\begin{equation}
  E_{z}= \exp{(-\frac{2 r^{2}}{w^{2}})} \left[\omega_{p} \tau_{0}
    \exp{(-\frac{\omega_{p}^{2}\tau_{0}^{2}}{4})}\right] a^{2} E_{0}
  \frac{\sqrt{\pi}}{4}
\end{equation} 

et :
\begin{equation}         
  \framebox{$\displaystyle \widehat{E}_{z}= a^{2} E_{0}
    \frac{\sqrt{\pi}}{2\mbox{\rm e}} $}
\end{equation} 

Pour la perturbation relative de densit\'e :
\begin{equation}         
  \framebox{$\displaystyle \widehat{\delta}_{\parallel}=
    \frac{\sqrt{\pi}}{2\mbox{\rm e}}a^{2} =
    \frac{\widehat{E}_{z}}{E_{0}} $}
\label{eq:a0:delta}
\end{equation}

Les trois ph\'enom\`enes suivants apparaissent donc en m\^eme temps :
\begin{itemize}
\item la non-lin\'earit\'e de l'onde
\item le plasma relativiste
\item un champ \'electrique longitudinal de l'onde plasma approchant sa
  limite au d\'eferlement, dans l'approximation non relativiste et de
  plasma froid.
\end{itemize}

\subsection{Mode lin\'eaire}

On  d\'efinit la limite du domaine lin\'eaire par
$\delta_{\parallel} =1$.  
L'intensit\'e maximale correspondante est $\check{I}_{max}$ :  
\begin{equation}
\check{I}_{max}=
 \mbox{\rm e} \sqrt{\pi}
\frac{m c^{3}}{r_{e} \lambda^{2}}
\end{equation}

correspondant \`a : 
\begin{equation}
  \check{a}^{2}=\frac{2 \mbox{\rm e}}{ \sqrt{\pi}} 
\end{equation}

et l'on rappelle que l'on a alors :
\begin{equation}
  (\check{E}_{z})_{max}=E_{0}
\end{equation}

soit $ \check{I}_{max} [\U{W}\UUc{m}{-2}] = 4.19~10^{18}/
(\lambda[\Uu{m}])^{2} $, ici $ \check{I}_{max}= 3.78
~10^{18}\U{W}\UUc{m}{-2}$, et $ \check{a}^{2}=3.07$.
L'\'energie laser correspondante est $\check{E}= 3.05\U{J}$, le gradient
maximal $(\check{E}_{z})_{max}=\widehat{E}_{0}=16.2\UG{V/m}$ et le gain en \'energie
maximal en n\'egligeant le d\'ephasage $\Delta\check{W}$=20.9~MeV.

\subsection{Mode non-lin\'eaire}

Dans le mode non-lin\'eaire, l'onde plasma n'est plus
sinuso\"{\i}dale. 
La pente est alors tr\`es raide et
un champ \'electrique sup\'erieur
\`a $E_{0}$ peut exister\cite{esarey}.  Il s'exprime approximativement
sous la forme ${E}_{z}= a^{2} \left[1+
  a^{2}/4\right]^{-1/2}\frac{\sqrt{\pi}}{2\mbox{\rm e}} E_{0} $.
Les valeurs maximales envisagées ici ( $E=10~\U{J}$, $\delta = 3.28$ sont dans le domaines non-linéaire, et  ${E}_{z}= 28  \UG{V/m}$.

Pour des valeurs extr\^emement \'elev\'ees, on finit par atteindre le
d\'eferlement. Dans l'approximation d'un plasma froid, et en th\'eorie
1D, celui-ci est
atteint pour $E=E_{WB}= \left[2(\gamma_{p}-1)\right]^{1/2} E_{0}$,
soit ici $E_{WB}= 19.3 E_{0} = 313 \UG{V/m}$.
Lorsque l'on prend en compte la temp\'erature du plasma, 
$E_{WB}$ est r\'eduit d'un facteur proche de 2.

Quand on prend en compte les effets tri-dimensionnels, la situation
est incertaine.  Lors de simulations, des champs sup\'erieurs \`a
$E_{0}$ ont \'et\'e ``observ\'es'' (Cf. la discussion pages 258-259
de la r\'ef\'erence \cite{esarey}).


Pour un acc\'el\'erateur de particules, ce mode non-lin\'eaire
appara\^{\i}t d\'efavorable.  La raison la plus importante est
certainement la variation de $\lambda_{p}$ avec l'intensit\'e laser :
$\lambda_{p}^{NL}= \lambda_{p} (2/\pi) E_{max}/E_{0}$,
pour  $E_{max}\gg E_{0}$.
L'\'electron voit la longueur d'onde de la
structure acc\'el\'eratrice augmenter jusqu'au foyer, puis diminuer en
s'\'eloignant, d'un facteur qui peut \^etre important.
De m\^eme, les colonnes de plasma situ\'ees \`a diff\'erentes
positions dans le plan transverse verront \`a $z$ donn\'e des ondes
plasma ayant des phases et des fr\'equences diff\'erentes, avec des
surfaces iso-densit\'es \'electronique en forme de bol.
De plus, tout cela se d\'ecale dans le temps.

%

\subsection{Amplitude de l'onde plasma transverse}

L'amplitude de l'onde plasma transverse est maximale sur l'axe :
\begin{equation} 
  \delta n_{\perp}= \delta n_{\parallel} \left( \frac{8}{
      k_{p}^{2}w^{2}} \right)= 2~\delta n_{\parallel}
  \left(\frac{\lambda_{p}}{\pi w} \right)^{2}
\end{equation}
  
soit :
\begin{equation} 
  \frac{ \delta n_{\perp}}{ \delta n_{\parallel}}=
\left( \widehat{\frac{E_{r}}{E_{z}} } \right)^{2}
\end{equation}
  



\clearpage 

\section{Effets du champ transverse sur le faisceau acc\'el\'er\'e}\label{sec:transverse}

Les champs transverse et longitudinal d'une EPW sont décalés de $\pi/2$
(eq. \ref{elongit}, \ref{etransverse}).
L'axe de phase peut \^etre segment\'e en quatre parties :
\begin{center}
\begin{tabular}{cccccccccc}
$\zeta$ & 0 & & $\pi/2$ & & $\pi$ & & $3\pi/2$ & & $2\pi$\\ \hline
  & $|$ &  acc\'el. & $|$ &  acc\'el. & $|$ & d\'ec\'el. & $|$ & d\'ec\'el. & $|$ \\ \hline
  & $|$ & d\'efoc. & $|$ & foc. & $|$ & foc. & $|$ &  d\'efoc. & $|$ \\
\end{tabular}
\end{center}

Seul le quart focalisant, accélérateur est d'intérêt ici.

\subsection{(D\'e)focalisation}

L'amplitude du champ transverse $E_{r}$ \'etant proportionnelle \`a $r$ pr\`es de
l'axe, l'angle de d\'eflection l'est aussi au premier ordre, et l'onde
plasma est \'equivalente \`a une lentille. A partir de la rigidit\'e
$K_{r}$ :
\begin{equation} 
K_{r}=\frac{F_{r}}{\gamma m c^{2} r}
\end{equation}

et de l'expression (\ref{etransverse}) de $E_{r}$, on obtient :
\begin{equation} 
 K_{r}=\frac{2\sqrt{\pi}}{\mbox{\rm e}}
\frac{a_{0}^{2}}{\gamma w_{0}^{2}}
\frac{\exp{(-2 r^{2}/w^{2})} \sin{(\omega_{p}t-k_{p}z)}}
{\left[1 + (z/z_{0})^{2}\right]^{2}}
\end{equation}

La  fonction bétatronique de l'onde plasma au waist vaut :

\begin{equation} 
\beta =
\frac{1}{\sqrt{K_{r}}}=
 \frac{w_{0}}{2}\sqrt{\frac{\gamma}{\delta\sin{\zeta}}},
\label{eq:fn_beta}
\end{equation}

où $\zeta=\omega_{p}t-k_{p}z$.


\clearpage 

\clearpage

\tableofcontents

\end{document}